\documentclass[a4paper,usenatbib]{mnras}
\usepackage{graphicx}	% Including figure files
\usepackage{amsmath}	% Advanced maths commands
\usepackage{amssymb}	% Extra maths symbols
\usepackage{bm}		    % Bold maths symbols, including upright Greek
\usepackage{pdflscape}	% Landscape pages

\usepackage{booktabs}
\usepackage{multicol}   % Multi-column entries in tables
\usepackage{multirow}

\usepackage{caption}
\usepackage{units}

\usepackage{hyperref}

\usepackage{placeins}

\usepackage{lineno}
\usepackage[T1]{fontenc}
\usepackage{ae,aecompl}

\usepackage{newtxtext,newtxmath}
\title[DESI LRG $\times$ Planck lensing]{Cross-Correlation of Planck CMB Lensing with DESI-Like LRGs}

\author[E. Kitanidis et al.]{Ellie Kitanidis$^{1,2}$ and Martin White$^{1,2}$ \\ \\ 
$^{1}$ Department of Physics, University of California, Berkeley, 366 LeConte Hall, Berkeley, CA 94720, USA\\
$^{2}$ Lawrence Berkeley National Laboratory, 1 Cyclotron Road, Berkeley, CA 93720, USA\\}

\date{Accepted 2020 December 14. Received 2020 November 24; in original form 2020 October 9.}

\pubyear{2020}

\begin{document}
\label{firstpage}
\pagerange{\pageref{firstpage}--\pageref{lastpage}}
\maketitle

\begin{abstract}
Cross-correlations between the lensing of the cosmic microwave background (CMB) and other tracers of large-scale structure provide a unique way to reconstruct the growth of dark matter, break degeneracies between cosmology and galaxy physics, and test theories of modified gravity. We detect a cross-correlation between DESI-like luminous red galaxies (LRGs) selected from DECaLS imaging and CMB lensing maps reconstructed with the Planck satellite at a significance of $S/N = 27.2$ over scales $\ell_{\rm min} = 30$, $\ell_{\rm max} = 1000$. To correct for magnification bias, we determine the slope of the LRG cumulative magnitude function at the faint limit as $s = 0.999 \pm 0.015$, and find corresponding corrections on the order of a few percent for $C^{\kappa g}_{\ell}, C^{gg}_{\ell}$ across the scales of interest. We fit the large-scale galaxy bias at the effective redshift of the cross-correlation $z_{\rm eff} \approx 0.68$ using two different bias evolution agnostic models: a HaloFit times linear bias model where the bias evolution is folded into the clustering-based estimation of the redshift kernel, and a Lagrangian perturbation theory model of the clustering evaluated at $z_{\rm eff}$. We also determine the error on the bias from uncertainty in the redshift distribution; within this error, the two methods show excellent agreement with each other and with DESI survey expectations.
\end{abstract}


%%%%%%%%%%%%%%%%%%%%%%%%%%%%%%%%%%%%%%%%%%%%%%%%%%

%%%%%%%%%%%%%%%%% BODY OF PAPER %%%%%%%%%%%%%%%%%%

\section{Introduction}

Modern cosmology hinges on observations of the large-scale structure of the Universe, which is rich with clues about gravity, dark energy, and the mechanisms of cosmic expansion. Next-generation galaxy surveys, including spectroscopic experiments such as the Dark Energy Spectroscopic Instrument (DESI, \citealt{DESI16}) and deep imaging experiments such as the Large Synoptic Survey Telescope (LSST, \citealt{LSST09}), will map billions of galaxies in the coming decade and tighten constraints on key fundamental parameters. While spectroscopic redshifts can be obtained for some subset of imaged galaxies, the majority will increasingly rely on photometric redshift estimates (see e.g. \citealt{Hogg98} and references contained therein) or clustering-based redshift estimates (see e.g. \citealt{Newman08} and references contained therein), enabling higher number density but noisier catalogs of galaxy positions.

Measurements of the cosmic microwave background (CMB) provide another window into the growth of large-scale structure, due to the lensing of the CMB photons as they free-stream through the Universe and are deflected (on the order of a few arcminutes) by the gravitational potentials of matter in their path. In the weak regime, gravitational lensing remaps the CMB temperature and polarization primary anisotropies in predictable ways that can be exploited to reconstruct high resolution maps of the projected matter density over the past 13 billion years (\citealt{ZaldarriagaSeljak99}, \citealt{HuOkamoto02}, \citealt{LewisChallinor06}). Detections of this mass lensing signal from the CMB have been made in a number of ways, including cross-correlations with other tracers of large-scale structure (see e.g.\ \citealt{PlanckI,Omori++19} for recent lists; also \citealt{Krolewski19}).

CMB lensing offers the advantage of directly probing the underlying distribution of dark matter, but suffers from information loss since it is a two-dimensional projection of the three-dimensional matter density integrated along the line of sight from the surface of last scattering $z \approx 1100$ to the present day. In contrast, galaxy samples with narrow redshift windows are relatively well localized in position but are biased tracers of dark matter due to the complex processes involved in galaxy formation. This leads to degeneracies between these galaxy bias parameters and cosmological parameters of interest such as $\sigma_8$ -- with recent surveys reporting a range of different inferences about the clustering amplitude \citep[e.g.][and references therein]{PlanckVI,Troxel18,Hikage19,Troster20,Philcox20,eBOSS20}. Cross-correlations between CMB lensing and galaxy catalogs thus provide a means to chart the growth of dark matter with time and break the degeneracy between galaxy physics and cosmology. Additionally, on a practical level, systematics in the galaxy sample are unlikely to be correlated to systematics in the CMB lensing maps, and a higher degree of uncertainty in the galaxy redshift distribution can also be tolerated due to the broad redshift kernel of the CMB lensing.

In this work, we leverage the high number density and completeness of the luminous red galaxy (LRG) target class as defined by DESI and selected from deep multi-band imaging, in combination with the all-sky CMB lensing convergence maps of the Planck collaboration \citep{PlanckVIII}, to detect a galaxy-matter cross-correlation at high significance out to small scales, $\ell_{\rm max} = 1000$. We jointly model the angular auto- and cross- spectra to probe the amplitude and evolution of the galaxy bias. In the absence of spectroscopic redshifts, we use a combination of photometric and clustering-based estimations of the galaxy redshift distribution. Within a simple linear bias model $P_{\rm gg}(k,z) \approx b_{\rm g}(z)^2 P_{\rm mm}(k,z)$, the advantage of the clustering-based method is that it allows us to measure an effective bias without assuming a bias evolution model. By comparing the results using photometric versus clustering redshift distributions, we also evaluate the impact of the uncertainty in the redshift distribution on the inferred parameters.

This paper is organized as follows: Section~\ref{sec:data} describes the lensing products and imaging data, and outlines the construction of the DESI-like LRG catalog. In Section~\ref{sec:dndz_pipe}, we characterize the redshift distribution of the galaxy sample based on angular cross-correlations with external spectroscopic catalogs, and present a framework for probing bias evolution using these results. Section~\ref{sec:master_pipe} outlines our methods for measuring and modelling angular power spectra and covariances on a partial sky. Section~\ref{sec:magbias} is devoted to determining and applying corrections for the effects of magnification bias. In Section~\ref{sec:results}, we present and model the resulting spectra, with Section~\ref{sec:results/halofit} fitting the linear Eulerian galaxy bias under the HaloFit \citep{Smith++03} prescription while Section~\ref{sec:results/lpt} interprets the results within a Lagrangian perturbation theory framework. Finally, in Section~\ref{sec:conclusions}, we summarize our findings and suggest future directions. 

Throughout, we work in co-moving coordinates and assume the fiducial cosmology of the Planck 2018 results (\citealt{PlanckVI}, Table 2, Column 7). All magnitudes are quoted as AB magnitudes, unless otherwise specified.

\section{Data}\label{sec:data}

\subsection{Planck CMB lensing maps}

Using the most recent reconstructed lensing convergence maps and analysis masks provided in the Planck 2018 release\footnote{\url{https://wiki.cosmos.esa.int/planck-legacy-archive}} \citep{PlanckVIII}, we focus mainly on the baseline estimates obtained from the SMICA DX12 CMB maps with a minimum-variance (MV) estimate determined from both the temperature and polarization maps. To gauge the impact of the thermal Sunyaev-Zeldovich (tSZ) effect, which has been shown to bias the lensing reconstruction and contaminate cross-correlations with other tracers of large-scale structure (see e.g. \citealt{Osborne++14, vanEngelen++14, Madhavacheril++18, Schaan++19}), we also repeat the analysis using the lensing estimate obtained from a temperature-only SMICA map where tSZ has been deprojected using multifrequency component separation. Throughout the remainder of this paper, these two lensing maps will be referred to as \texttt{BASE} and \texttt{DEPROJ}, respectively. 

The spherical harmonic coefficients of the reconstructed lensing convergence maps are provided in HEALPix\footnote{\url{http://healpix.sf.net}} \citep{Gorski++05} format with maximum order $\ell_{\text{max}}=4096$, and the associated analysis masks are given as HEALPix maps with resolution $N_{\text{SIDE}} = 2048$. The approximate lensing noise power spectrum for the fiducial cosmology used in \cite{PlanckVIII} is also provided up to $\ell_{\rm max} = 4096$. To minimize information loss, we use the resolution of the Planck mask, $N_{\text{SIDE}}=2048$, as the resolution for our analysis. We consider the full lensing harmonics up to $\ell_{\rm max} = 4096$ and do not encounter any numerical issues associated with the noise spike at high $\ell$, which may become more significant when attempting to downgrade the map to lower resolution while there is significant power at the pixel level.

\subsection{Photometric DESI LRGs}

The Dark Energy Spectroscopic Instrument (DESI; \citealt{DESI16}) is an upcoming Stage IV\footnote{As defined in the Dark Energy Task Force report \citep{DarkEnergyTaskForce}.} dark energy experiment, installed on the Mayall 4m telescope at Kitt Peak. DESI aims to produce the largest ever three-dimensional map of the universe, with a massively multiplexed spectrograph that uses robotic fiber positioners to measure as many as 5000 spectra in parallel. Among the four main classes targeted by DESI are luminous red galaxies (LRGs) out to $z \approx 1$. LRGs, as their name suggests, are luminous and intrinsically red due to their high stellar mass and lack of recent star formation activity. LRGs are excellent tracers of large-scale structure; as early-type galaxies with generally old populations of stars, they are expected to reside in massive halos and therefore cluster strongly. Furthermore, their inherent brightness and the strong 4000\AA \ feature in their spectral energy distributions enable the efficient selection of a homogeneous sample using photometry.

\subsubsection{DECaLS imaging data}

%https://arxiv.org/pdf/1906.00970.pdf

The DECam Legacy Survey (DECaLS) is a deep, wide-field survey providing the optical imaging used to conduct targeting for approximately two-thirds of the DESI footprint, covering the region bounded by $\delta \lesssim 32^{\circ}$. Through the DECam instrument \citep{Flaugher15} on the Blanco 4m telescope, DECaLS observes in three optical and near-IR bands ($g$, $r$, $z$), with four additional mid-IR bands ($W1$, $W2$, $W3$, $W4$) provided by the Wide-field Infrared Survey Explorer (WISE; \citealt{Wright10}). DECam images are processed and calibrated though the National Optical Astronomy
Observatory (NOAO) Community Pipeline, then fed into \textit{The Tractor}\footnote{\url{https://github.com/dstndstn/tractor}} \citep{Lang16a}, which uses forward-modeling to perform source extraction and produce probabilistic inference of source properties.

%Raw DECam images are processed through the NOAO Community Pipelines, with astrometric calibration and photometric characterization based on Pan-STARRS-1 measurements. The calibrated images are then run through \textit{The Tractor}\footnote{\url{https://github.com/dstndstn/tractor}} \citep{Lang16a}, which produces an inference-based catalog by optimizing the likelihood for source properties, given the data and a noise model.

Our analysis is based on Data Release 8 (DR8), the latest data release of the Legacy Survey \citep{Dey18}, which contains DECaLS observations from August 2014 through March 2019 (NOAO survey program 0404). DR8 also includes some non-DECaLS observations from the DECam instrument, mainly from the Dark Energy Survey (DES; \citealt{DES05}). In total, the DECaLS +DES portion of DR8 covers approximately 14,996 square degrees in the $g$-band, 15,015 square degrees in the $r$-band, 15,130 square degrees in the $z$-band, and 14,781 square degrees in all three optical bands jointly.\footnote{Estimated from using randoms distributed uniformly across the footprint to sum up the areas with at least one exposure in each band.} 

\subsubsection{Galaxy selection}\label{sec:data:ts}

\begin{figure}
\centering
\includegraphics[width=0.7\columnwidth, trim={0.35cm 0 0 0},clip]{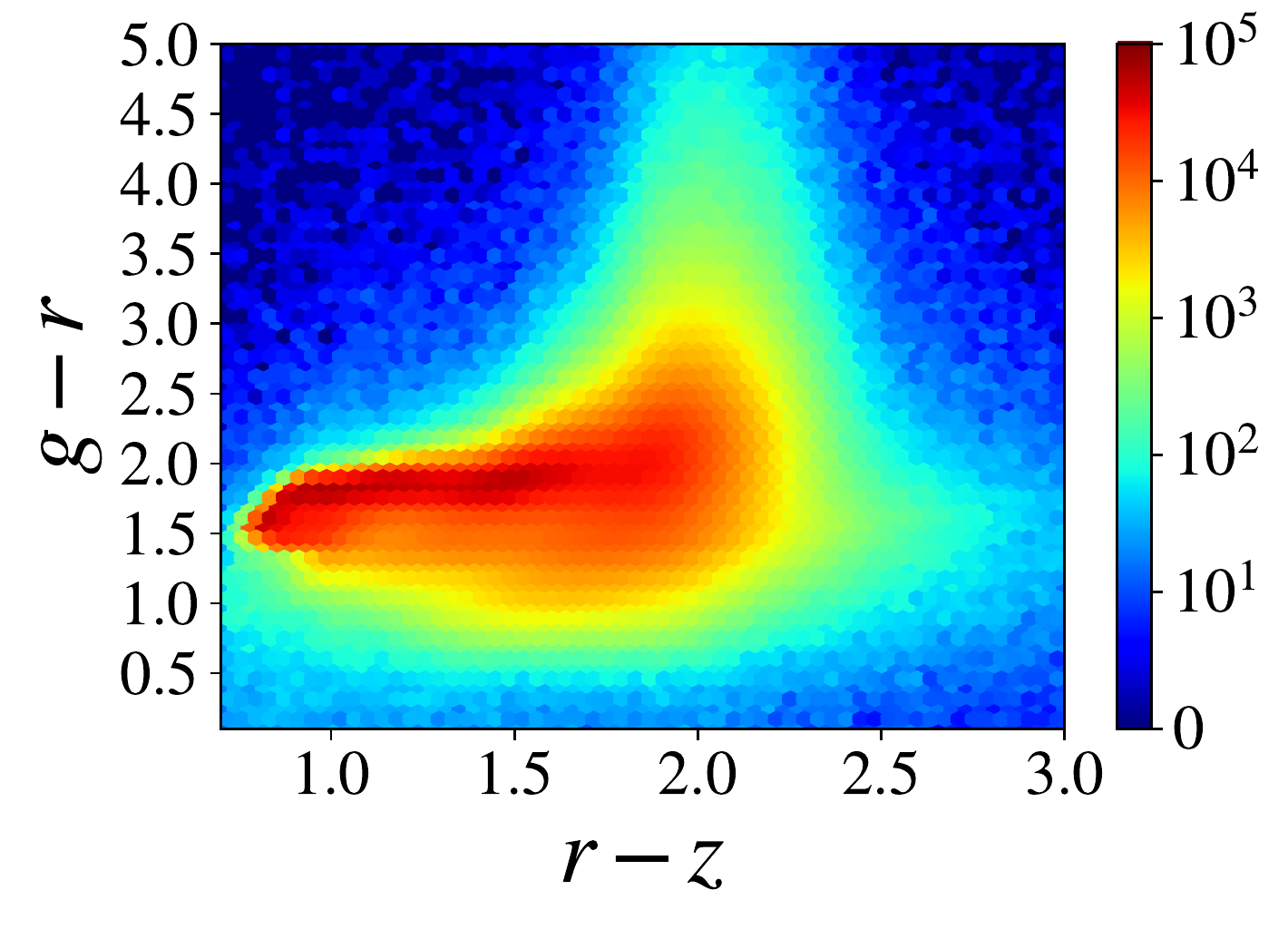}
\includegraphics[width=0.7\columnwidth, trim={0.35cm 0 0 0},clip]{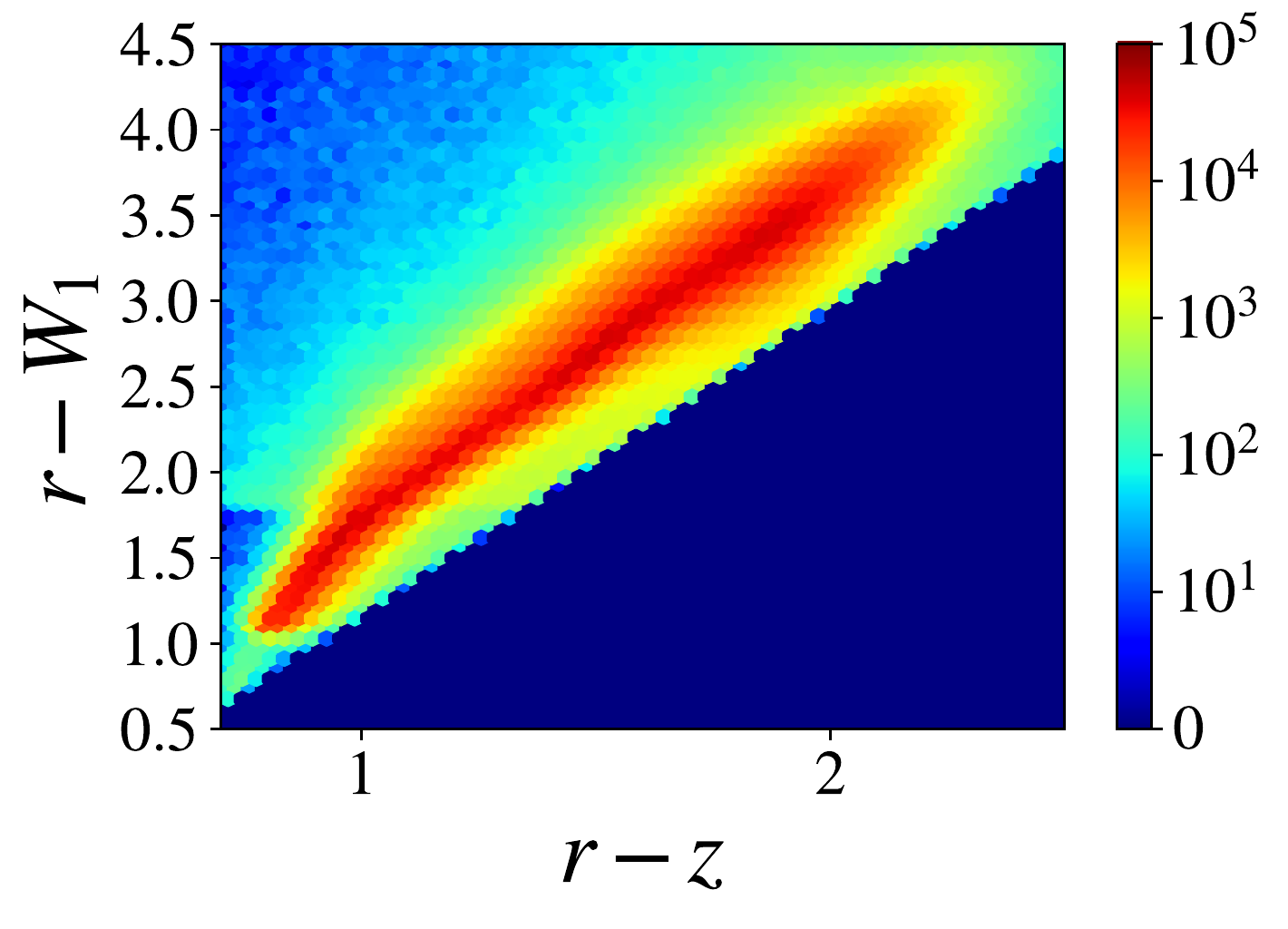}
\caption{Color-color plots of the LRG target selection in DECaLS DR8, with the color bar representing the total number of targets.}
\label{fig:colorcolor}
\end{figure}

DESI LRGs are selected from DECaLS by applying a complex series of color cuts on extinction-corrected magnitudes in $g$, $r$, $z$, and $W1$ bands:

\begin{equation}
\begin{split}
z_{\rm fiber} &< 21.5 \\ %
r - z &> 0.7 \\ %
(z - W1) > 0.8 &\ (r - z) - 0.6 \\ %
((g - W1 > 2.6) \text{ AND } (g - r > &1.4)) \text{ OR } (r - W1 > 1.8) \\ % 
(r - z > (z - 16.83) \ 0.45) \text{ AND } &(r - z > (z - 13.80) \ 0.19) \\%
\end{split}
\end{equation} \\ 
We note that the faint magnitude limit uses fiber flux, which is defined as the flux within a 1.5 arcsec diameter circular aperture centered on the model convolved with a 1.0 arcsec FWHM Gaussian. Color-color plots of the resulting sample are displayed in Figure~\ref{fig:colorcolor}.
%
%We can further split the LRGs into two subsamples with different mean redshifts by applying an additional photometric cut. Motivated by eBOSS studies, \textcolor{red}{Kitanidis et al. 2019} divided the DESI LRG catalog along $r-W1=2.6$ and calculated the clustering $dN/dz$ for each, showing that the result is two subsamples of approximately equal size, with one at $\bar{z} \approx 0.5$ and the other at $\bar{z} \approx 0.8$. This binning is also helpful for minimizing the effect of bias evolution in the clustering $dN/dz$ estimation \citep{Menard13, Schmidt13, Rahman15, Gatti18}.

\subsubsection{Masks}
\label{sec:data:masks}

Instrument effects and transients create artifacts in the images which may impact the detection or fitting of sources. Additionally, bright foregrounds, including point sources such as stars and extended sources such as large galaxies, globular clusters, and planetary nebulae, can contaminate the pixels around them with false targets, thereby affecting the apparent angular distribution of the target sample. DR8 provides bitmasks which leverage the NOAO Community Pipeline's data quality map, as well as several external catalogs, to reject bad pixels and mask around foregrounds. The bits we use in our analysis are summarized in Table~\ref{tab:masks} and briefly described below:

The \texttt{ALLMASK\_X} bits are set for pixels that touch a bad pixel (as flagged by the NOAO Community Pipeline) in all of the overlapping $X$-band images. The \texttt{WISEM1} and \texttt{WISEM2} bits are set for pixels that touch a pixel in a mask around bright stars from the WISE catalog, with the two masks using the $W1$ and $W2$ bands, respectively. The \texttt{MEDIUM} bit is set for pixels that touch a pixel containing a medium-bright ($\texttt{phot\_g\_mean\_mag} < 16$) star from the Gaia DR2 catalog \citep{GaiaDR2} or a bright ($VT < 13$) star from the Tycho-2 catalog \citep{Hog00}. The \texttt{GALAXY} bit is set for pixels that touch a pixel containing a large galaxy, where the source catalog used for this mask is taken from John Moustakas' Legacy Survey Large Galaxy Atlas\footnote{\url{https://github.com/moustakas/LSLGA}} work with Dustin Lang. Finally, clusters and nebulae from OpenNGC\footnote{\url{https://github.com/mattiaverga/OpenNGC}} are masked around using a circular mask whose diameter is equal to the major axis of the object being masked, and the \texttt{CLUSTER} bit is set for pixels touching this mask.

As demonstrated in Table~\ref{tab:masks}, masking near foreground stars causes the largest cut in observed objects. To determine whether any additional stellar masking is warranted, we measure the density of targets as a function of proximity to stars after the above bitmasks have been applied. Using the Tycho-2 and WISE catalogs, we first bin the stars by their magnitudes (using the $VT$ and $W1$ bands, respectively), and then determine the density of LRGs in annular bins around these stacks of stars. We find that there are still residual effects near Tycho-2 stars, particularly for the brightest bins, that are not entirely captured by the bitmasks. We find even more significant effects around WISE stars, with the LRG density peaking beyond the radius of the bitmasks. We fit a magnitude-dependent masking radius for each star catalog to apply as additional geometric masks: 
\begin{align}\label{eq:mr-veto}
R = 
\begin{cases}
    10^{\ 3.41 \ - \ 0.16 \ \times \ VT} \ \text{arcsec}, \hspace{0.5cm} \text{Tycho-2} \\
    10^{\ 2.87 \ - \ 0.13 \ \times \ W1} \ \text{arcsec}, 
    \hspace{0.5cm} \text{WISE}
\end{cases}
\end{align}
The addition of the geometric masks results in a slight increase in the total masked area.
\begin{table}
    {\centering
\noindent\begin{tabular}{p{1.0cm}p{1.5cm}p{1.3cm}p{1.4cm}p{0.6cm}}
\toprule
\multicolumn{2}{c}{Mask} & Number & Area (deg$^2$) & $f_{\rm survey}$ \\
\hline
\multicolumn{2}{c}{\textbf{no masks}} & 9003243 & 14610.72 & 1.000 \\
\hline
\multirow{8}{*}{bits}    & \texttt{ALLMASK\_G}  & 9002762 & 14610.72 & 1.000 \\
                          & \texttt{ALLMASK\_R}  & 9002742 & 14610.72 & 1.000 \\
                          & \texttt{ALLMASK\_Z}  & 9002458 & 14610.72 & 1.000 \\
                          & \texttt{WISEM1}  & 8578461 & 14230.96 & 0.974 \\
                          & \texttt{WISEM2}  & 8679070 & 14406.05 & 0.986 \\
                          & \texttt{MEDIUM}  & 8566358 & 13945.27 & 0.954 \\
                          & \texttt{GALAXY}  & 8996317 & 14599.17 & 0.999 \\
                          & \texttt{CLUSTER} & 9003232 & 14609.73 & 1.000 \\
\cmidrule{2-5}
                          & all bits     & 8559863 & 13933.29 & 0.954 \\
\hline
\multirow{2}{*}{geometric} & Tycho-2  & 8675511 & 14181.29 & 0.971 \\
                           & WISE    & 8488111 & 14094.18 & 0.965 \\
\cmidrule{2-5}
                           & all geometric    & 8399015 & 13859.42 & 0.949 \\
\hline
\multicolumn{2}{c}{\textbf{all masks}} & 8390823 & 13851.50 & 0.948 \\
\bottomrule
\end{tabular}}
    \caption{Summary of foreground masks.}
    \label{tab:masks}
\end{table}

\subsubsection{Tests of potential systematics}

Astrophysical foregrounds, poor observing conditions, and systematic errors in instrument calibration or data reduction can introduce non-cosmological density variations in the galaxy sample, which may in turn bias cosmological analyses (see e.g.\ \citealt{Myers06}, \citealt{Crocce11}, \citealt{Ross11}, \citealt{Suchyta++16}, \citealt{Crocce++16}, \citealt{Leistedt++16}, \citealt{ElvinPoole18}, \citealt{Ross20}, \citealt{Weaverdyck20} for studies of imaging systematics in the context of other surveys). A full analysis of the effect of imaging systematics  on the clustering of DESI main targets using data from DECaLS DR7 is presented in \citealt{Kitanidis++19}. Here, we briefly perform tests of the LRG density dependence on these potential systematics using DR8 data and target selection.  

\begin{figure*}
\includegraphics[width=0.24\linewidth, trim={1.5cm 0 1.2cm 1.5cm},clip]{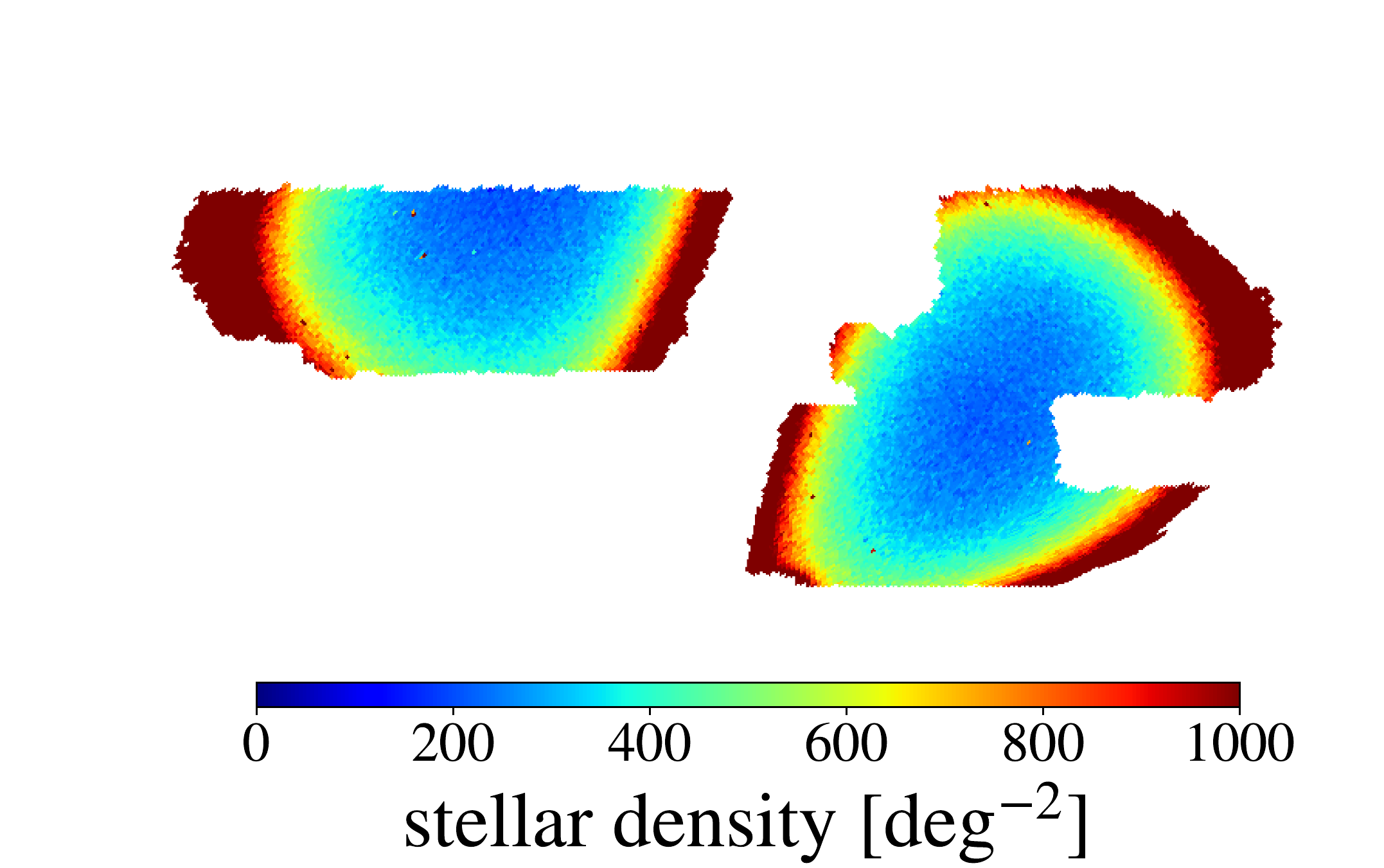}
\includegraphics[width=0.24\linewidth, trim={1.5cm 0 1.2cm 1.5cm},clip]{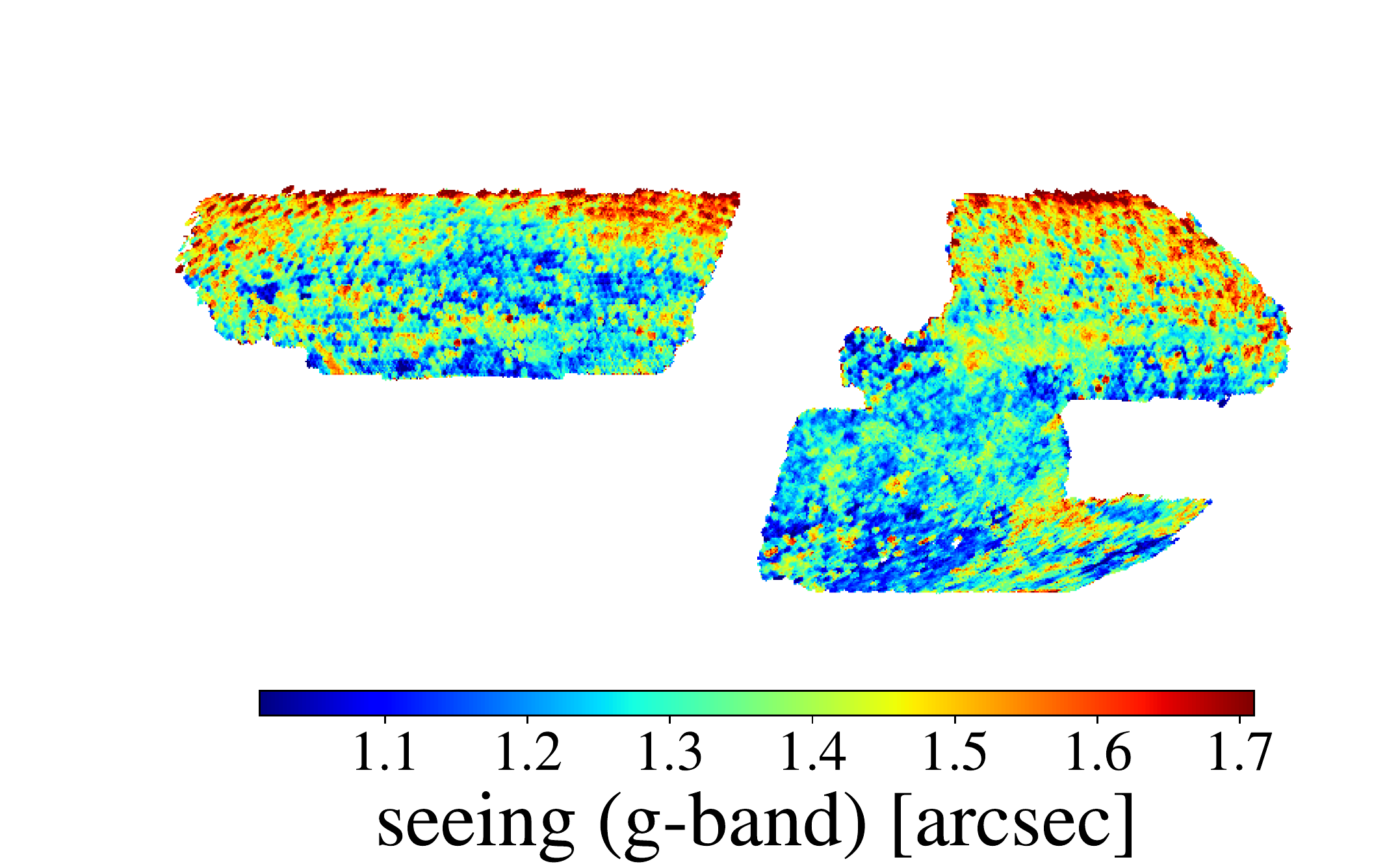}
\includegraphics[width=0.24\linewidth, trim={1.5cm 0 1.2cm 1.5cm},clip]{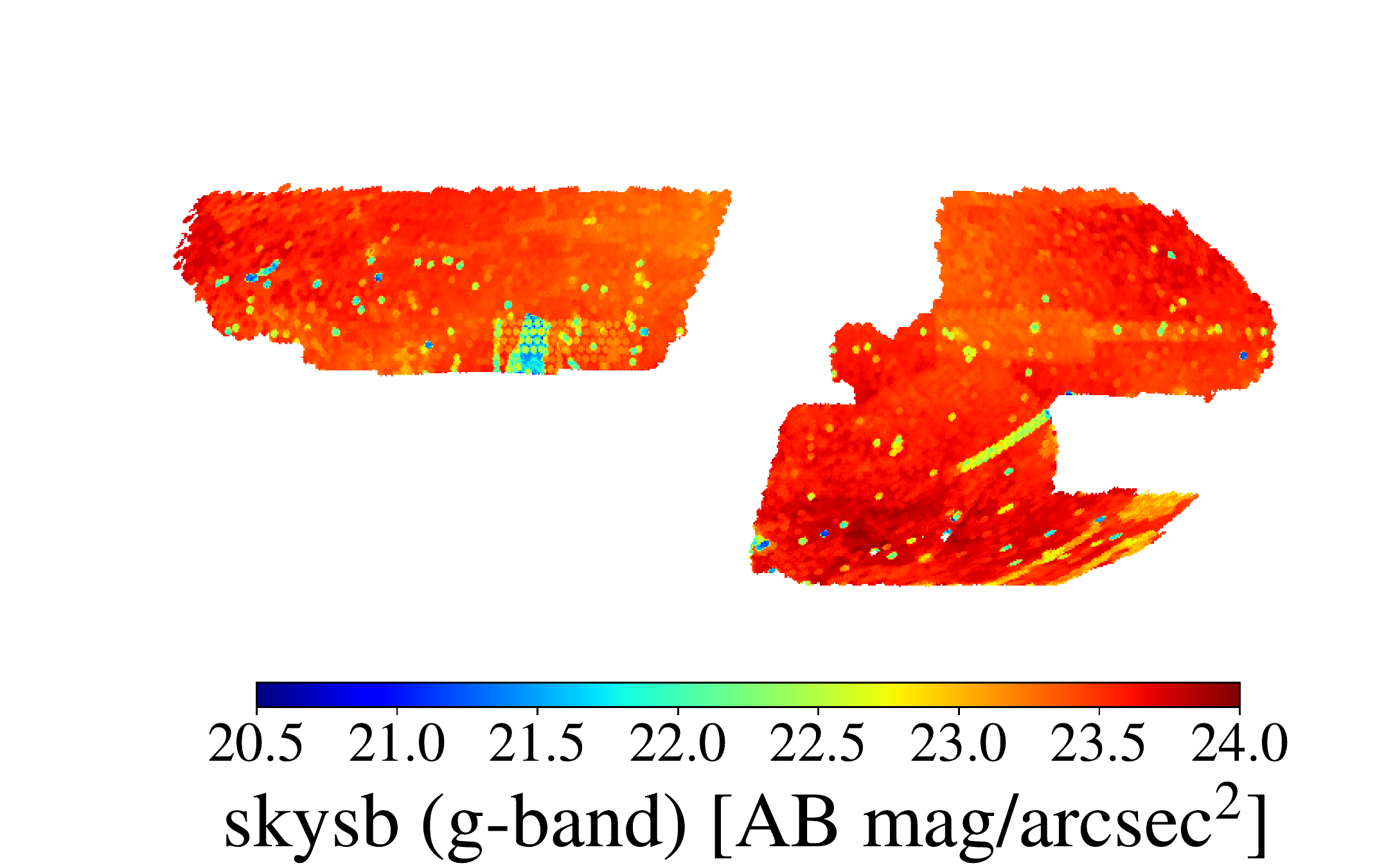}
\includegraphics[width=0.24\linewidth, trim={1.5cm 0 1.2cm 1.5cm},clip]{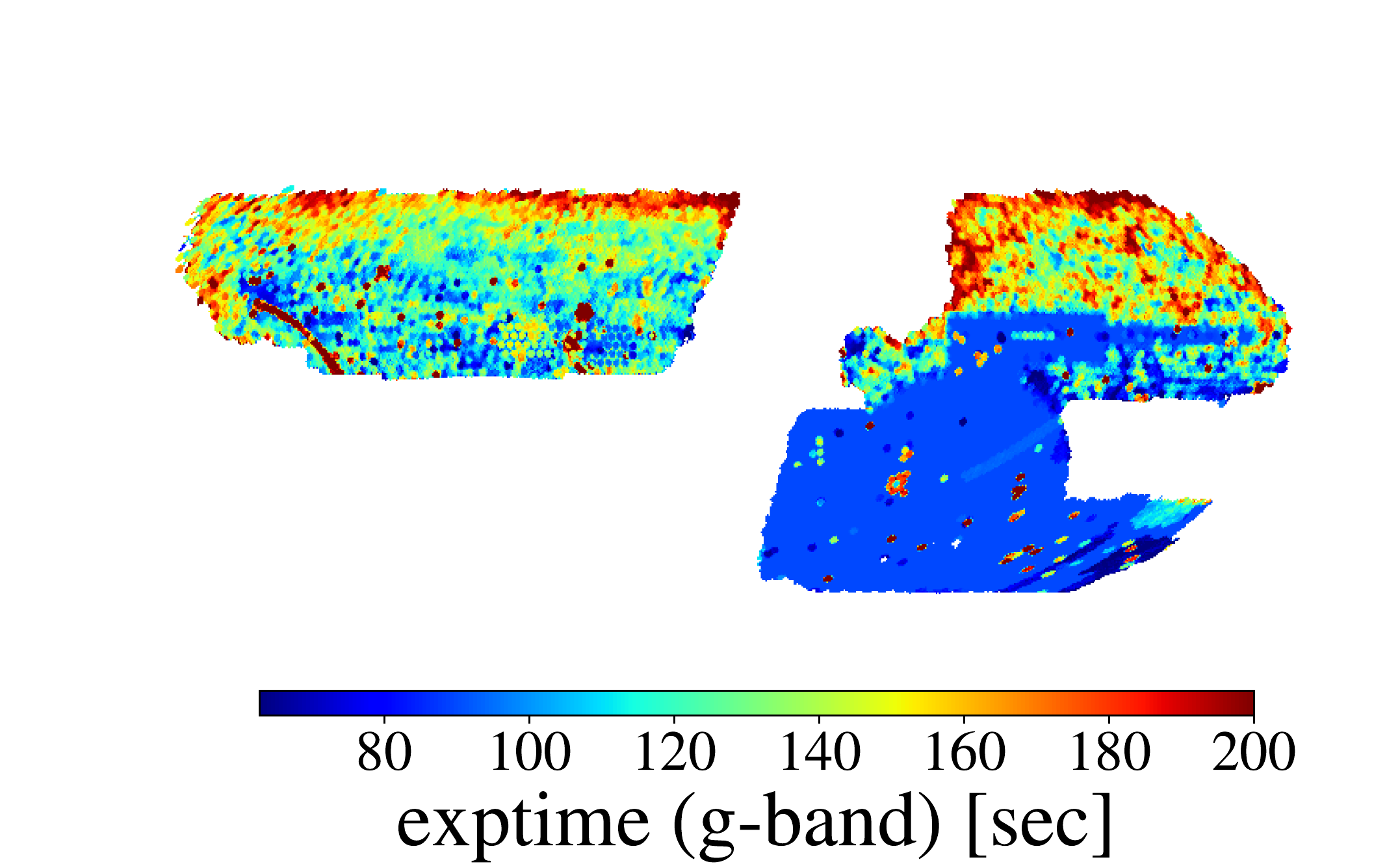}
\includegraphics[width=0.24\linewidth, trim={1.5cm 0 1.2cm 1.5cm},clip]{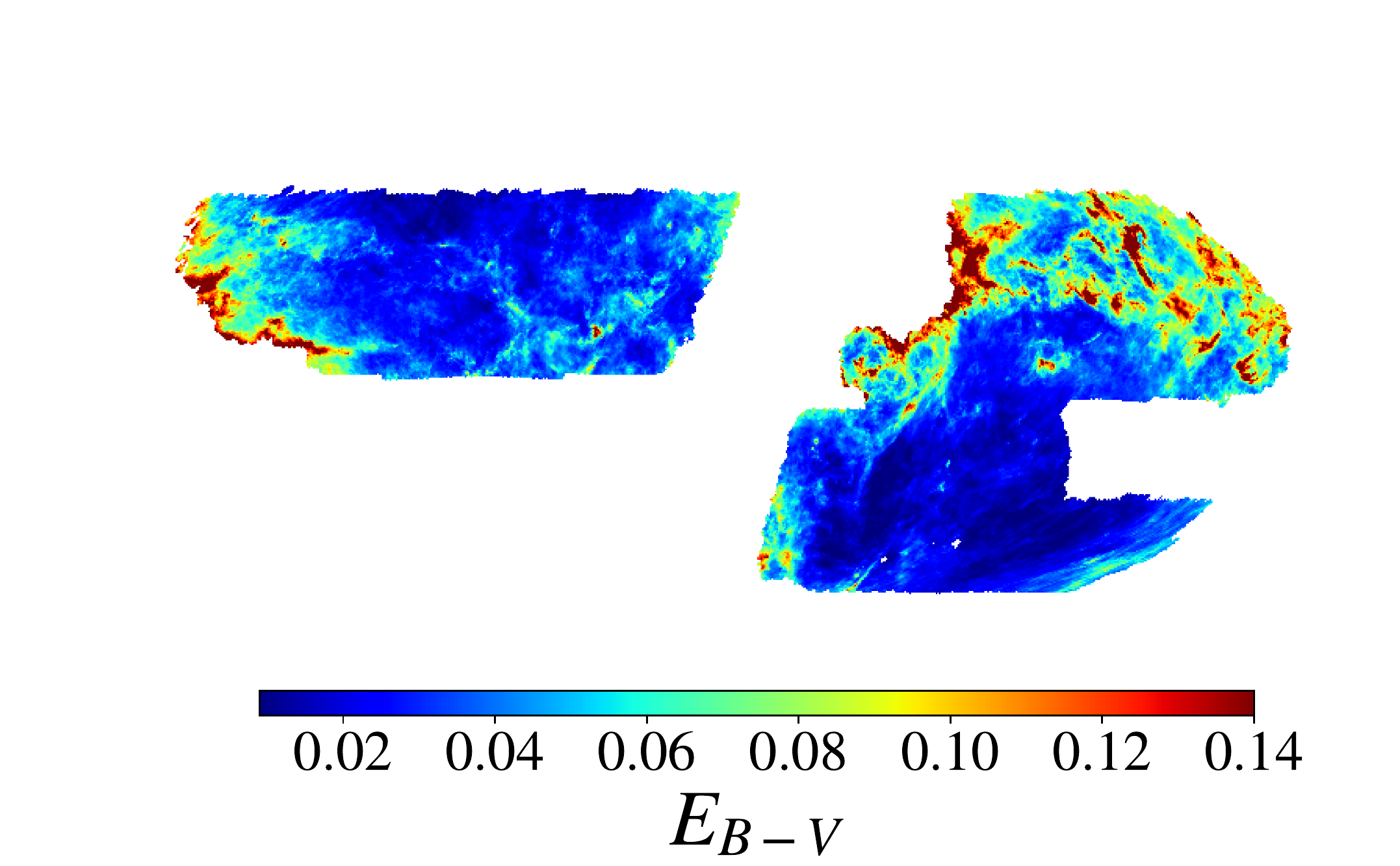}
\includegraphics[width=0.24\linewidth, trim={1.5cm 0 1.2cm 1.5cm},clip]{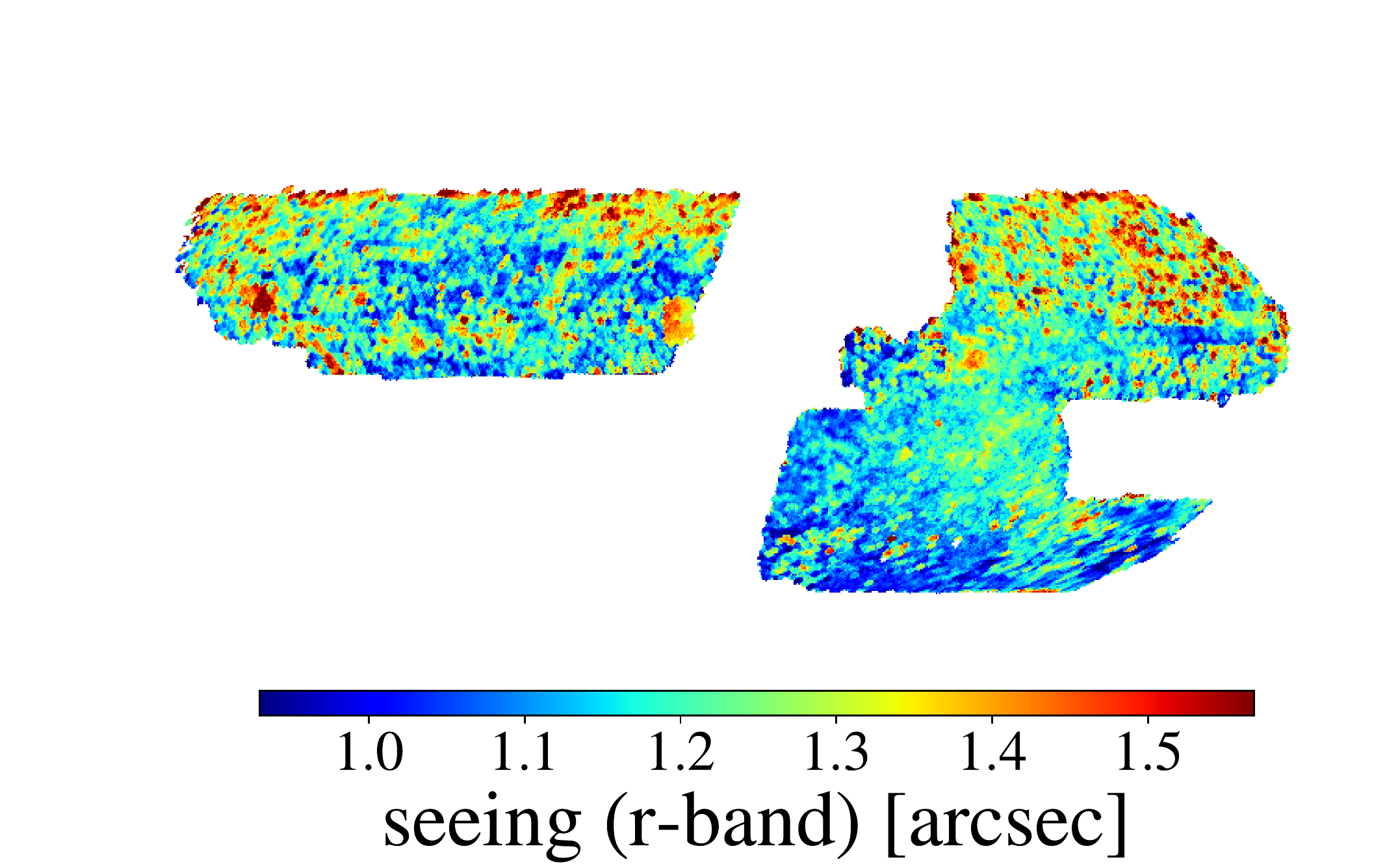}
\includegraphics[width=0.24\linewidth, trim={1.5cm 0 1.2cm 1.5cm},clip]{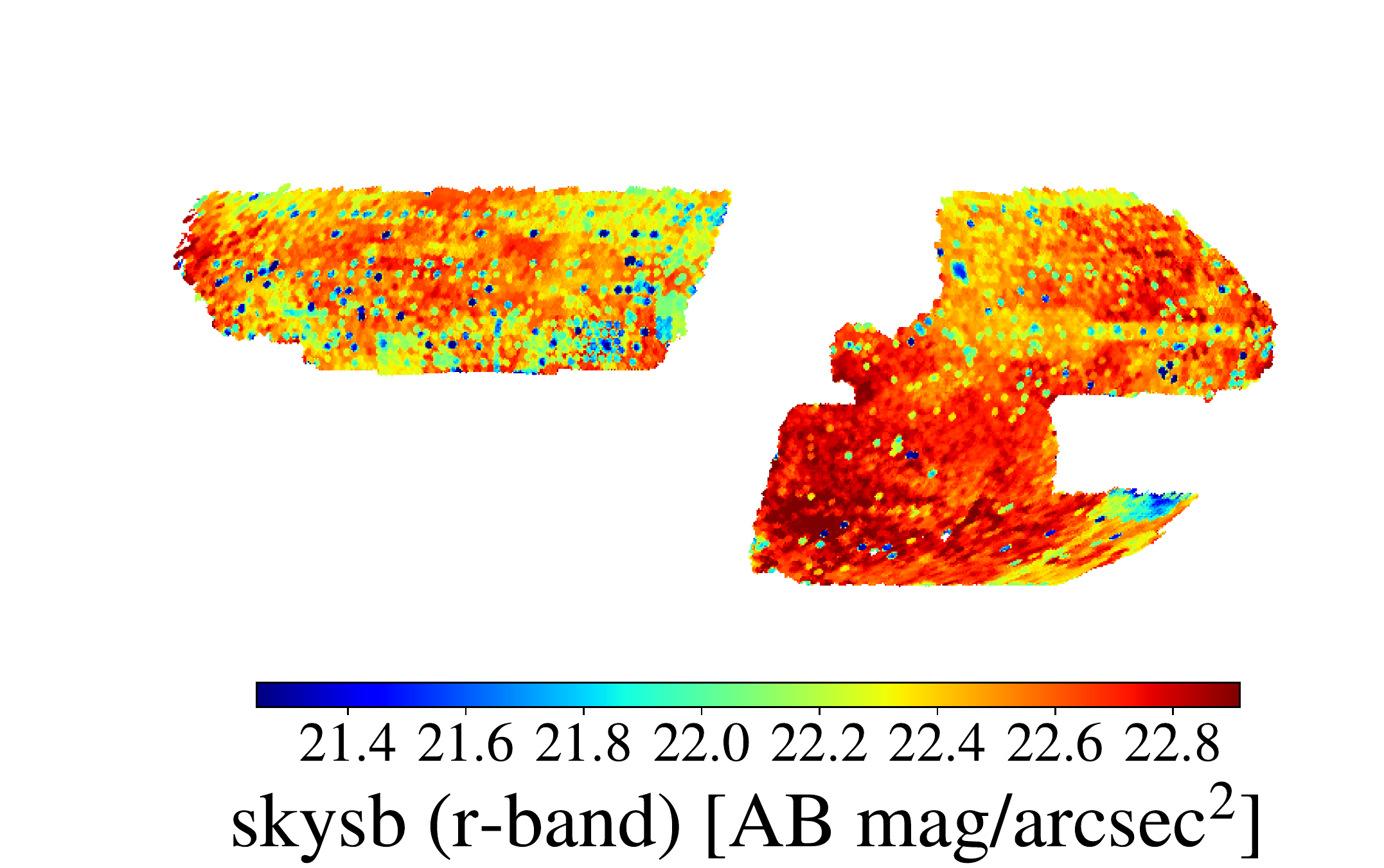}
\includegraphics[width=0.24\linewidth, trim={1.5cm 0 1.2cm 1.5cm},clip]{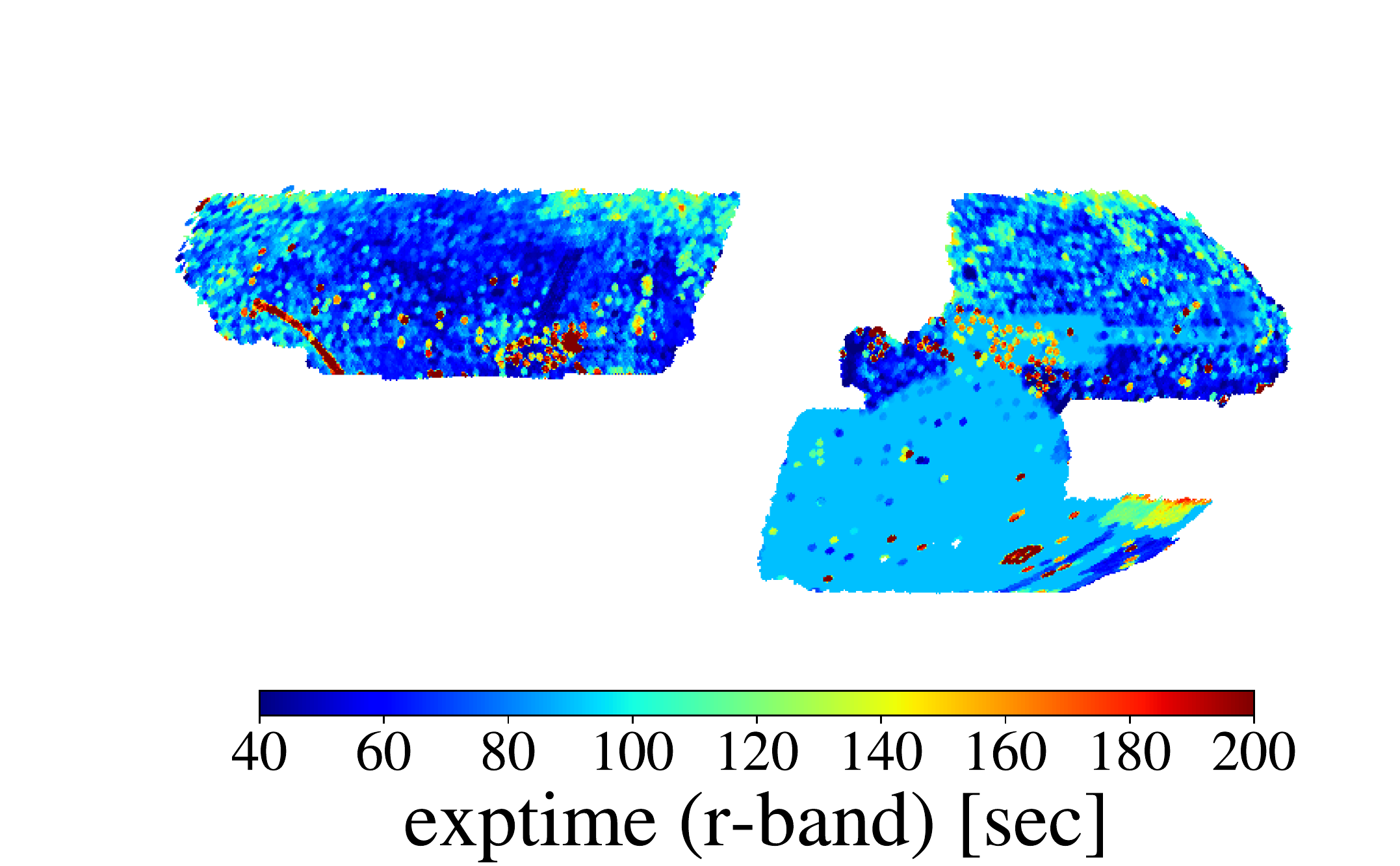}
\includegraphics[width=0.24\linewidth, trim={1.5cm 0 1.2cm 1.5cm},clip]{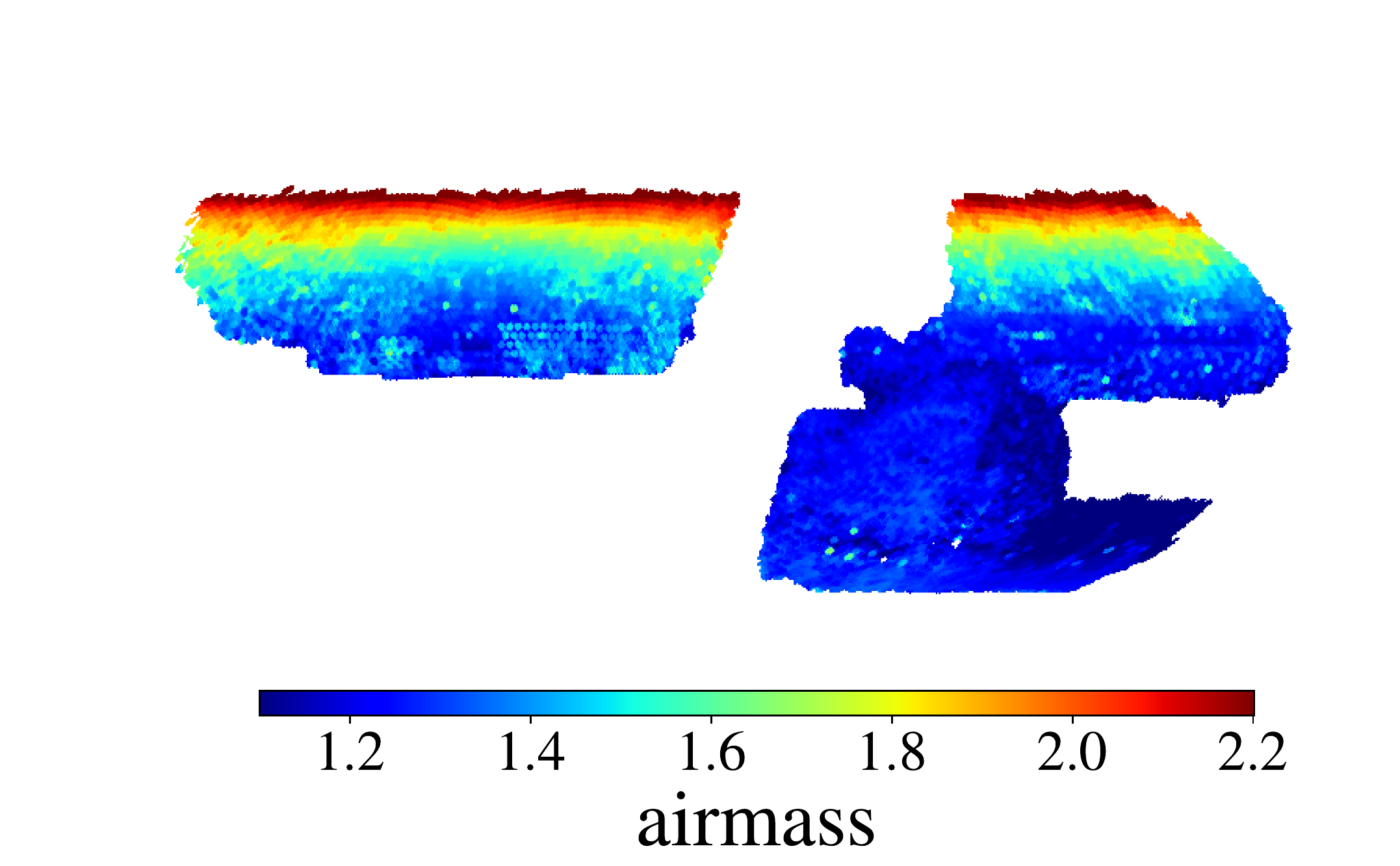}
\includegraphics[width=0.24\linewidth, trim={1.5cm 0 1.2cm 1.5cm},clip]{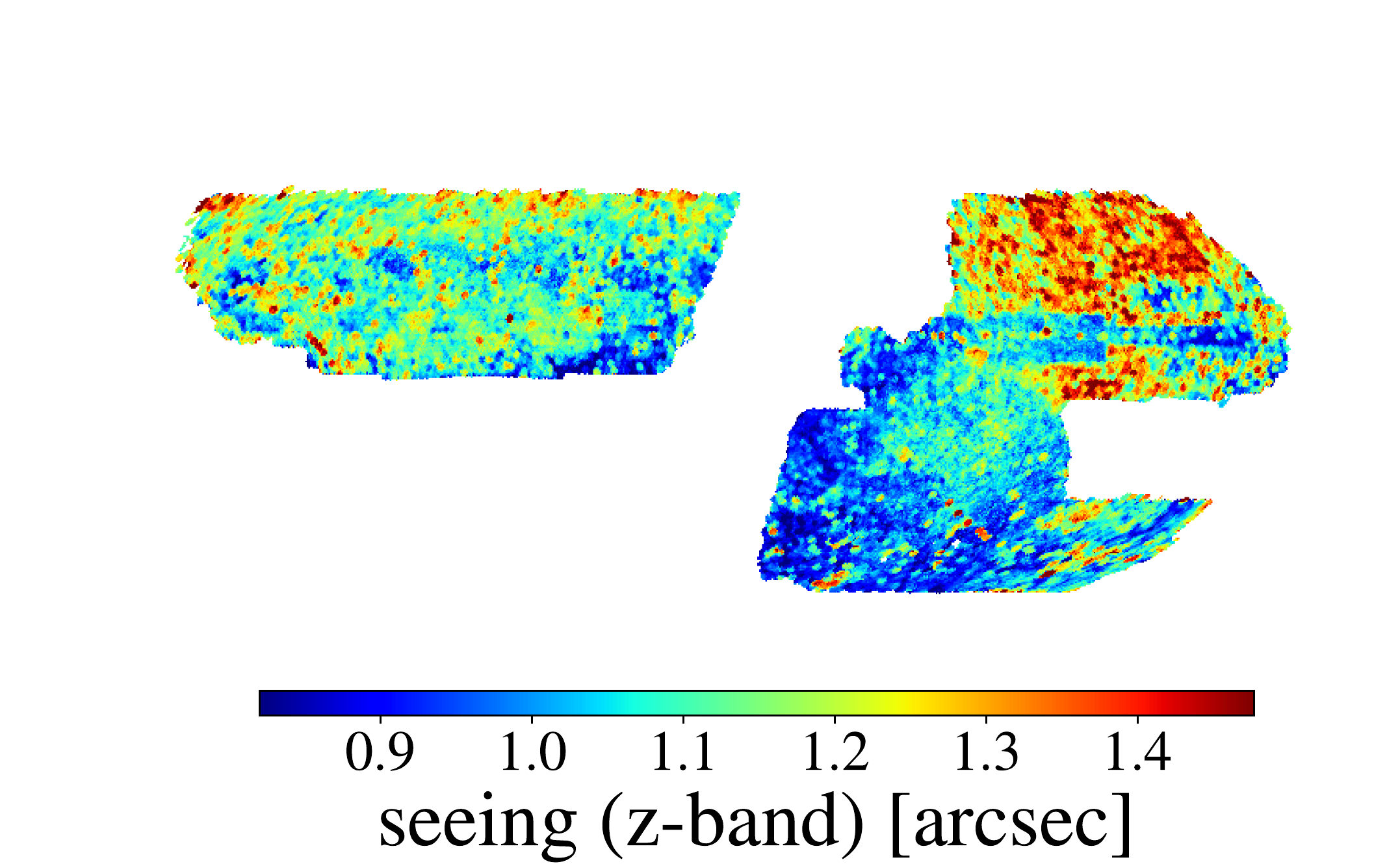}
\includegraphics[width=0.24\linewidth, trim={1.5cm 0 1.2cm 1.5cm},clip]{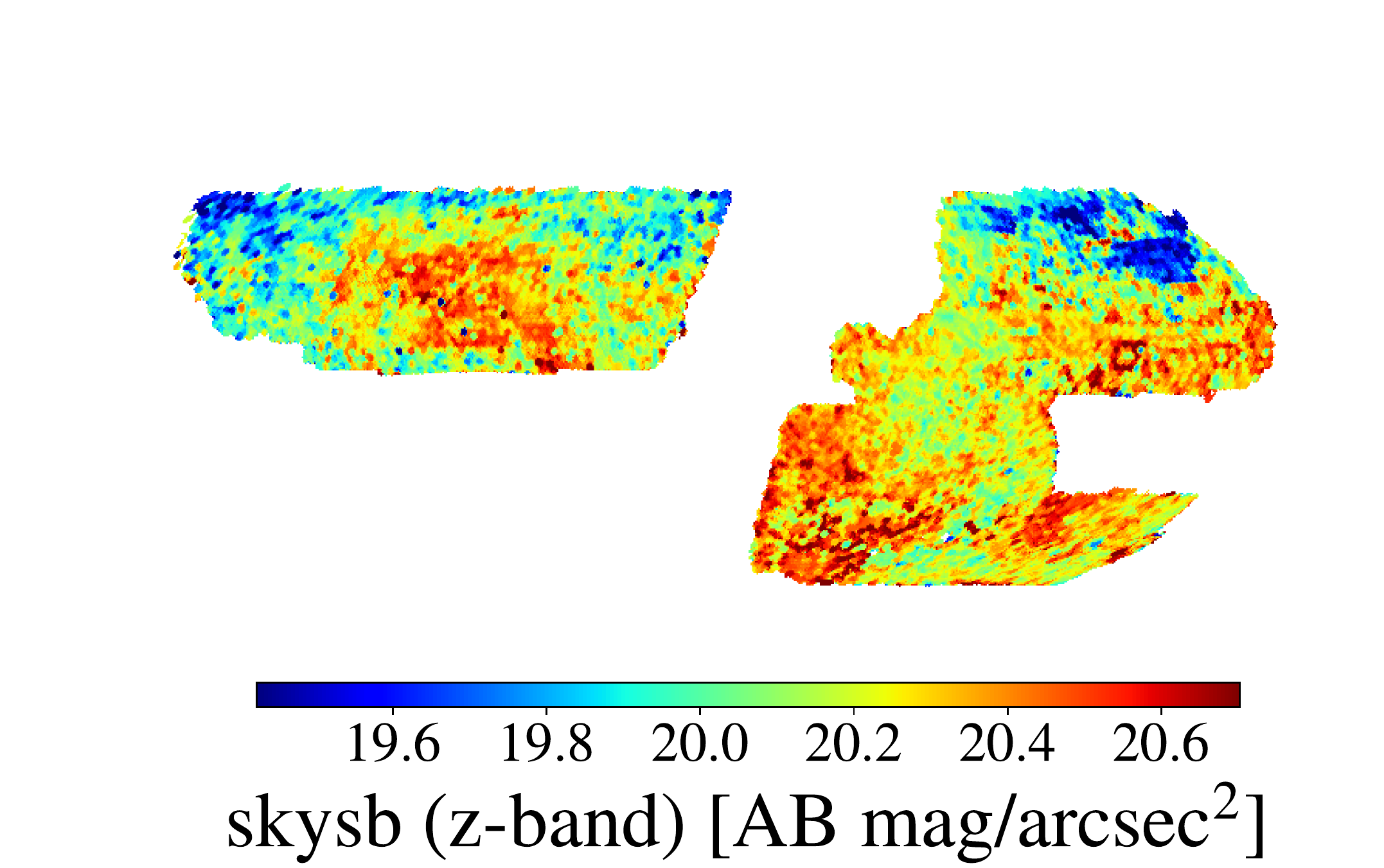}
\includegraphics[width=0.24\linewidth, trim={1.5cm 0 1.2cm 1.5cm},clip]{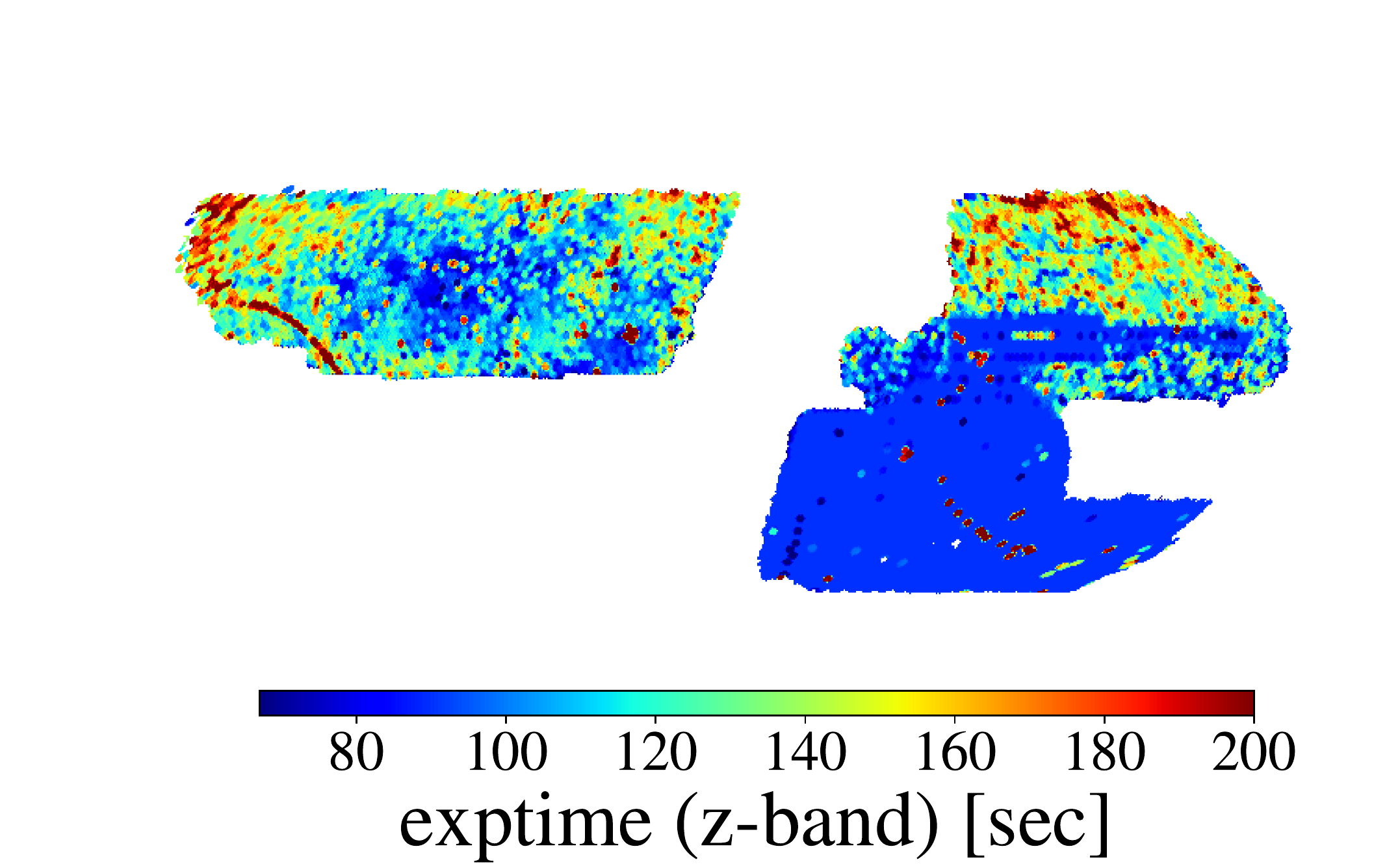}
\caption{Maps of spatially-varying potential systematics in equatorial coordinates with Mollweide projection and the astronomy convention (east towards left).}
\label{fig:systematic_maps}
\end{figure*}

We use the \texttt{HEALPix} scheme with $N_{\text{SIDE}} = 256$ to divide the footprint into pixels of equal area, over which we average each systematic. This resolution is chosen to ensure most pixels contain $>10$ galaxies, for better statistics. These pixelised maps are shown in Figure~\ref{fig:systematic_maps}. The survey properties we look at are stellar density, galactic extinction, airmass, seeing, sky background, and exposure time. For full descriptions of these survey properties, how they are calculated, and why they are included in the analysis, see Section 6 of \citealt{Kitanidis++19}. 

\begin{figure*}
\includegraphics[width=0.24\linewidth]{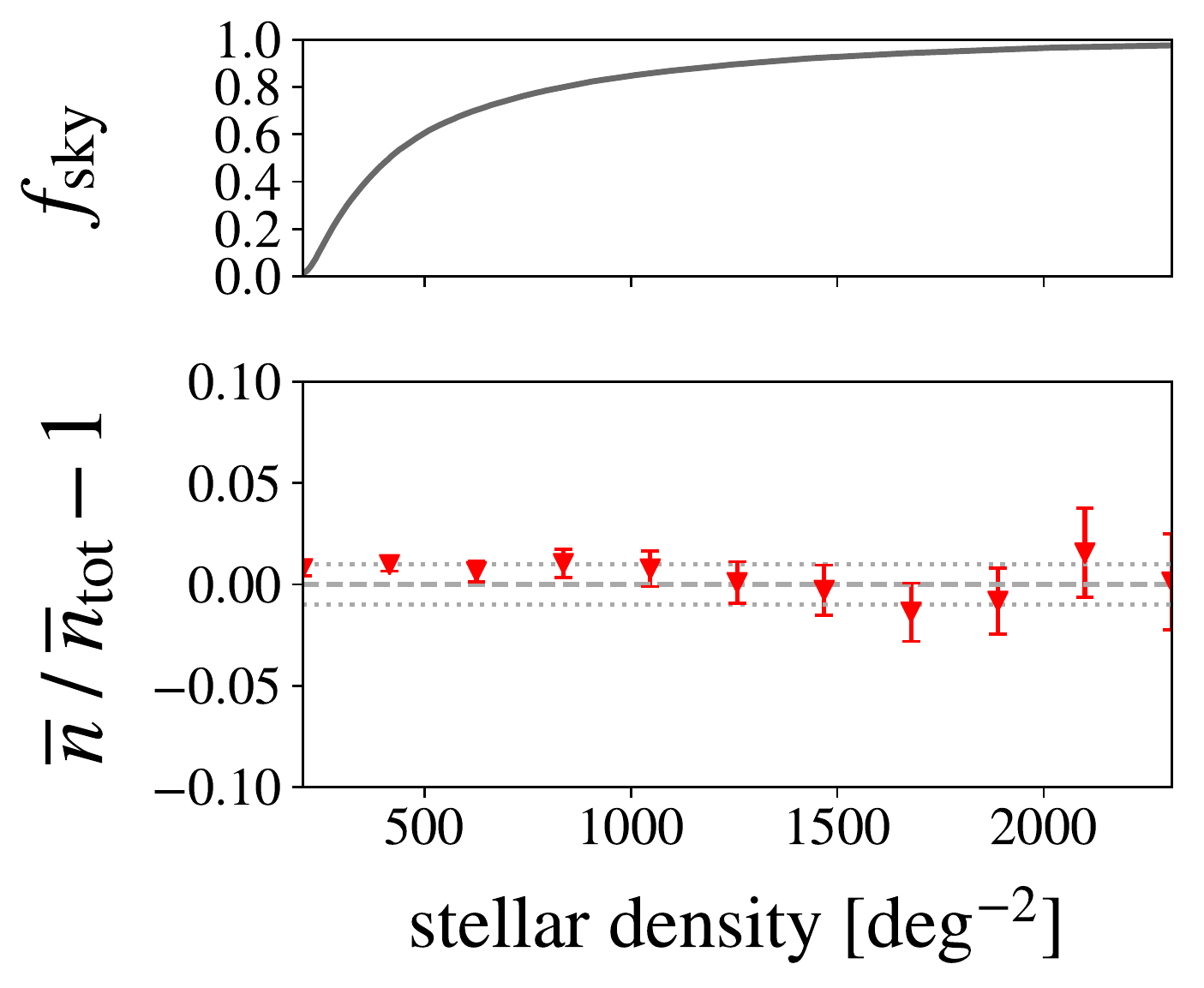}
\includegraphics[width=0.24\linewidth]{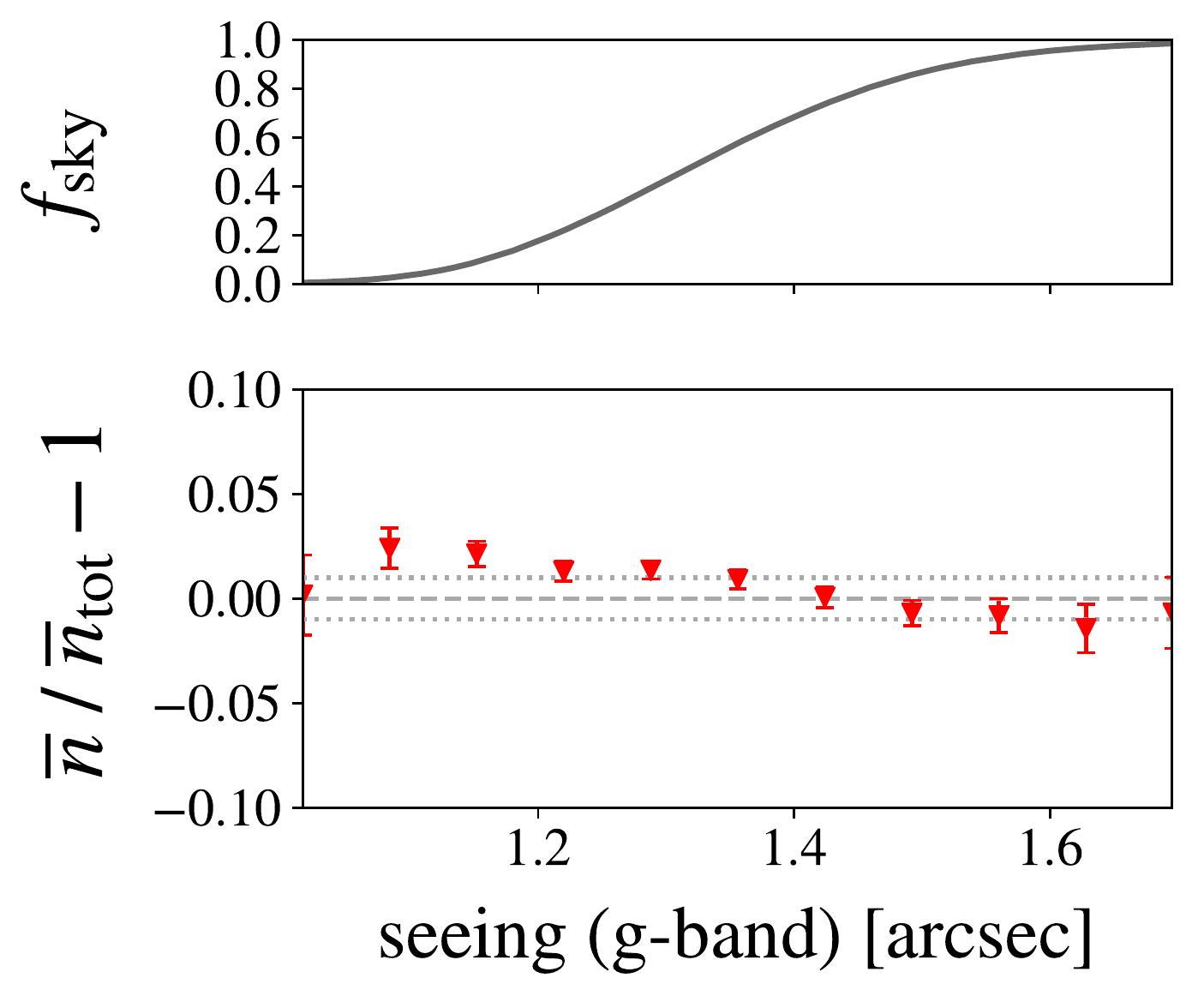}
\includegraphics[width=0.24\linewidth]{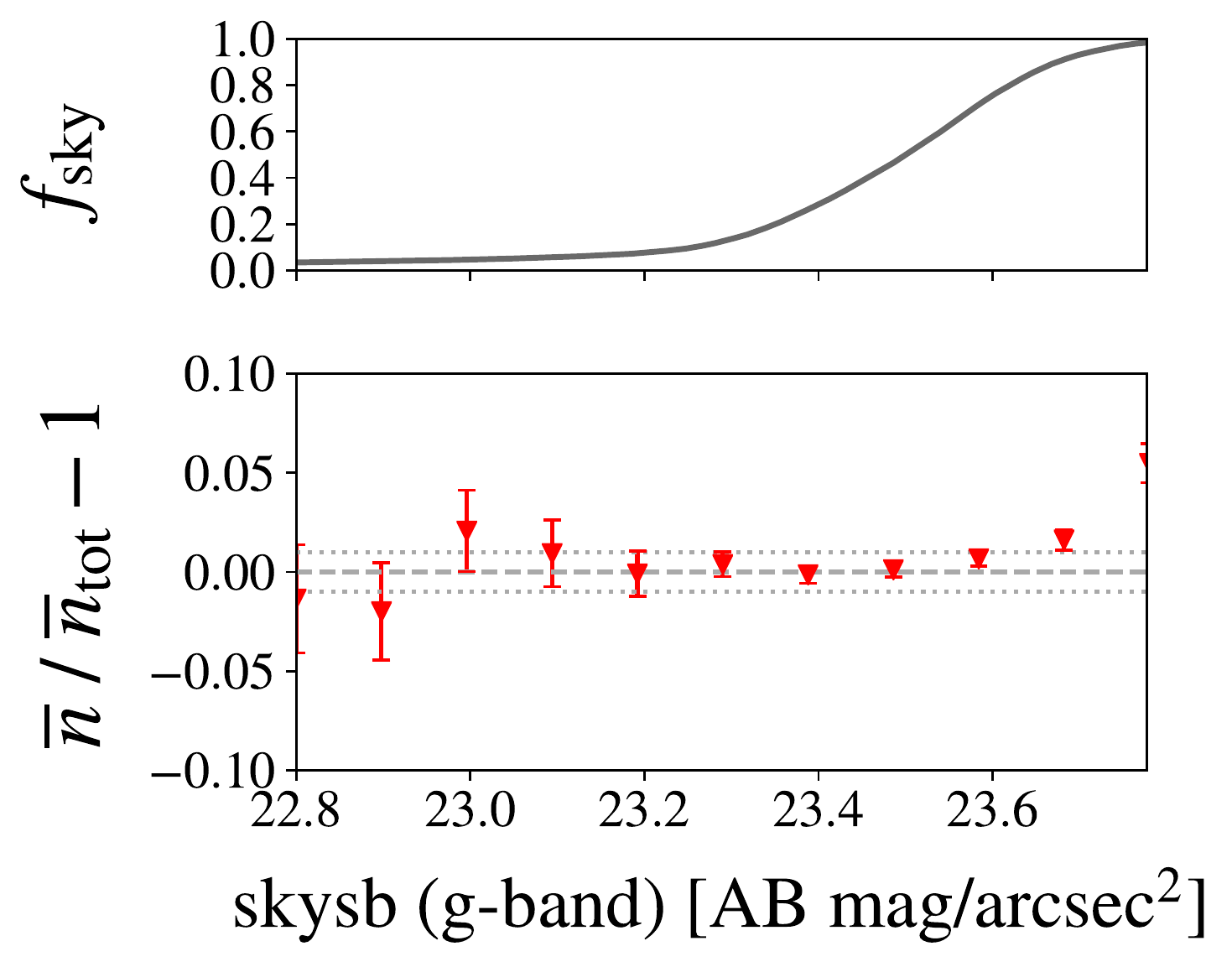}
\includegraphics[width=0.24\linewidth]{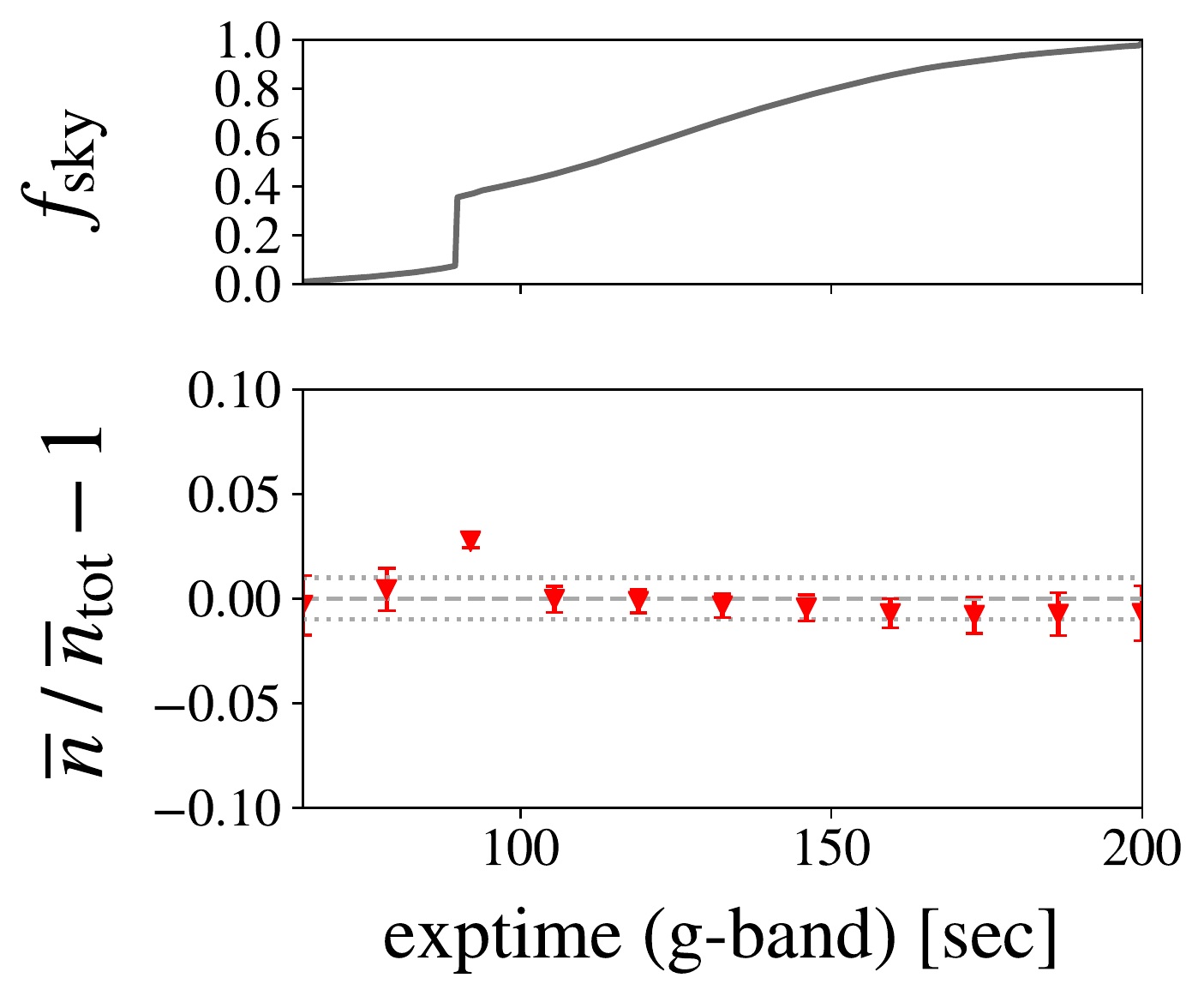}
\includegraphics[width=0.24\linewidth]{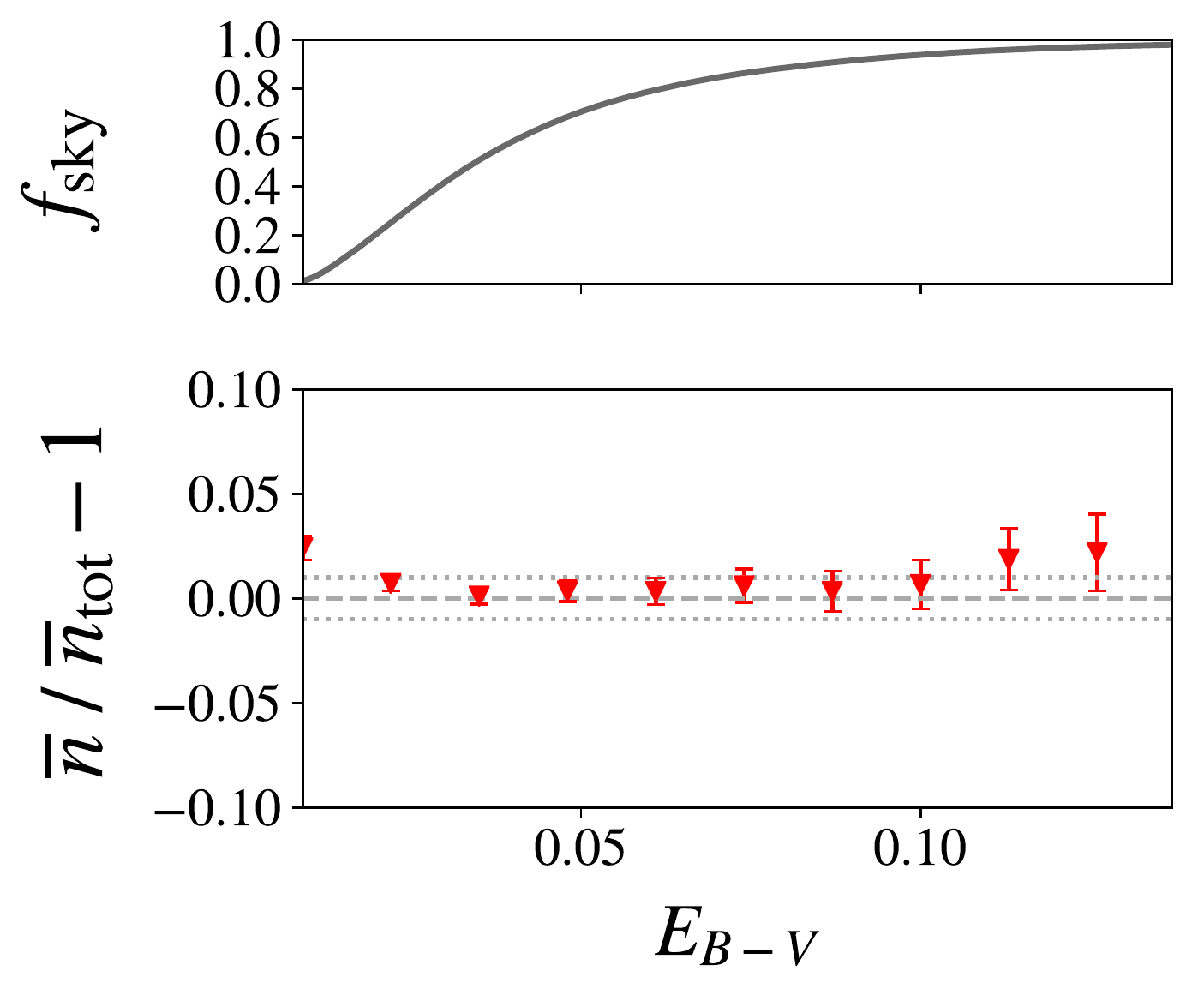}
\includegraphics[width=0.24\linewidth]{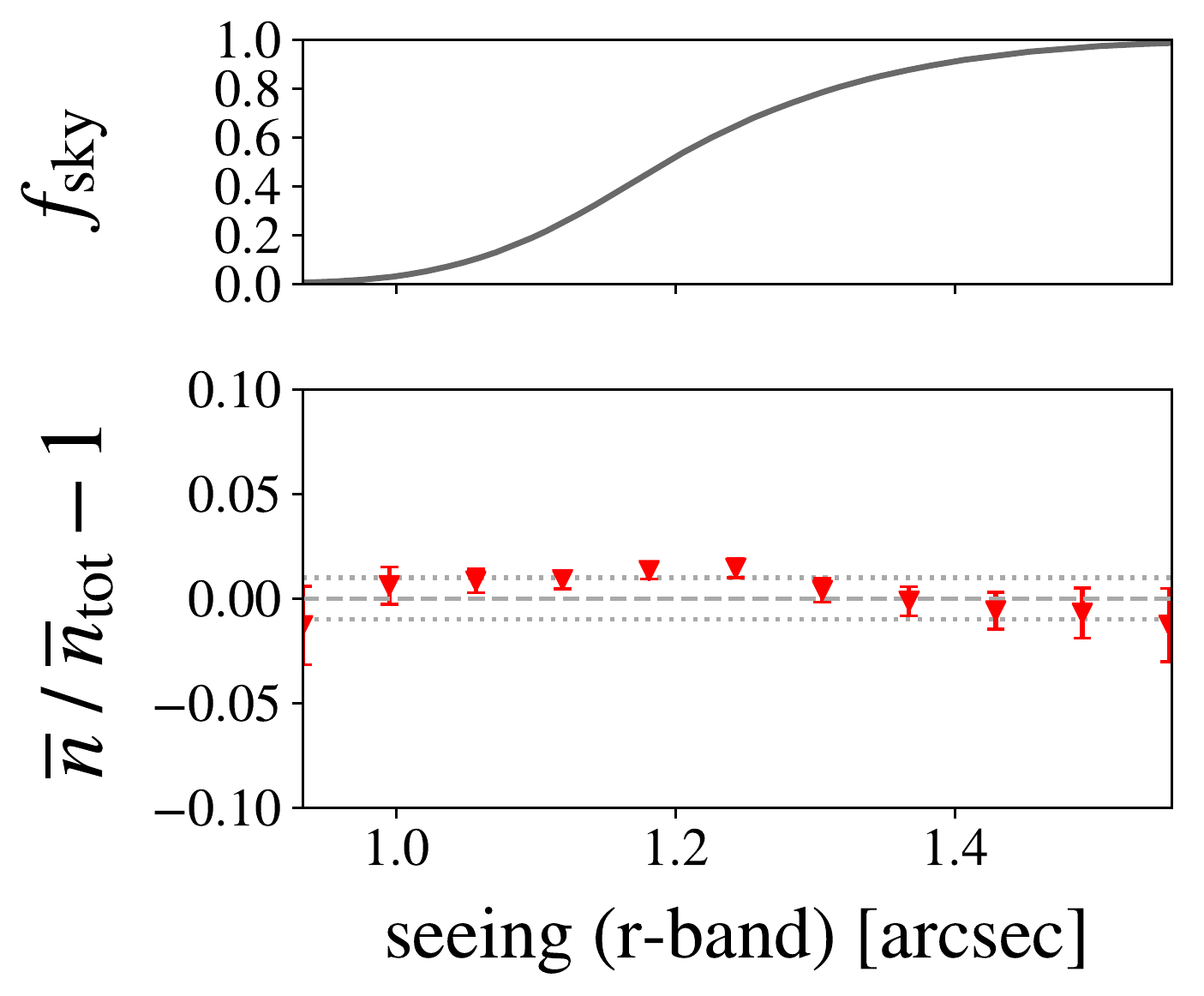}
\includegraphics[width=0.24\linewidth]{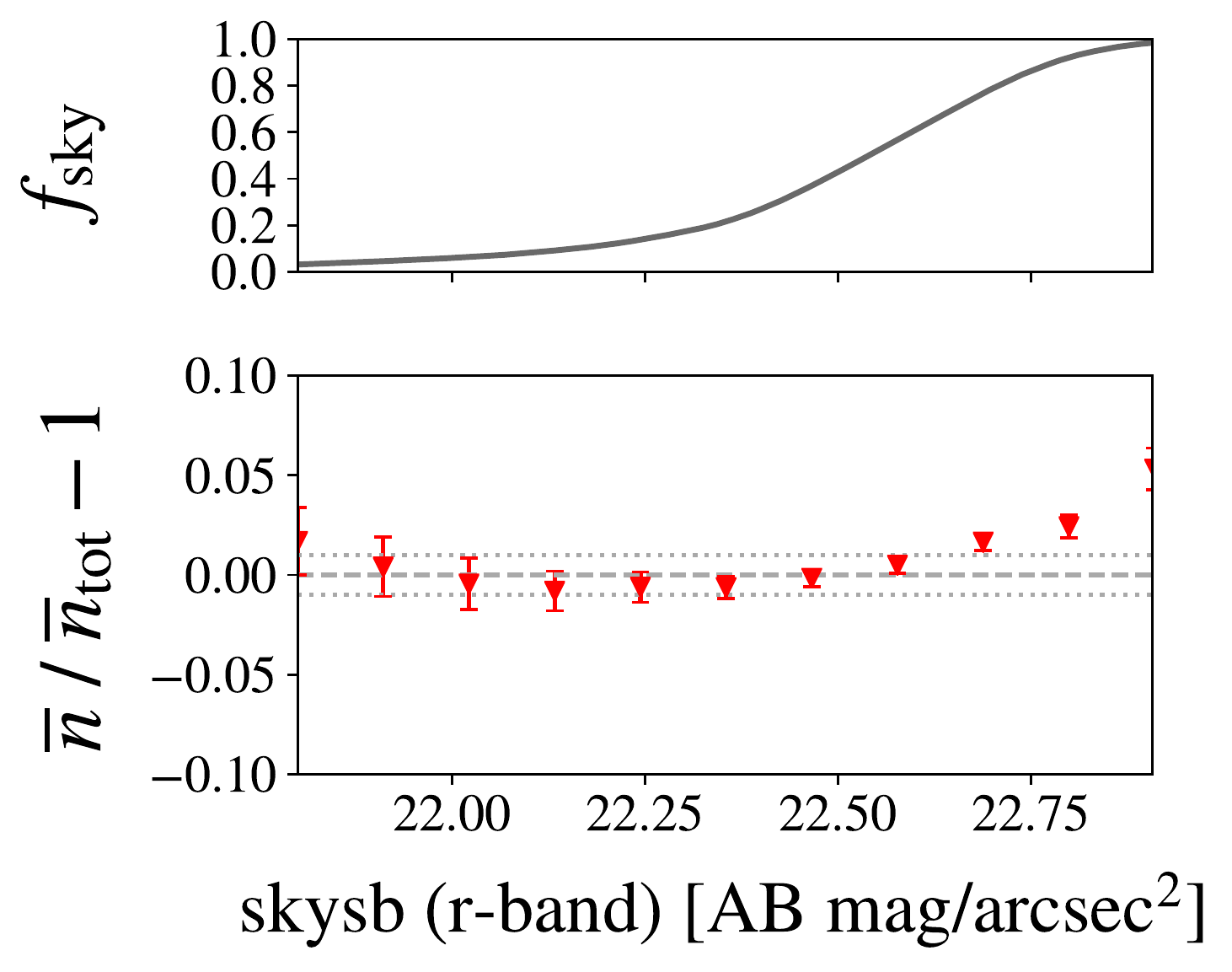}
\includegraphics[width=0.24\linewidth]{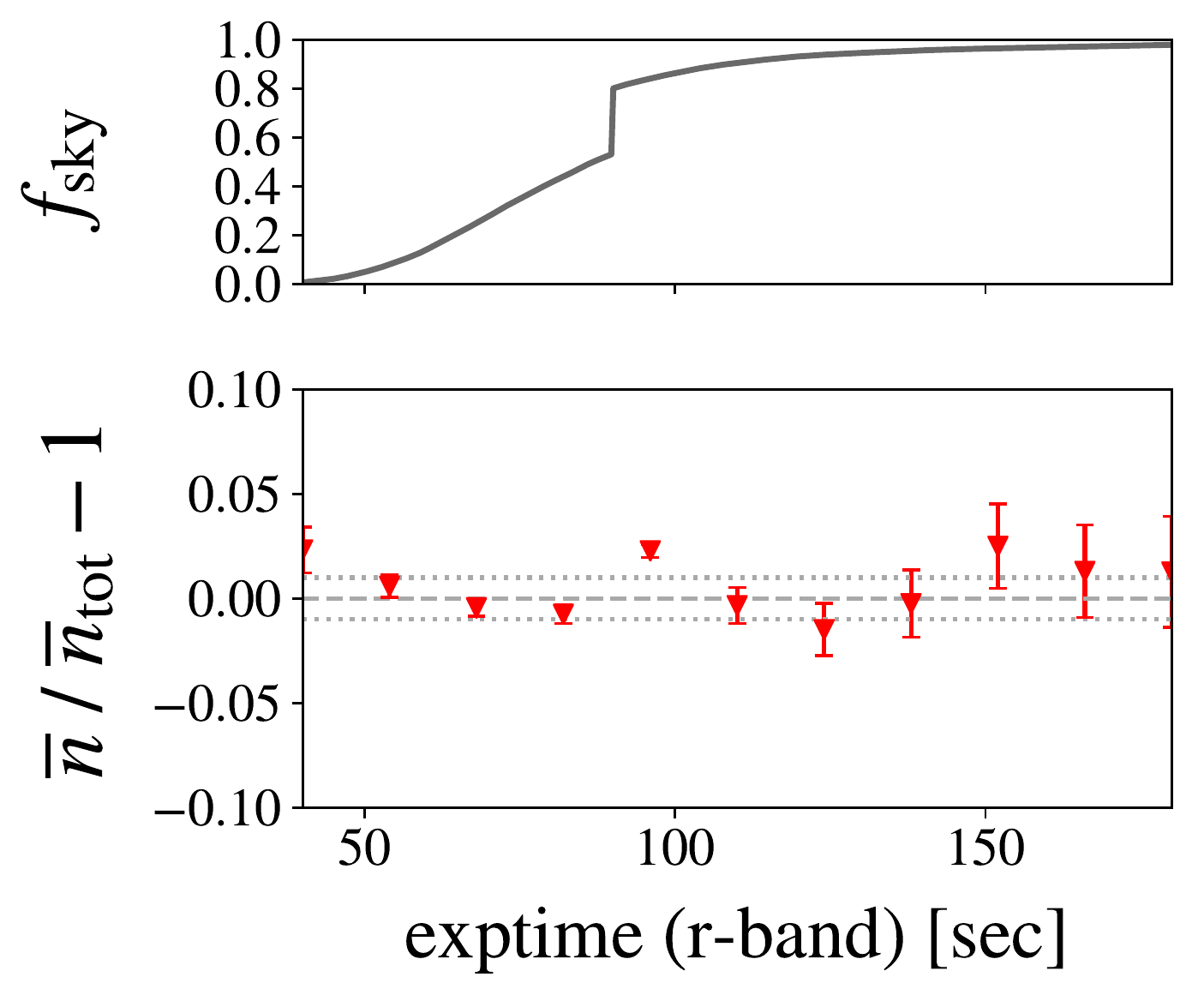}
\includegraphics[width=0.24\linewidth]{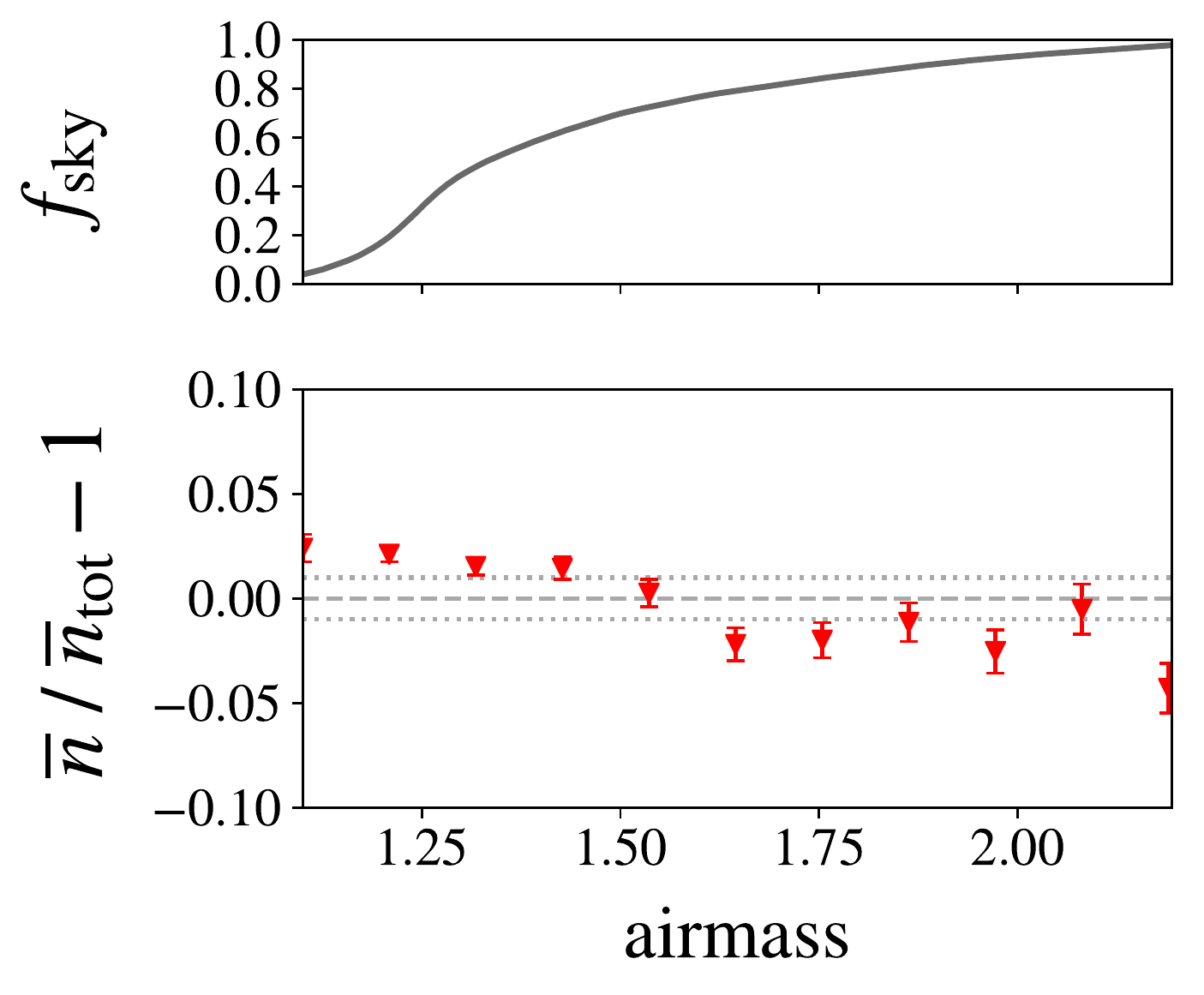}
\includegraphics[width=0.24\linewidth]{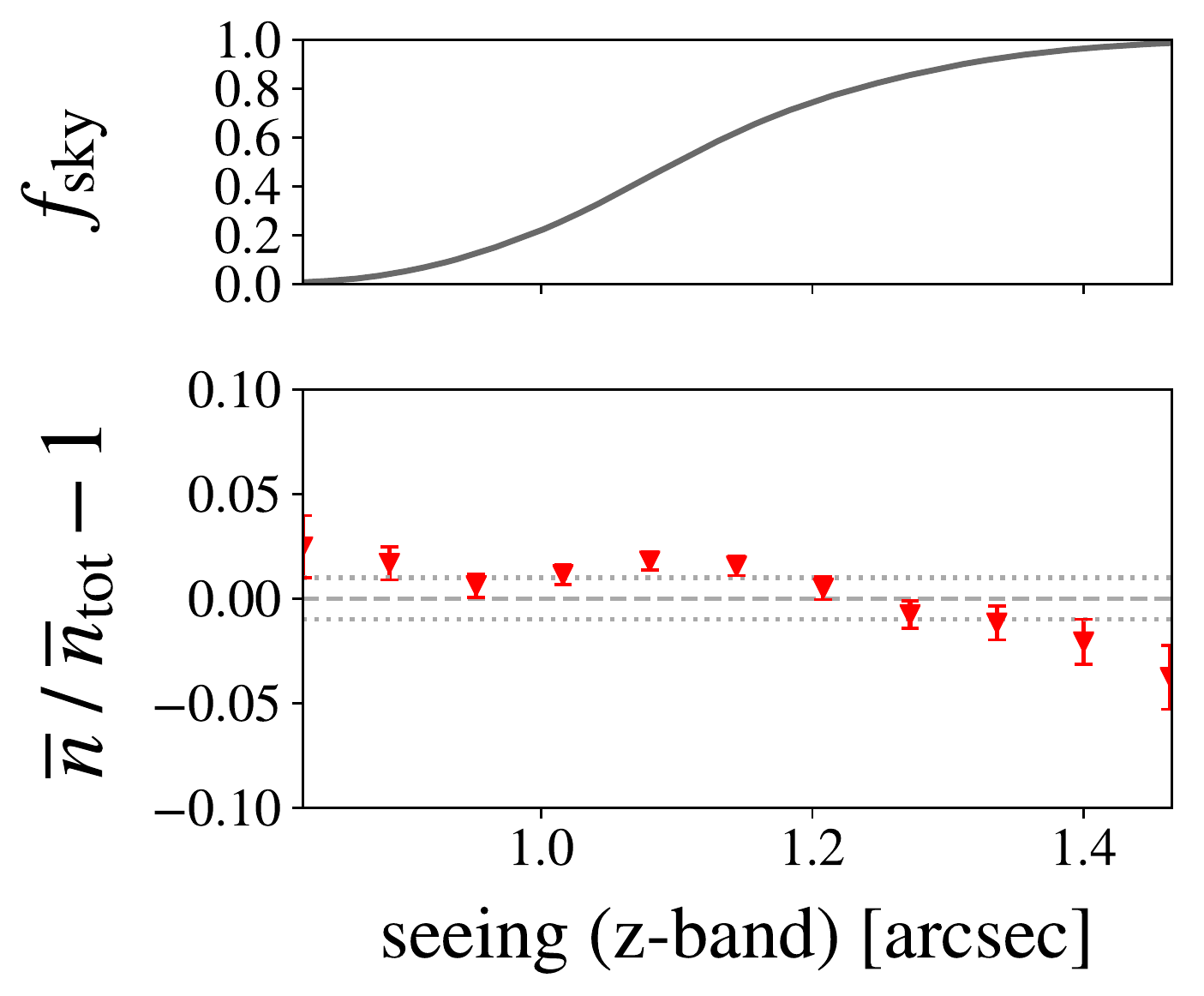}
\includegraphics[width=0.24\linewidth]{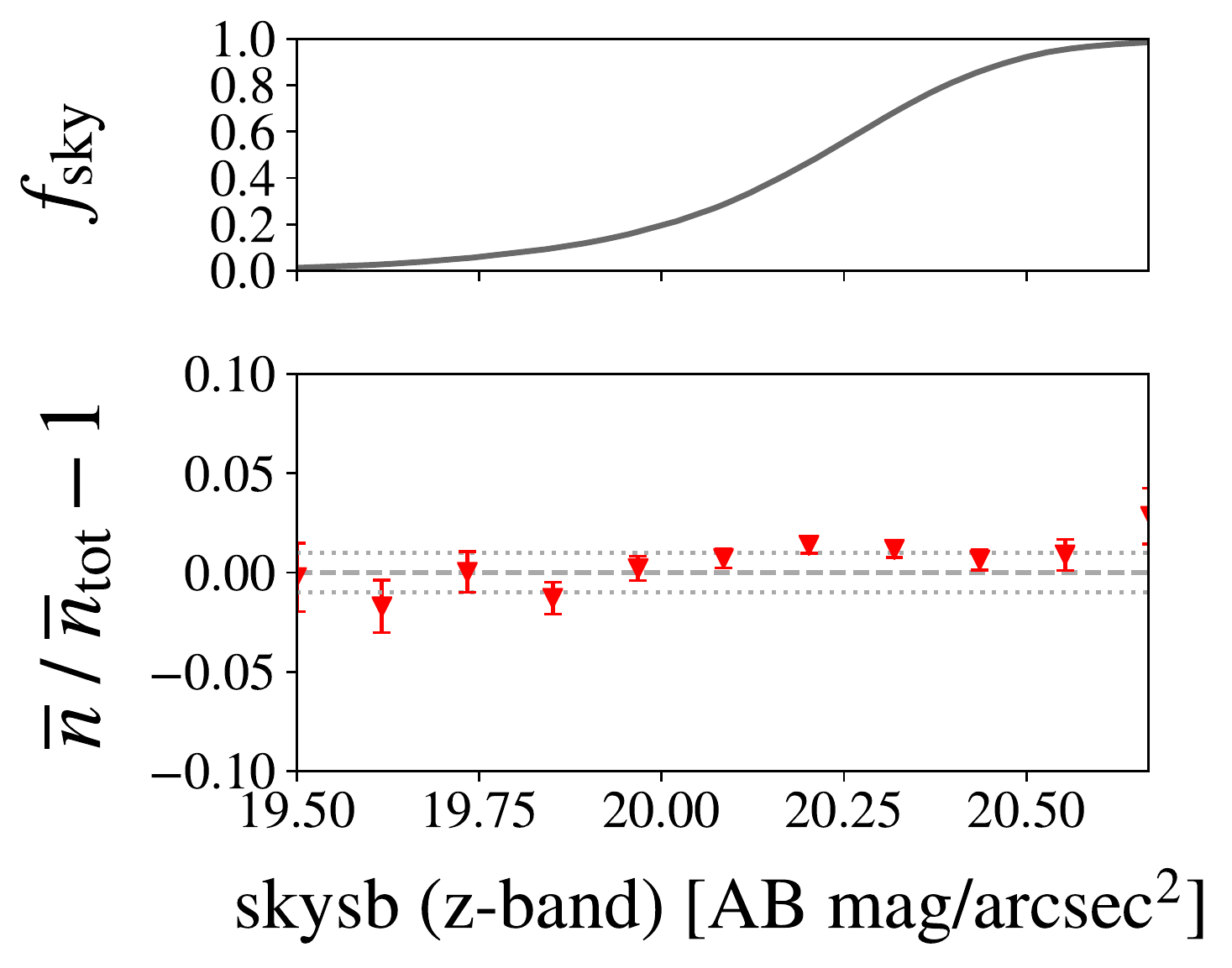}
\includegraphics[width=0.24\linewidth]{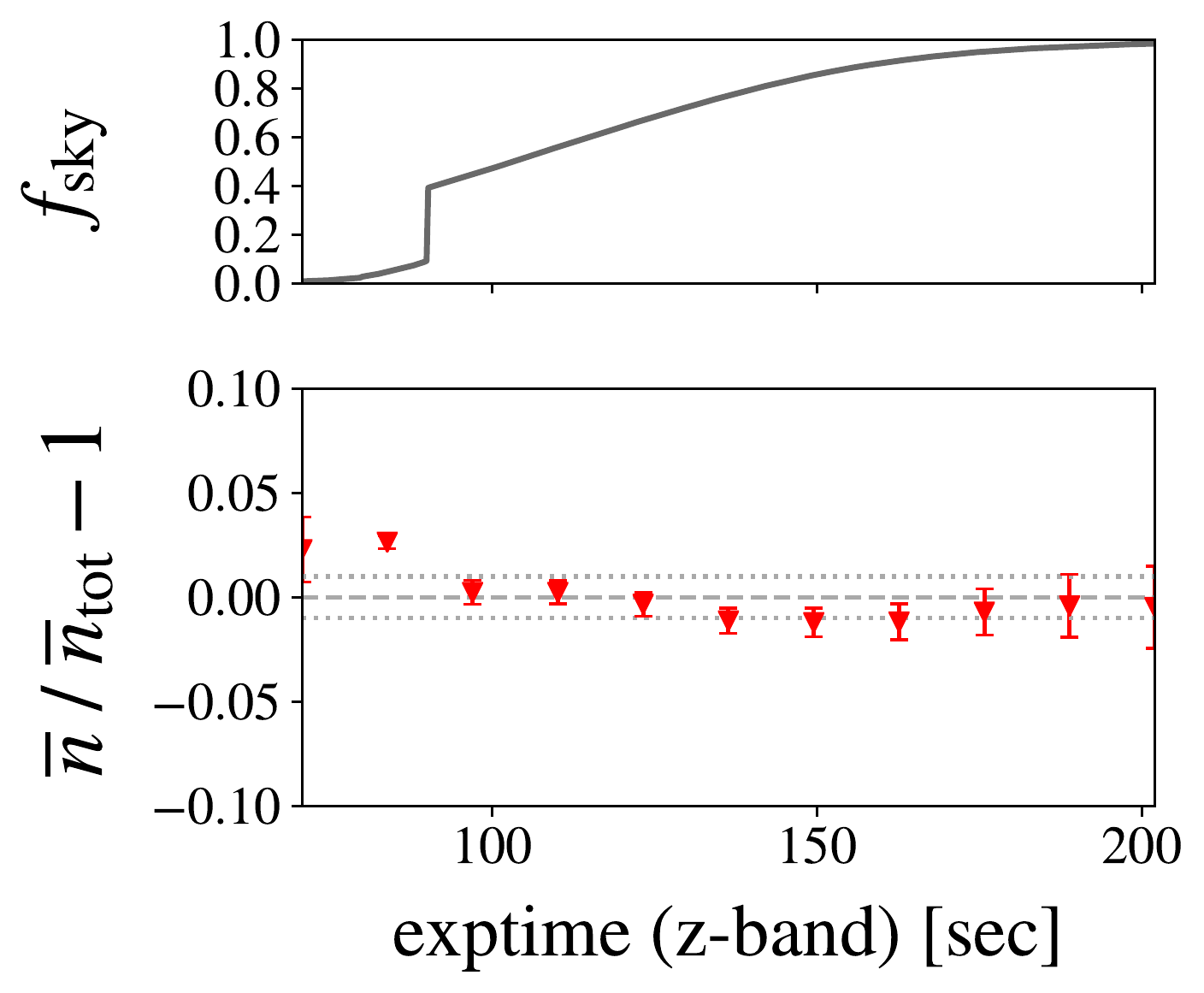}
\caption{Density of LRGs as a function of stellar density, galactic extinction (color excess), airmass, seeing in each optical band, sky subtraction in each optical band, and exposure time in each optical band. Densities and survey properties are smoothed over the scale of the pixelised maps in Figure~\ref{fig:systematic_maps}. The upper panels show the cumulative sky fractions for each survey property, and the dotted lines correspond to $\pm 1\%$ density fluctuations.}
\label{fig:systematic_trends}
\end{figure*}

For each map, we bin the pixels by the value of the survey property, and then determine the average density per bin. The resulting plots of LRG density contrast $\delta = n/\bar{n} - 1$ as a function of survey properties are shown in Figure~\ref{fig:systematic_trends}, with the cumulative sky fractions shown in the upper panels and the dotted lines corresponding to 1\% fluctuations. We show that LRG density variation due to systematic sources of error are controlled to within 5\% and, more often than not, 1\%. As such, we conclude that imaging systematics should not significantly affect our cross-correlation measurements.

%\EK{Based on feedback from other paper, maybe we want to split this by NGC vs SGC. Also based on feedback from previous paper, maybe write a note about our choice of resolution.}

\section{Galaxy Redshift Distribution}\label{sec:dndz_pipe}

In order to interpret the 2D measurements, information about the distribution of the redshifts of the photometrically selected galaxies is required. One option is to use photometrically determined redshifts (photo-z's) for this purpose; for instance, \citealt{Zhou++20} outlines a method for determining photo-z's for DESI LRGs selected from DECaLS DR7 using a machine learning method based on decision trees. We use the DR8 version of the resulting $dN/dz$ provided by Rongpu Zhou in private communications. 

However, such methods have intrinsic scatter due to photometric errors and can be biased if the distribution of galaxies used in the training set is not representative of the overall population. It is thus useful to have an alternative method for estimating the redshift distribution, if only as a proxy to gauge the effect of errors in $dN/dz$ on the desired parameter estimation. We also apply a clustering-based redshift method, as described in the following sections.

\subsection{Clustering redshift formalism}

As modern deep imaging surveys probe ever greater volumes, they detect many more sources than can realistically be targeted for spectroscopy. The idea of leveraging cross-correlations between a spectroscopic sample and a photometric sample to infer redshift information about the latter is not a new one (see e.g. \citealt{SeldnerPeebles79}, \citealt{PhillippsShanks87}, \citealt{Landy++96}, \citealt{Ho08}, \citealt{Newman08}). Since clustering-based redshift estimation presents an attractive alternative to photometric redshift methods, it has experienced a recent resurgence in popularity. Over the last decade or so, a number of clustering $dN/dz$ estimators have been presented and analyzed (\citealt{MatthewsNewman10}, \citealt{Schulz10}, \citealt{MatthewsNewman12}, \citealt{McQuinnWhite13}, \citealt{Menard13}) and tested on real or simulated data (\citealt{Schmidt13}, \citealt{Scottez++16}, \citealt{Hildebrandt++17}, \citealt{Scottez++18}, \citealt{Davis++18}, \citealt{Gatti18}, \citealt{Chiang18}, \citealt{Krolewski19}, \citealt{Kitanidis++19}). 

We use a version of the estimator proposed by \cite{Menard13}, which exploits small-scale clustering information and avoids using autocorrelation functions since they are necessarily more impacted by systematic errors than cross-correlations. We provide a detailed derivation of our formalism and its assumptions in Appendix~\ref{app:dndz}, and simply state the key result here:
\begin{align}\label{eqn:dndz_estimator}
    {w}_{\rm ps}(\theta, z_{\rm i}) &\propto \phi_{\rm p}(z_{\rm i})\frac{H(z_{\rm i})}{c}b_{\rm p}(z_{\rm i})b_{\rm s}(z_{\rm i})I(\theta, z_{\rm i}) 
\end{align}
where ${w}_{\rm ps}$ is the angular cross-correlation, $\phi_{\rm p}(z_{\rm i})$ is the photometric redshift distribution, $b_{\rm p}(z_{\rm i})$ and $b_{\rm s}(z_{\rm i})$ are the large-scale biases of the two samples, and
\begin{equation}\label{I_term}
    I(\theta, z_{\rm i}) \equiv \int_{\chi_{\rm min}}^{\chi_{\rm max}}d\chi \ \xi_{\rm mm}{\Bigg (}\sqrt{\chi_{\rm i}^2\theta^2 + (\chi-\chi_{\rm i})^2},z_{\rm i}{\Bigg )}
\end{equation}
can be computed directly from Hankel transforming the theoretical dark matter power spectrum,
\begin{align}
    \xi_{\rm mm}(r,z) = \int_0^{\infty}\frac{dk}{2\pi^2}k^2P_{\rm mm}(k,z)j_0(kr)
\end{align}
Here, $\chi_{\rm min}$ and $\chi_{\rm max}$ are the co-moving distances corresponding to the minimum and maximum redshifts of the photometric sample.

\subsection{Bias evolution}

Note that we do not need to know the amplitudes of the biases $b_{\rm p}$ and $b_{\rm s}$ in order to leverage Equation~\ref{eqn:dndz_estimator}, since they are degenerate with the overall normalization of $\phi_{\rm p}$. We only need to know the shapes of the bias evolutions. For the spectroscopic catalog, this can be determined directly. For the photometric catalog, we use two complementary methods for modelling $b_{\rm p}(z) = b_0 \times f(z)$ for some unknown evolution $f(z)$:
\begin{enumerate}
    \item Fit ``effective'' bias $b_{\rm eff} \equiv \int dz \ b_{\rm p}(z)\phi_{\rm p}(z)$.
    \item Assume parametric form for $f(z)$, fit present day bias $b_0$.
\end{enumerate}
These two methods are explained in detail in the subsections below.

\subsubsection{Fit $b_{\rm eff}$ without parametric $f(z)$}\label{sec:dndz_pipe/q}

In principle, we do not need to know the evolution of $b_{\rm p}$ in order to model the angular power spectra $C_{\ell}$, since $b_{\rm p}(z)\phi_{\rm p}(z)$ is the quantity that enters the $C_{\ell}$ integrals for a linear bias model (see e.g. Equation~\ref{eq:simplified_cells}). Equations~\ref{eqn:dndz_estimator}-\ref{I_term} allow us to constrain $f(z)dN_{\rm p}/dz$ times some unknown proportionality constant.\footnote{We are using $dN/dz$ to refer to the un-normalized redshift distributions, whereas $\phi(z)$ is normalized.} After normalization, we obtain the quantity
\begin{equation}
q(z) \equiv \frac{f(z)dN_{\rm p}/dz}{\int dz^{\prime} \ f(z^{\prime})dN_{\rm p}/dz^{\prime}}
\end{equation}
Meanwhile, in the $C_{\ell}$ equations\footnote{In order to model the galaxy-convergence bias with this approach, we assume that it has the same evolution as the galaxy-galaxy bias.}, the term $b_{\rm p}(z)\phi_{\rm p}(z)$ can be rewritten 
\begin{align}
    b_{\rm p}(z)\phi_{\rm p}(z) &= \frac{b_0 f(z) dN_{\rm p}/dz}{\int dz^{\prime} \ dN_{\rm p}/dz^{\prime}} = \frac{b_0 q(z) \int dz^{\prime} \ f(z^{\prime})dN_{\rm p}/dz^{\prime}}{\int dz^{\prime} \ dN_{\rm p}/dz^{\prime}} \\
    &= b_{\rm eff}q(z)
\end{align}
where $b_{\rm eff}$ is the effective bias term
\begin{align}
    b_{\rm eff} &\equiv \frac{b_0\int dz \ f(z)dN_{\rm p}/dz}{\int dz \  dN_{\rm p}/dz} \\
    &= \int dz \ b_0 f(z) \phi_{\rm p}(z) = \int dz \ b_{\rm p}(z) \phi_{\rm p}(z)
\end{align}
Thus, by not assuming a shape for the bias evolution, we are fitting an integrated effective bias $b_{\rm eff}$ rather than the present day bias $b_0$. This $b_{\rm eff}$ essentially represents the bias weighted by the redshift distribution; for a sharply peaked redshift distribution and weakly evolving bias, as expected in the LRG sample, $b_{\rm eff} \approx b(z_{\rm eff})$.

\subsubsection{Fit $b_0$ with parametric $f(z)$}

Working with a parametric form (e.g. $b_p(z) = b_0 / D(z)$ based on DESI's Final Design Report, where $D(z)$ is the linear growth function), $b_0$ can be measured directly. Equations~\ref{eqn:dndz_estimator}-\ref{I_term} constrain $dN_{\rm p}/dz$ times some unknown proportionality constant. After normalizing to get $\phi_{\rm p}(z)$, we insert this into the $C_{\ell}$ integrals, along with the parametric $f(z)$. Thus by ``floating'' $b_0$ until theory matches observation, we obtain a value for $b_0$.

\subsection{Integrating over scales}

Following the method of \citealt{Menard13}, we integrate $w_{\rm ps}$ over a range of angular scales as the sensitivity of the estimator is improved by encoding information from many clustering scales. In order to maximize the signal-to-noise, we weight each point by $\theta^{-1}$, which gives equal amounts of clustering information per logarithmic scale: 
\begin{align}
    \bar{w}_{\rm ps}(z_{\rm i}) = \int_{\theta_{\rm min}}^{\theta_{\rm max}}d\theta \ \frac{1}{\theta} w_{\rm ps}(\theta, z_{\rm i})
\end{align}
Hence, we have
\begin{align}
    \bar{w}_{\rm ps}(z_{\rm i}) &\propto \phi_{\rm p}(z_{\rm i})\frac{H(z_{\rm i})}{c}b_{\rm p}(z_{\rm i})b_{\rm s}(z_{\rm i})\bar{I}(z_{\rm i})
\end{align}
where
\begin{equation}\label{Ibar}
    \bar{I}(z_{\rm i}) = \int_{\theta_{\rm min}}^{\theta_{\rm max}}d\theta \ \frac{1}{\theta} \int_{\chi_{\rm min}}^{\chi_{\rm max}}d\chi \ \xi_{\rm mm}{\Bigg (}\sqrt{\chi_{\rm i}^2\theta^2 + (\chi-\chi_{\rm i})^2},z_{\rm i}{\Bigg )}
\end{equation}
In order to integrate over the same range of physical scales for each redshift bin, we take the following approach: for each photometric-spectroscopic pair, we assume that the photometric object is at the same redshift as the spectroscopic object, allowing us to convert from angle $\theta$ to projected distance $r_p = \chi(z_{\rm i})\theta$. Thus, we obtain an $r_p$-binned $w_{\rm ps}$ measurement. Then, in our equations, we perform a change of variables from $\theta$ to $r_p$:
\begin{align}
    \bar{w}_{\rm ps}(z_{\rm i}) = \int_{r_{\rm p, min}}^{r_{\rm p, max}}dr_{\rm p} \ \frac{1}{r_{\rm p}} w_{\rm ps}(r_{\rm p}, z_{\rm i})
\end{align}
\begin{equation}\label{Ibar_rp}
    \bar{I}(z_{\rm i}) = \int_{r_{p,\rm min}}^{r_{\rm p, max}}dr_{\rm p} \ \frac{1}{r_{\rm p}} \int_{\chi_{\rm min}}^{\chi_{\rm max}}d\chi \ \xi_{\rm mm}{\Bigg (}\sqrt{r_{\rm p}^2 + (\chi-\chi_{\rm i})^2},z_{\rm i}{\Bigg )}
\end{equation}
Note that in this section we have implicitly assumed scale-independent biases. In Appendix~\ref{app:dndz}, we explore how scale-dependent bias can make the shape of the estimated redshift distribution sensitive to the choice of $\theta_{\rm min}$, $\theta_{\rm max}$.

\subsection{Measurement}

\begin{figure}
\centering
\includegraphics[width=\linewidth]{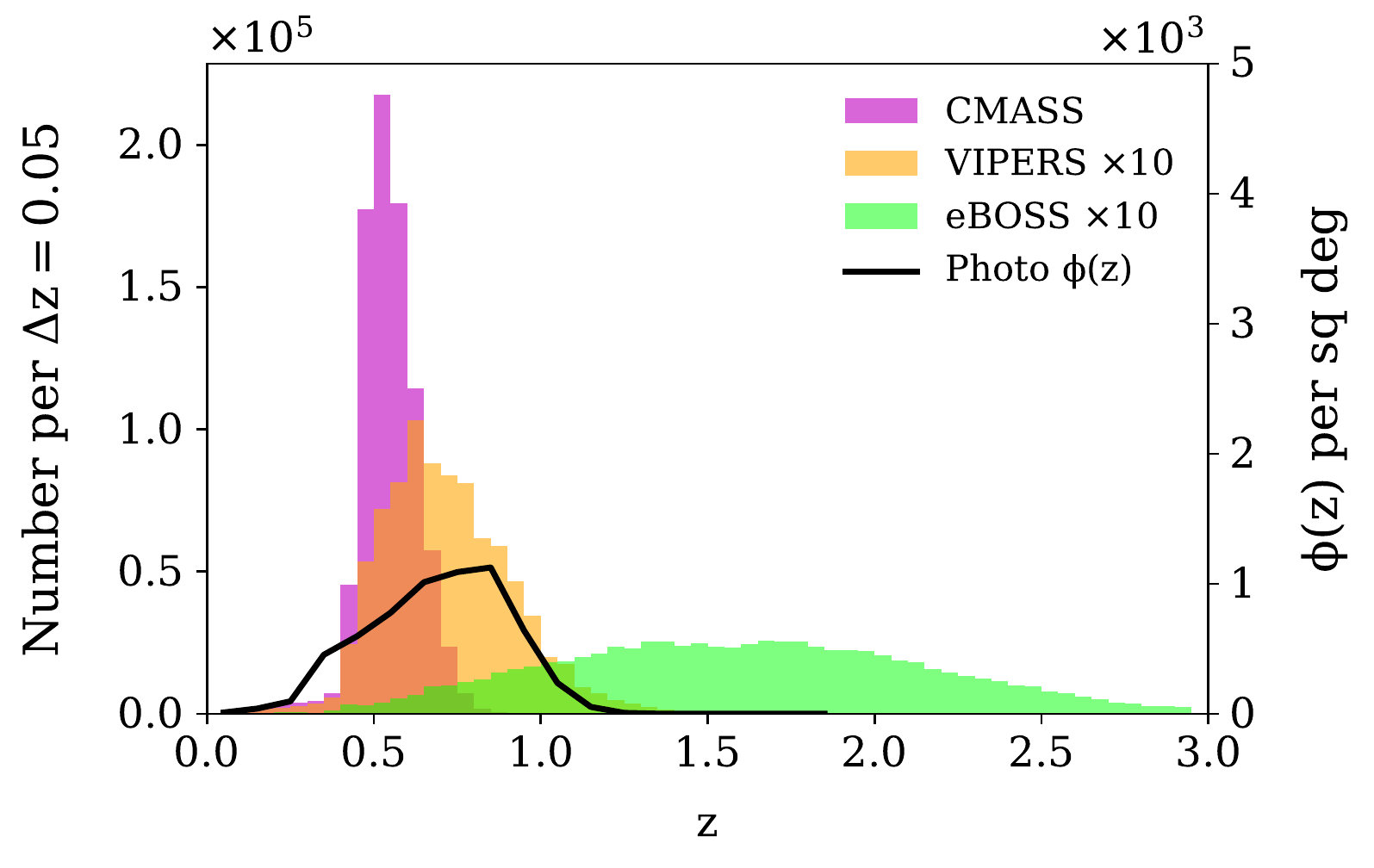}
\caption{Visualizing how the redshifts of the external spectroscopic catalogs (histograms) overlap with the redshift distribution of DESI LRGs selected from DECaLS, as estimated using photometric redshifts (solid line).}
\label{fig:spectro_overlap}
\end{figure}

We use three well-defined spectroscopic samples that overlap significantly with our LRG sample and span its full redshift range (see Figure~\ref{fig:spectro_overlap}): CMASS galaxies from Data Release 12 of the Baryon Oscillation Spectroscopic Survey (BOSS; \citealt{Eisenstein11}, \citealt{BOSS13}); galaxies from the the final data release of the VIMOS Public Extragalactic Redshift Survey (VIPERS; \citealt{VIPERS18}); and the main sample of quasars (QSOs) from Data Release 14 of eBOSS (\citealt{Dawson++16}) in the South Galactic Cap. We assume passive bias evolution for the CMASS and VIPERS galaxies, based on previous clustering studies of these samples (see e.g. \citealt{Torres++16} and \citealt{Marulli++13}, respectively). For the eBOSS QSOs, we assume the functional fit to $b(z)$ published in \citealt{Laurent++17} (and further validated using finer redshift bins in \citealt{Krolewski19}).

To measure the angular cross-correlation $w_{\rm ps}(\theta, z_{\rm i})$ between photometric sources and spectroscopic sources, with the latter first divided into narrow redshift bins $z_{\rm i} \pm \delta z_{\rm i}$, we use the Landy-Szalay pair-count estimator \citep{LandySzalay93}, 
\begin{equation}\label{eqn:LSestimator}
\hat{w}_{LS}(\theta) = \frac{D_1D_2 - D_1R_2 - D_2R_1 + R_1R_2}{R_1R_2}
\end{equation}
where $DD$, $DR$, and $RR$ are the counts of data-data, data-random, and random-random pairs at average separation $\theta$, within annular bins $\theta \pm \delta\theta$. We use 16 logarithmically spaced angular bins from $\theta = 0.001^{\circ}$ to $\theta = 1^{\circ}$. For each redshift bin, we convert the angular bins into bins of projected distance $r_p$ using the mean redshift of the bin. If we make the modest approximation that every photometric object is at the same redshift as the spectroscopic object it is being correlated with, we can obtain the angular correlation function binned in $r_p$ rather than $\theta$, $w_{\rm ps}(r_p, z_{\rm i})$.  

We estimate the errors on $w_{\rm ps}$ using bootstrapping \citep{Efron79}. Rather than resampling on individual objects, which has been shown to lead to unreliable errors (\citealt{Mo92}, \citealt{Fisher94}), we partition the sky into equal area sub-regions, using the \texttt{HEALPix} scheme with coarse resolution $N_{\rm SIDE} = 4$. We discard any sub-regions that are fully disjoint from either the photometric or spectroscopic survey, then randomly select (with replacement) from the remaining sub-regions until the number of randoms in each bootstrap realization is similar to the total number of randoms in the overlapping part of the footprint\footnote{Since the randoms are uniformly distributed and massively oversampled, the number of randoms can be treated as as a proxy for the effective area.}. The mean and variance are estimated from 500 bootstrap realizations, and are found to be highly robust to increasing or decreasing the number of bootstrap realizations.

As Table~\ref{tab:xcorr} and Figure~\ref{fig:spectro_overlap} show, the three spectroscopic catalogs vary widely in  their available overlapping area, their number density, and the widths of their redshift distributions. In order to maximize the signal-to-noise of each cross-correlation, the hyper parameters are adjusted individually. For instance, since VIPERS is made up of two very small windows, we must use a higher resolution of $N_{\rm SIDE} = 16$ to create the sub-regions for bootstrapping. The greatest signal-to-noise is achieved from the cross-correlation with the CMASS sample, whose large overlapping area and high number density near the peak of the LRG distribution allows us to use finer redshift bins of $\delta z = 0.05$ compared to $\delta z = 0.1$ used for the other two samples. 

\begin{table}
    %\noindent\begin{tabular}{p{3.0cm}|p{1.0cm}p{1.0cm}p{1.0cm}}
\begin{tabular}{|c|ccc|}
\toprule
\textbf{Spectroscopic Catalog} & \texttt{CMASS} & \texttt{VIPERS} & \texttt{eBOSS QSO} \\ 
\midrule
%\multirow{2}{*}{\textbf{Overlapping Area (deg$^2$)}} & %x (NGC) & \multirow{2}{*}{x (SGC)}& \multirow{2}{*}{x (SGC)} \\[-3pt] & x (SGC) &  & \\ 
\textbf{Overlapping Area (deg$^2$)} & $\sim$7461 & $\sim$23.5 & $\sim$940 \\
\textbf{Overlapping Number} & 615,056 & 68,022 &  19,266 \\
%\textbf{Mean Surface Density (deg$^{-2}$)} & $\sim$82.6 & $\sim$2915 & $\sim$83.5\\ 
\textbf{Redshift Bin Size Used} & 0.05 & 0.1 & 0.1 \\ 
\textbf{\bm{$N_{\rm SIDE}$} Resolution Used} & 4 & 16 & 4 \\ 
\textbf{\# Sub-Regions for Bootstrap} & 66 & 8 & 18 \\ 
\textbf{\# Bootstrap Ensembles} & 500 & 500 & 500 \\ 
$\pmb{r_{\rm p, min}}$, $\pmb{r_{\rm p, max}}$ \textbf{($h^{-1}$ Mpc)} & 0.5, 5 & 0.005, 1 & 0.5, 5 \\ 
\bottomrule
\end{tabular}
    \caption{Summary of the external spectroscopic catalogs and the parameters of the cross-correlation analysis. ``Overlapping Area'' is the approximate intersection of the spectroscopic and DESI-DECaLS DR8 footprints. ``Overlapping Number'' is the number of spectroscopic objects falling within this overlap with redshifts in the range $0.1 < z < 1.2$ (see Figure~\ref{fig:spectro_overlap} for a visualization of the overlap in redshift distributions). For bootstrapping, we reject any pixels lying entirely outside either survey; the remaining sub-regions are sampled with replacement to create the bootstrap ensembles.}
    \label{tab:xcorr}
\end{table}

\subsection{Results}

\begin{figure}
\includegraphics[width=\linewidth]{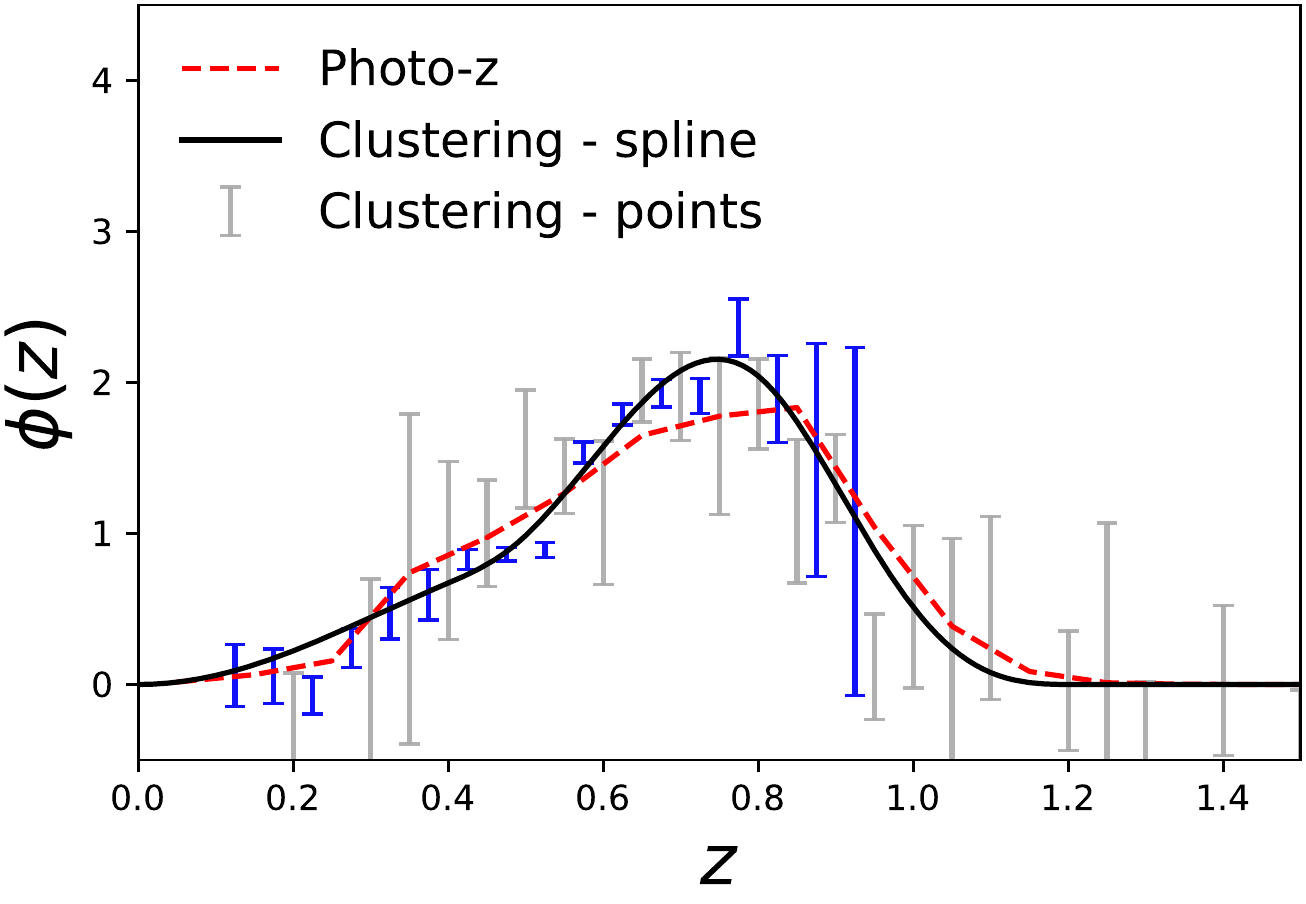}
\caption{The normalized redshift distribution derived from cross-correlations with external spectroscopy (gray error bars) and resulting cubic B-spline fit (black solid line). The normalized redshift distribution derived from photo-z's is shown for comparison (red dashed line). The spline fit is dominated by the cross-correlation with CMASS galaxies (blue highlight).}
\label{fig:clustering_dndz}
\end{figure}

Following Equation~\ref{eqn:dndz_estimator}, we obtain an estimate for $dN_{\rm p}/dz$ from each cross-correlation. Bootstrap errors from $w_{\rm ps}$ are propagated to $\phi(z)$ by performing the full calculation, including normalization, with each bootstrap separately, and then determining the standard deviation in $\phi(z)$. We use a cubic B-spline to fit the combined results, where each value $\phi_i$ is weighed by the inverse of its standard deviation, $w_i = 1/\sigma_i$. 
%We apply a smoothness parameter $s$ such that $\sum_i{w_i^2 (\phi_i - g(z_i))^2} \leq s$ where $g(z)$ is the smoothed interpolation of $(z,\phi)$. 
A common rule of thumb recommends using a value of the smoothness parameter $s$ in the range $m \pm \sqrt{2m}$ where $m$ is the number of data points being fit. Based on this, we choose a value of $s = 41$, which results in 6 interior knots. In order the respect the physicality of $\phi(z) \ge 0$ for all $z$, we force any negative spline coefficients to be zero. The clustering-based $\phi(z)$ points and fit are shown in Figure~\ref{fig:clustering_dndz}, along with the photo-z derived $\phi(z)$. Unsurprisingly, the spline fit is dominated by the CMASS cross-correlations (highlighted in blue in the figure) due to the comparatively high signal-to-noise of these cross-correlations. The photo-z and clustering redshift distributions are qualitatively similar but not identical, with the clustering $\phi(z)$ having a sharper peak.

\section{Measuring Angular Power Spectra}\label{sec:master_pipe}

\subsection{Angular power spectra in the Limber approximation}
%
% Given that we are interested in sub-degree and degree angular scales (\ell >> 10), we can safely adopt the so-called Limber approximation
%
% Our focus will be on small angular scales (high \ell), where the signal to noise is highest and the effects of quasi-linear evolution become important. This allows us to make the Limber approximation, which in our context is
%
% On small angular scales (high \ell) one may make the Limber approximation [44], under which C_{\ell} reduces to a single integral of the equal-time, real-space power spectrum:
%
% Because the relevant angular scales are much smaller than 1 radian (multipoles \ell > 100), the theoretical angular cross-correlation can be computed using the Limber approximation (Limber 1953) as
%

\begin{figure}
\includegraphics[width=\linewidth]{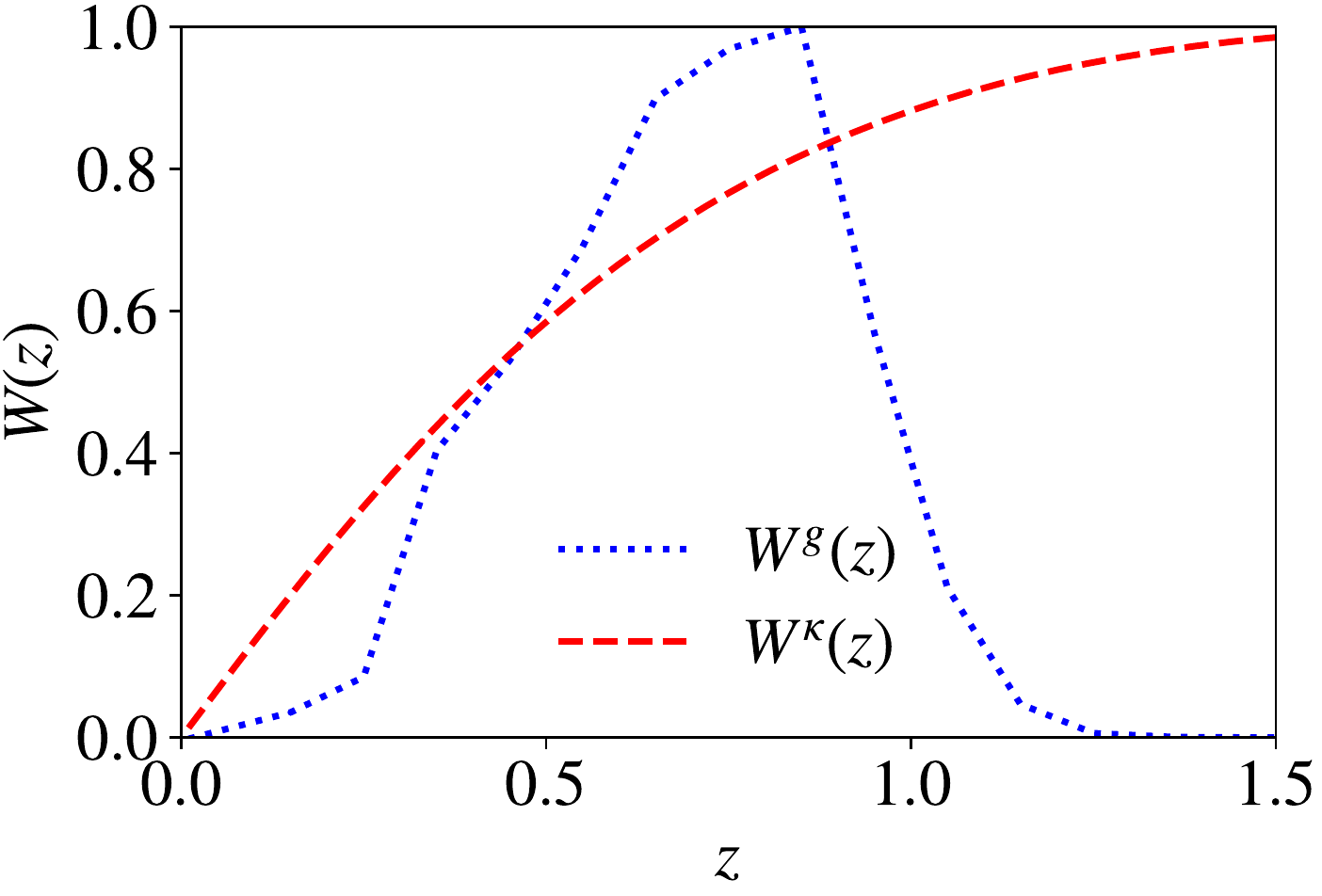}
\caption{Projection kernels for the galaxy sample (dashed blue line) and CMB lensing (dotted red line), both normalized to a unit maximum.}
\label{fig:kernels}
\end{figure}

Galaxy overdensity $\delta_g$ and CMB lensing convergence $\kappa$ are both projections of 3D density fields, expressed as line-of-sight integrals over their respective projection kernels. The angular cross-spectrum between two such fields $X$ and $Y$ is given by
\begin{align}
C_{\ell}^{XY} = \int & d\chi_1 \int d\chi_2 \ W^X(\chi_1) \ W^Y(\chi_2) \nonumber \\ & \int \frac{2}{\pi}k^2dk \ P_{XY}(k;z_1,z_2) \ j_{\ell}(k\chi_1) \ j_{\ell}(k\chi_2)
\end{align}
where $W^X$ and $W^Y$ are the projection kernels, $P_{XY}$ is the real-space power cross-spectrum, and $j_{\ell}$ are spherical Bessel functions of the first kind. As we are primarily interested in angular scales $\lesssim 1^{\circ}$ ($\ell \gtrsim 100$), we can adopt the Limber approximation \citep{Limber53, Rubin54} and its first order correction \citep{LoverdeAfshordi08}, under which the $k$ integral evaluates to
\begin{equation}
    P_{XY}{\Big (}k = \frac{\ell + 
\nicefrac{1}{2}}{\chi_1}; z{\Big )}\frac{1}{\chi_1^2}\delta^{D}(\chi_1 - \chi_2)
\end{equation}
Hence the angular cross-spectrum may be expressed as a single integral over line-of-sight co-moving distance,
\begin{align}
C_{\ell}^{XY} &= \int d\chi \ W^X(\chi) W^Y(\chi) \frac{1}{\chi^2}P_{XY}{\Big (}k = \frac{\ell + 
\nicefrac{1}{2}}{\chi}; z{\Big )} \\
&= \int dz \ \frac{H(z)}{c} W^X(z) W^Y(z) \frac{1}{\chi^2}P_{XY}{\Big (}k = \frac{\ell + 
\nicefrac{1}{2}}{\chi}; z{\Big )}
\end{align}
The projection kernels for galaxy overdensity and CMB lensing convergence are, respectively,
\begin{gather}
\begin{aligned}
    W^{g}(z) &= \phi(z) = \frac{c}{H(z)}\phi(\chi) = \frac{c}{H(z)} W^{g}(\chi) \\
    W^{\kappa}(z) &= \frac{3}{2c} \Omega_{m0} \frac{H_0^2}{H(z)} (1+z) \frac{\chi(\chi_{\rm *}-\chi)}{\chi_{\rm *}} = \frac{c}{H(z)} W^{\kappa}(\chi)
\end{aligned}
\end{gather}
where $\chi_{\rm *} = \chi(z_{\rm *}{\approx}1100) \approx 9400$ $h^{-1}$Mpc is the distance to the surface of last scattering, and $\phi(z)$ is the normalized redshift distribution of the galaxy sample. These kernels are plotted in Figure~\ref{fig:kernels}.

Plugging in and simplifying the expressions for the spectra,
\begin{align}\label{eq:simplified_cells}
    C_{\ell}^{\rm \kappa g} &= \int d\chi \ \frac{3\Omega_{m0}H_0^2}{2c^2} \frac{1+z}{\chi^2}\frac{\chi(\chi_{\rm *}-\chi)}{\chi_{\rm *}}\phi(\chi)P_{\rm mg}{\Big (}k = \frac{\ell + 
\nicefrac{1}{2}}{\chi}; z{\Big )} \nonumber \\ %%%
    &= \int dz \ \frac{3\Omega_{m0}H_0^2}{2cH(z)} \frac{1+z}{\chi^2}\frac{\chi(\chi_{\rm *}-\chi)}{\chi_{\rm *}}\phi(z)P_{\rm mg}{\Big (}k = \frac{\ell + 
\nicefrac{1}{2}}{\chi}; z{\Big )}\nonumber \\ %%%
    C_{\ell}^{\rm gg} &= \int d\chi \ \phi(\chi)^2 \frac{1}{\chi^2} P_{\rm gg}{\Big (}k = \frac{\ell + 
\nicefrac{1}{2}}{\chi}; z{\Big )} \\ %%%
    &= \int dz \ \frac{H(z)}{c} \phi(z)^2 \frac{1}{\chi^2} P_{\rm gg}{\Big (}k = \frac{\ell + 
\nicefrac{1}{2}}{\chi}; z{\Big )} \nonumber \\ %%%
    C_{\ell}^{\kappa\kappa} &= \int d\chi \ {\Big (}\frac{3\Omega_{m0}H_0^2}{2c^2} \frac{1+z}{\chi^2}\frac{\chi(\chi_{\rm *}-\chi)}{\chi_{\rm *}}{\Big )}^2P_{\rm mm}{\Big (}k = \frac{\ell + 
\nicefrac{1}{2}}{\chi}; z{\Big )} \nonumber \\
    &= \int dz \ \frac{H(z)}{c}{\Big (}\frac{3\Omega_{m0}H_0^2}{2cH(z)} \frac{1+z}{\chi^2}\frac{\chi(\chi_{\rm *}-\chi)}{\chi_{\rm *}}{\Big )}^2P_{\rm mm}{\Big (}k = \frac{\ell + 
\nicefrac{1}{2}}{\chi}; z{\Big )} \nonumber
\end{align}

\subsection{Estimating angular power spectra}

Many different approaches for estimating angular power spectra from cosmological maps exist in the literature, including maximum likelihood estimators (\citealt{Bond++98}, \citealt{WandeltHansen03}), the optimal quadratic estimator (\citealt{Tegmark97}, \citealt{Tegmark++01}), and Bayesian sampling techniques (e.g. \citealt{Eriksen++04}, \citealt{Taylor++08}). While these methods have the advantage of recovering the unbiased power spectrum directly, they are computationally expensive to implement, particularly for the high resolution maps produced by modern experiments, since they scale as $\mathcal{O}(\ell_{\rm max}^6)$. Sub-optimal but numerically efficient pseudo-$C_{\ell}$ algorithms \citep{MASTER} are a popular alternative when dealing with multipoles $\ell > 30$ \citep{Efstathiou04a}, as they take advantage of speedy spherical harmonics transforms to recover the power spectrum in $\mathcal{O}(\ell_{\rm max}^3)$ time. Below, we briefly outline the pseudo-$C_{\ell}$ approach.

Any scalar function, $T(\hat{n})$, defined on a sphere may be expanded into spherical harmonics, $Y_{\ell m}$, with expansion coefficients $a_{\ell m}$ as
\begin{align}
    T(\hat{n}) &= \sum_{\ell=0}^{\infty} \sum_{m=-\ell}^{\ell} a_{\ell m} Y_{\ell m}(\hat{n}) \\ 
    a_{\ell m} &= \int_{4\pi}d\Omega \ T(\hat{n}) \ Y^{*}_{\ell m}(\hat{n})
\end{align}
The angular power spectrum $C_{\ell}$ measures the amplitude as a function of wavelength averaged over direction,
\begin{equation}
    C_{\ell} = \frac{1}{2\ell + 1}\sum_{m=-\ell}^{\ell} |a_{\ell m}|^2
\end{equation}
This is the observed angular power spectrum of a given Gaussian realization; the average over an ensemble of universes, $\langle C_{\ell} \rangle \equiv C_{\ell}^{\rm th}$, is specified by the physics (primordial perturbations, galaxy formation, etc.) with uncertainty due to cosmic variance given by
\begin{align}
    \sigma_{\ell}^2 = \frac{C^{\rm XX}_{\ell}C^{\rm YY}_{\ell} + (C^{\rm XY}_{\ell})^2}{2\ell + 1}
\end{align}

However, in practice, we are not dealing with measurements over the full sky, but rather a masked and weighted partial sky. We must account for the effect of the masking window function $W(\hat{n})$, which couples different $\ell$ modes and biases the estimator. Naive calculation of the spherical harmonics transform on a partial sky map produces the pseudo angular power spectrum, whose coefficients are a convolution of the mask and the true coefficients,
\begin{align}
    \tilde{C}_{\ell} &= \frac{1}{2\ell + 1}\sum_{m=-\ell}^{\ell} |\tilde{a}_{\ell m}|^2 \\
    \tilde{a}_{\ell m} &= \int_{4\pi}d\Omega \ T(\hat{n}) \ W(\hat{n}) \ Y^{*}_{\ell m}(\hat{n}) %\\
%    &= \sum_{\ell^{\prime}>0} \sum_{m^{\prime}=-\ell^{\prime}}^{\ell^{\prime}} a_{\ell^{\prime} m^{\prime}} K_{\ell m \ell^{\prime} m^{\prime}}
\end{align}
Fortunately, their ensemble averages are related simply as
\begin{equation}
    \langle \tilde{C}_{\ell} \rangle = \sum_{\ell^{\prime}=0}^{\infty} M_{\ell \ell^{\prime}} \langle C_{\ell^{\prime}} \rangle
\end{equation}
where the mode-mode coupling matrix $M$ can be determined purely from the geometry of the mask.
This $\ell$-by-$\ell$ matrix is generally singular in the case of large sky cuts. In order to perform matrix inversion, a common method is to use a set of discrete bandpower bins $L$ and assume the angular power spectrum is a step-wise function in each bin. Using this approach, the MASTER algorithm \citep{MASTER} is able to efficiently calculate and invert the $L$-by-$L$ mode-mode coupling matrix to extract the binned angular power spectrum from the binned pseudo angular power spectrum,
\begin{equation}
    \langle C_{L} \rangle = \sum_{L^{\prime}} M_{L L^{\prime}}^{-1} \langle \tilde{C}_{L^{\prime}} \rangle
\end{equation}
We use the implementation \texttt{NaMaster} \citep{Alonso++19} to calculate the mode-mode coupling matrix and decoupled angular power spectra in bandpower bins. Multipole resolution is limited by $\Delta \ell \approx 180^{\circ} / \varphi$, where $\varphi$ is the smallest dimension of the angular patch, and the minimum multipole that can be meaningfully constrained is the wavelength corresponding to this angular scale \citep{Peebles80}. Since the angular power of the mask is concentrated at large modes, dropping to below 10\% power at $\ell \sim 20$, we choose a conservative binning scheme with linearly spaced bins of size $\Delta \ell = 20$ from $\ell_{\rm min} = 30$ to $\ell_{\rm max} = 1500$. However, following the approach of \cite{Krolewski19}, we run \texttt{NaMaster} out to $\ell_{\rm max} = 6000$ to avoid power leakage near the edge of the measured range.

Evaluating the observational results requires consistent application of the same binning scheme to the theory curves. Since the theory curves are not necessarily piecewise constant, they must first be convolved with the mode-mode coupling matrix $M_{\ell\ell^{\prime}}$, then binned into the appropriate bandpowers, and then finally decoupled.

Additionally, the observed auto-spectra will be a combination of signal plus noise,
\begin{align}
    C_{L}^{\rm gg} = S_{L}^{\rm gg} + N_{L}^{\rm gg} \\
    C_{L}^{\kappa \kappa} = S_{L}^{\kappa \kappa} + N_{L}^{\kappa \kappa}
\end{align}
Here, $N^{\rm gg}$ is the shot noise of the galaxy field, approximately equal to $1/\bar{n}$ (where $\bar{n}$ is the mean number of galaxies per square steradian)\footnote{We have checked that our fitting results are insensitive to the amplitude of the shot-noise term we subtract from the galaxy-galaxy spectrum, and since it is somewhat degenerate with the counter-term $\alpha_{a}$ in the perturbation theory model, we elect to fix the noise term to the Poisson expression.}, while an estimate of the lensing noise $N^{\kappa \kappa}_{\ell}$ due to e.g. instrumental and atmospheric effects is provided by the Planck collaboration and binned into bandpowers using the method discussed above. In subsequent analysis, we have subtracted the noise terms from the observed auto-spectra, as well as dividing out the appropriate pixel window functions.

\subsection{Estimating covariance matrices}

The Gaussian or ``disconnected'' part of the covariance matrix, i.e. the covariance for perfectly Gaussian fields, dominates the total covariance matrix on linear and weakly nonlinear scales. While trivial to compute for full-sky fields, the exact correlations between different modes induced by a partial sky are computationally expensive to calculate, requiring $\mathcal{O}(\ell_{\rm max}^6)$ operations (\citealt{Efstathiou04b}, \citealt{Garcia++19}). A common approximation assumes that the off-diagonal elements remain negligible after mode coupling and simply modifies the diagonal elements by rescaling the number of degrees of freedom,
%\begin{align}
%    \text{Cov}(C^{\rm XY}_{\ell},C^{\rm XY}_{\ell^{\prime}})  &= (\sigma^{\rm XY}_{\ell})^2 \delta_{\ell \ell^{\prime}} \\ 
%    (\sigma^{\rm XY}_{\ell})^2 &= \frac{[C^{\rm XX}_{\ell}C^{\rm YY}_{\ell} + (C^{\rm XY}_{\ell})^2]_{\rm theory}}{2\ell + 1}\frac{1}{f_{\rm sky}}\frac{w_4}{w_2^2}
%\end{align}
\begin{align}\label{eq:cov_gen}
    \Sigma^{\rm XY}_{\ell\ell^{\prime}}  &= (\sigma^{\rm XY}_{\ell})^2 \delta_{\ell \ell^{\prime}} \\ 
    (\sigma^{\rm XY}_{\ell})^2 &= \frac{[(C^{\rm XX}_{\ell}+N^{\rm XX}_{\ell})(C^{\rm YY}_{\ell}+N^{\rm YY}_{\ell}) + (C^{\rm XY}_{\ell} + N^{\rm XY}_{\ell})^2]_{\rm th}}{f_{\rm sky}(2\ell + 1)}\frac{w_4}{w_2^2} \nonumber
\end{align}
where $f_{\rm sky}$ is the fraction of the sky masked,
\begin{equation}
    f_{\rm sky} = \int_{4\pi}d\Omega \ W(\hat{n})
\end{equation}
and $w_i$ is related to the $i$th moment as
\begin{equation}
    w_i = \frac{1}{f_{\rm sky}}\int_{4\pi}d\Omega \ W^{i}(\hat{n})
\end{equation}
The factor $f_{\rm sky}w_2^2/w_4$ accounts for the loss of modes induced by masking. This analytic expression has been shown to reproduce errors that are nearly identical to those obtained from Monte Carlo simulations \citep{MASTER}.

We average over the bandpower bins with the inverse weighting
\begin{align}
    \frac{1}{(\sigma_{L}^{\rm XY})^2} &= \frac{1}{\Delta \ell} \sum_{\ell \in L} \frac{1}{(\sigma_{\ell}^{\rm XY})^2}
\end{align}
where $\Delta \ell$ is the width of the bandpower bin.

\subsection{Pixelised maps and masks}

To create the galaxy density map, we pixelise the sky using the \texttt{HEALPix} scheme with resolution $N_{\rm SIDE} = 512$, corresponding to a pixel area of approximately $0.013$ square degrees. This resolution was chosen to avoid the shot noise limit in which most pixels contain zero or one galaxies (for our sample with mean density $\approx 610$ per square degree, it produces an average of 5-10 galaxies per pixel), while still probing the scales of interest, $\ell_{\rm max} \sim 3 \times N_{\rm SIDE} \approx 1500$. Using galaxy and random catalogs with the masks of Section~\ref{sec:data:masks} applied to both, we calculate the density contrast $\delta = n/\bar{n} - 1$ within each pixel. Under the \texttt{HEALPix} scheme, pixels have identical areas; however, the effective area of some pixels may be less than this if they straddle the irregular shape of the footprint boundary or overlap with masked regions around bright stars, large galaxies, etc. Since our masks are applied to both the galaxy and random catalogs in a consistent manner, we can use the random catalog to estimate the effective area of each pixel, and thus calculate accurate mean galaxy densities even in pixels that are partially masked.

\begin{figure}
    \centering
    \includegraphics[width=0.85\columnwidth, trim={0.1cm 0 0 0},clip]{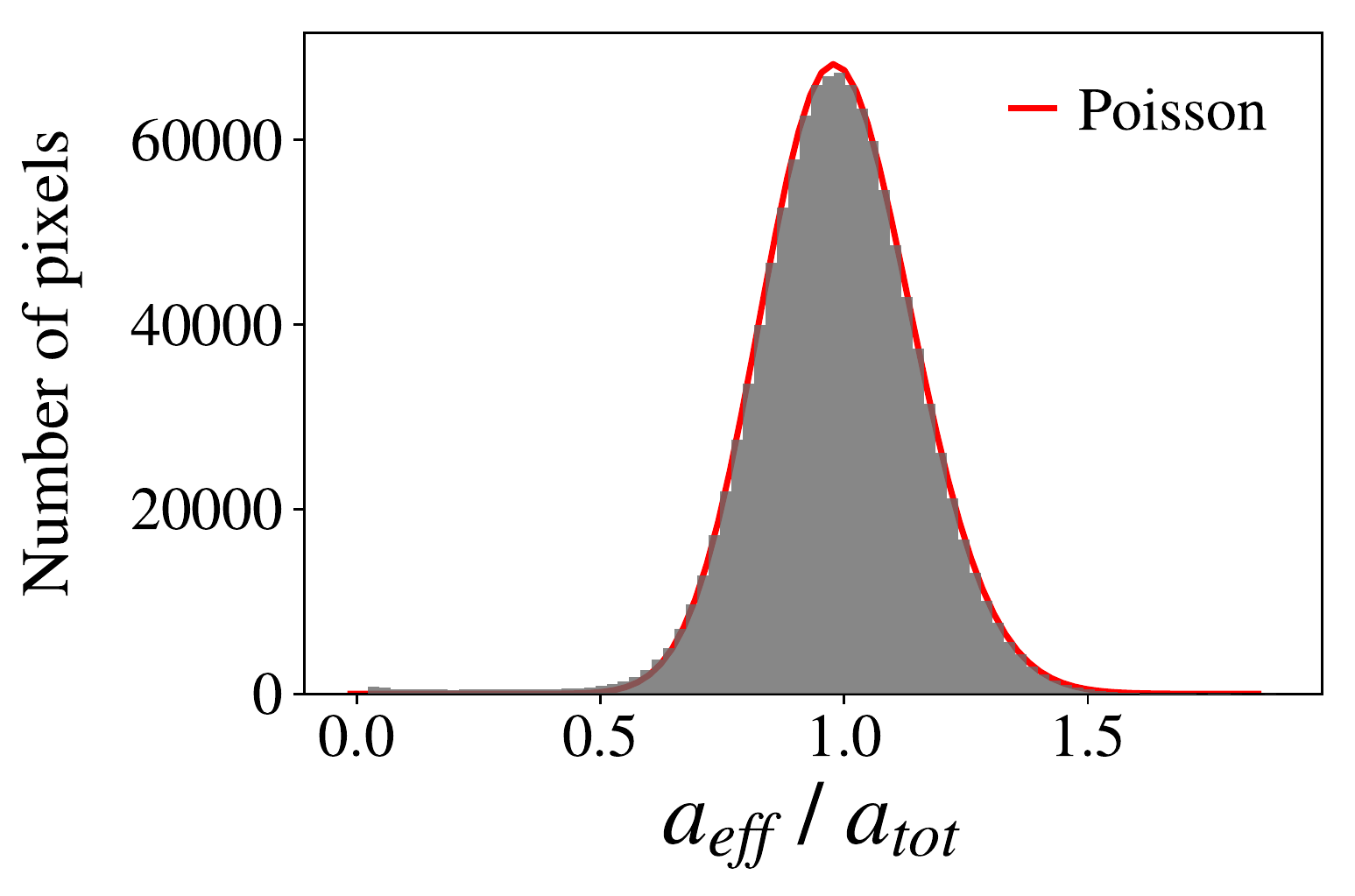}  
    \includegraphics[width=\columnwidth]{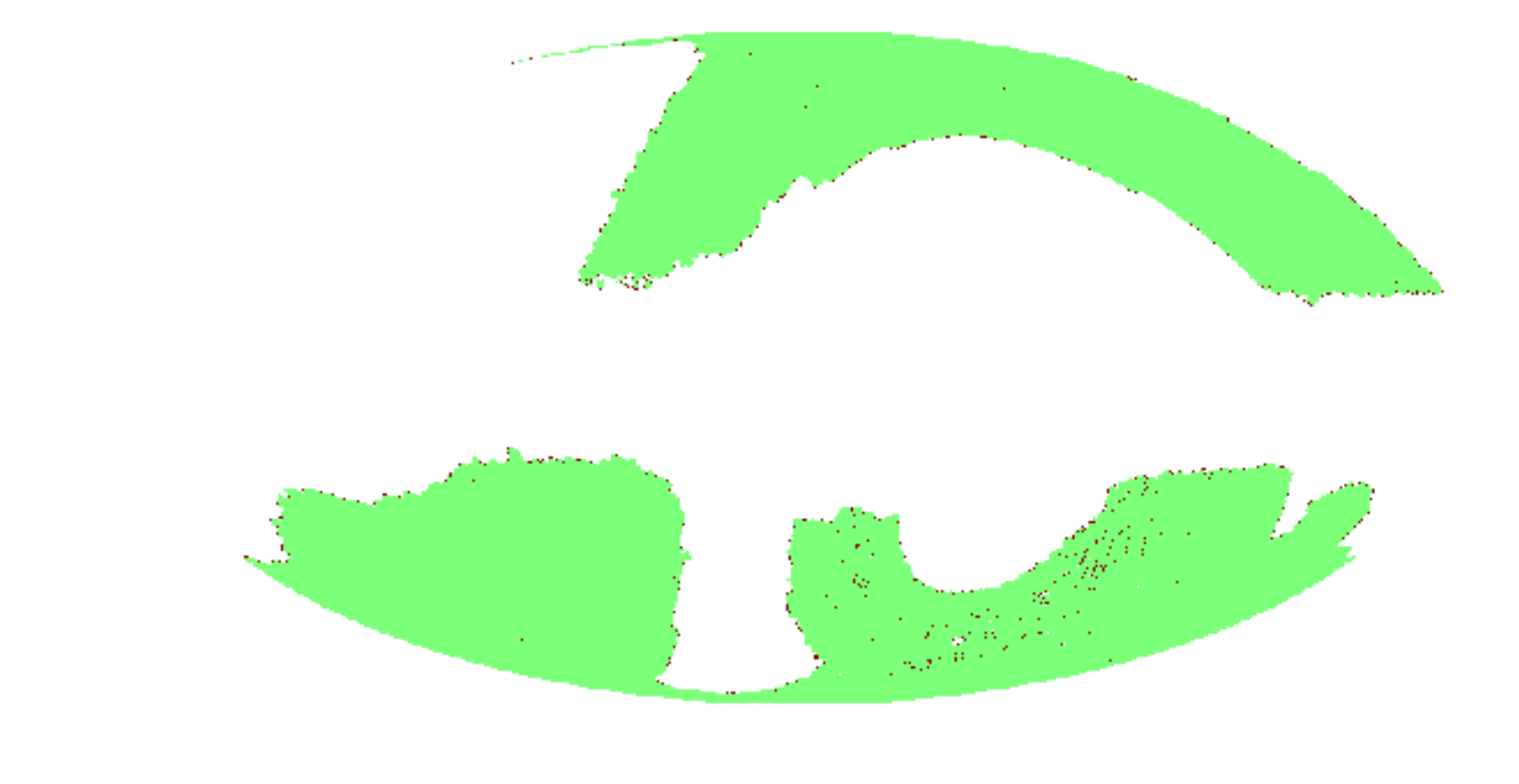}
    \caption{Upper: Histogram of the effective areas of pixels created with \texttt{HEALPix} resolution $N_{\rm SIDE} = 512$, showing a slight deviation from a Poisson distribution at the low end due to pixels straddling the footprint boundary or holes from the galaxy mask. Lower: Binary map showing that pixels selected as $a_{\rm eff}/a_{\rm tot} < 0.5$ (red pixels) lie predominately on the edges of the footprint. Most pixels have $a_{\rm eff}/a_{\rm tot} \geq 0.5$ and are shown in green.}
    \label{fig:desi_mask}
\end{figure}

To construct the pixelised galaxy mask, we measure where the distribution of effective areas deviates from a Poisson distribution, since the effective areas are estimated directly from the number of randoms per pixel, which is a Poisson process. We determine a cutoff of $a_{\rm eff}/a_{\rm tot} = 0.5$, and confirm that the pixels below this cutoff lie mainly along footprint boundary, as shown in Figure~\ref{fig:desi_mask}. Here, the effective area is calculated by using the random catalog \textit{pre}-masking, hence why the distribution is centered at $a_{\rm eff}/a_{\rm tot} \approx 1$. The equivalent distribution calculated using masked randoms results in a slightly lower mean $\tilde{a}_{\rm eff}/a_{\rm tot} \approx 0.95$ (matching the masked sky fraction of Table~\ref{tab:masks}) and an enhanced left tail since a substantial fraction of pixels are now partially masked. However, we do not necessarily need to discard these partially masked pixels as long as we are able to accurately estimate the density within them, since the pixelisation smooths the density on scales smaller than the pixel size. Hence, for our binary pixel mask, we use the cutoff calculated using the unmasked randoms, with the mask set to $1$ for $a_{\rm eff}/a_{\rm tot} > 0.5$ and 0 otherwise.

\begin{figure}
\includegraphics[width=\linewidth]{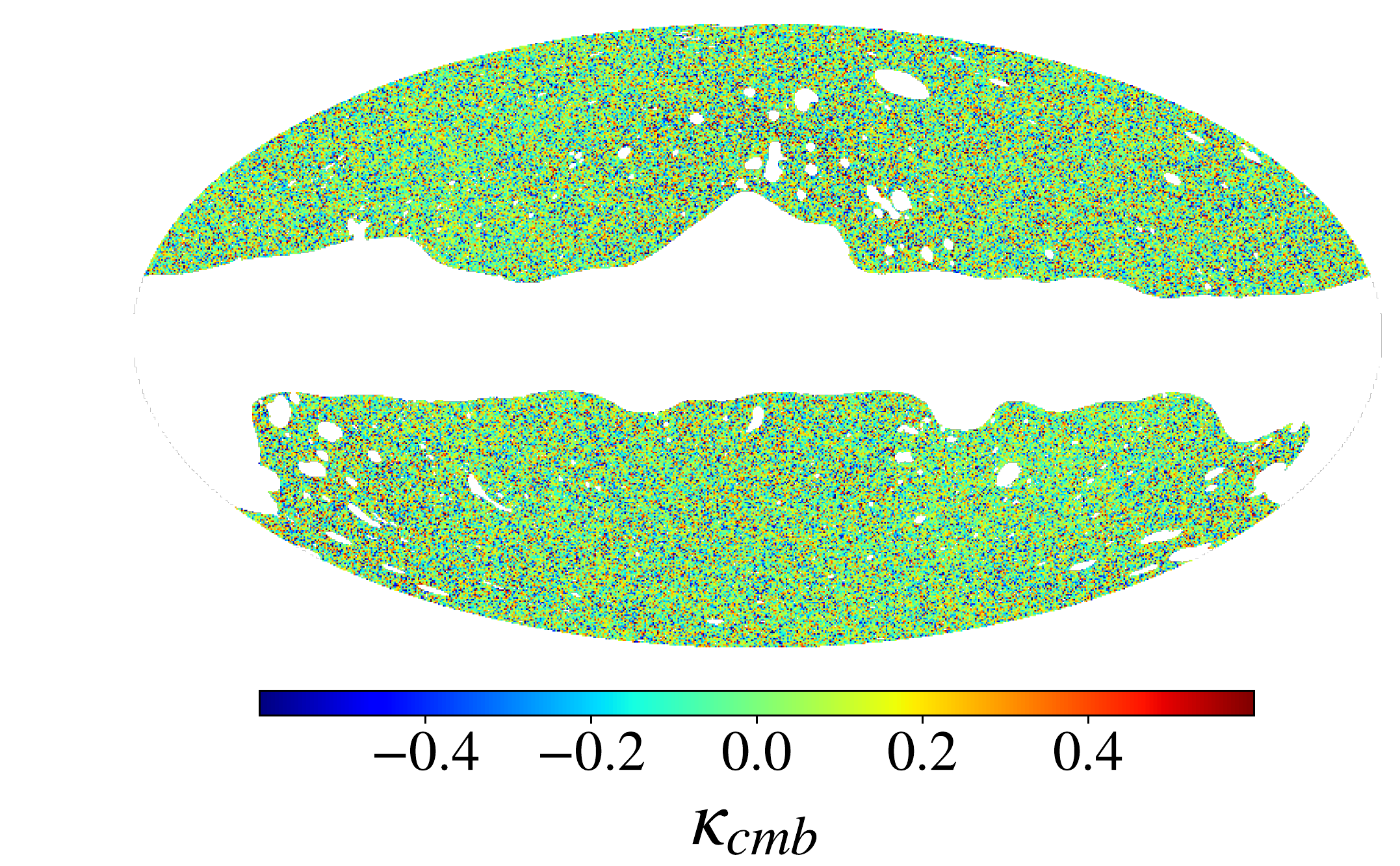} 
\includegraphics[width=\linewidth]{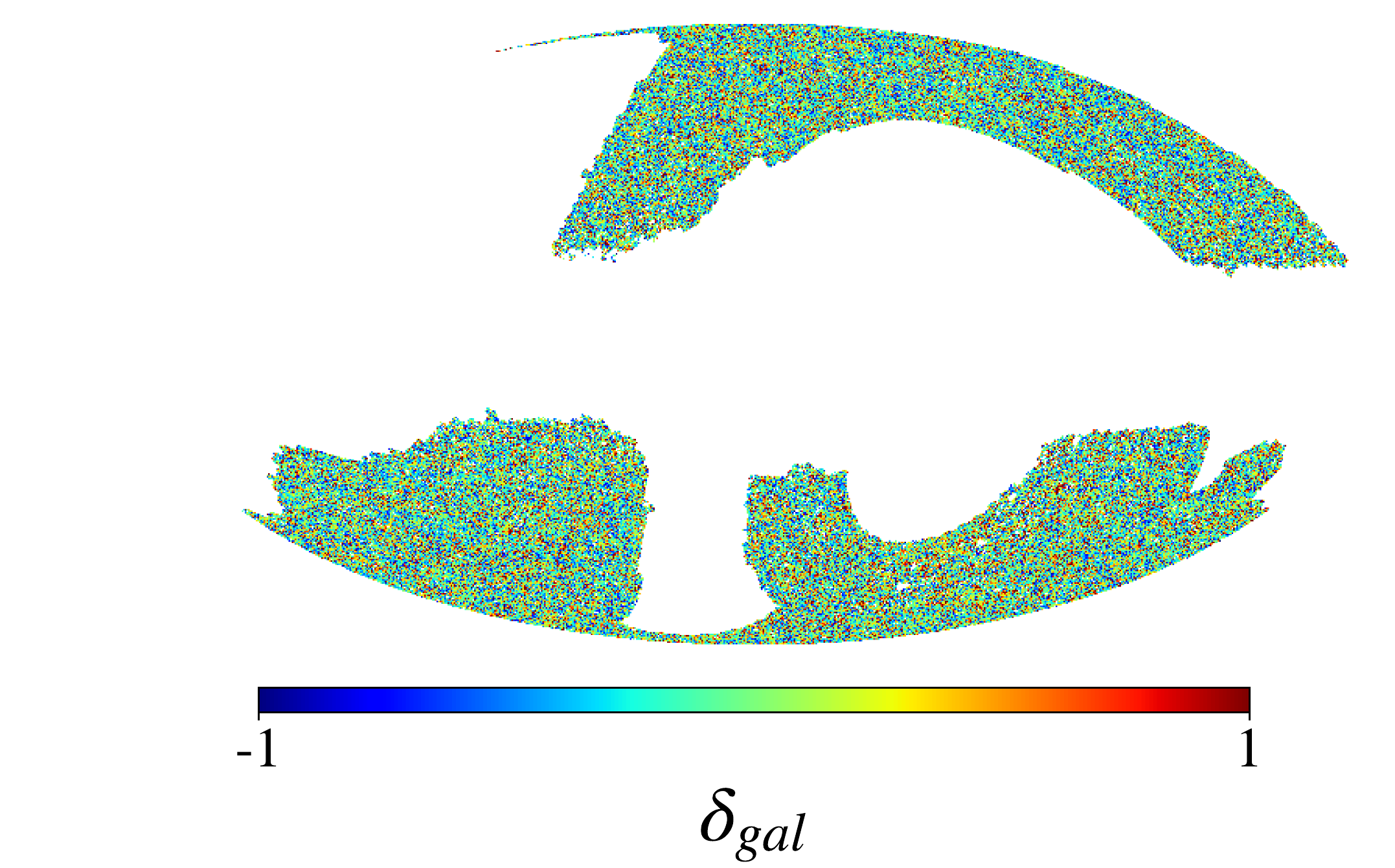}
\caption{Maps of Planck \texttt{BASE} CMB lensing convergence (upper) and DESI LRG galaxy overdensity (lower) in galactic coordinates, using \texttt{HEALPix} scheme with resolution $N_{\rm SIDE} = 2048$, Mollweide projection, and the astronomy convention (east towards left). Both maps are multiplied by their corresponding masks. The CMB lensing convergence is additionally smoothed on a scale of 10 arcmin for visual clarity.}
\label{fig:pixel_maps}
\end{figure}

The galaxy density map and mask are then upgraded to $N_{\rm SIDE} = 2048$ to match the resolution of the Planck CMB lensing map and mask, and converted from equatorial to galactic coordinates. To improve the stability of the matrix inversion, the Planck mask is apodized using a $1^{\circ}$ FWHM Gaussian\footnote{\citealt{Krolewski19} determined this to be the optimal smoothing scale for the Planck mask by testing on Gaussian simulations.}. The resulting masked galaxy density and CMB lensing convergence maps are shown in Figure~\ref{fig:pixel_maps}.

\section{Magnification Bias}\label{sec:magbias}

Magnification bias is a well-known weak lensing effect (for a review of weak lensing, we refer the reader to \citealt{BartelmannSchneider01}) that modulates the number density of galaxies in a flux-limited survey. When distant galaxies are magnified by gravitational lenses along the line-of-sight, their observed number per unit area is decreased due to the apparent stretching of space between and around them. At the same time, there is a corresponding increase in their observed brightness. As a consequence, the lensed galaxies are drawn from a fainter source population than the unlensed galaxies, leading to an increase in the number count as galaxies that would normally fall below the limiting magnitude of the survey become detectable with magnification. Through these two competing effects, magnification induces correlations between the galaxies and intervening matter in their foreground, and thus can bias the galaxy-galaxy and galaxy-convergence angular power spectra (see e.g. \citealt{Loverde++08}, \citealt{Ziour++08}, and references contained therein).

In practice, the magnification bias introduces an additional term in the galaxy window function,
\begin{align}
    W^{g}(z)  &\xrightarrow{ } W^{g}(z) + W^{\mu}(z)
\end{align}
which, to first order, is given by
\begin{align}
    W^{\mu}(z) &= (5s-2)\frac{3}{2c} \Omega_{m0} \frac{H_0^2}{H(z)} (1+z) \int_z^{z^{*}} dz^{\prime}g(z^{\prime}) \\
    g(z^{\prime}) &= \frac{\chi(z)(\chi(z^{\prime})-\chi(z))}{\chi(z^{\prime})} \phi(z^{\prime})
\end{align}
where $s$ is the slope of the cumulative magnitude function, i.e.\ the response of the number density of the sample to a multiplicative change in brightness at the limiting magnitude of the survey,
\begin{align}
s = \frac{d\log_{10}n(m<m_{\rm lim})}{dm}|_{m=m_{\rm lim}}
\end{align}
This $W^{\mu}$ term in the galaxy window function leads to additional terms in the galaxy-convergence and galaxy-galaxy angular power spectra,
\begin{align}
    C_{\ell}^{\rm \kappa g} &\xrightarrow{} C_{\ell}^{\rm \kappa g} + C_{\ell}^{\kappa \mu} \label{eq:mag-bias-corrections1} \\
    C_{\ell}^{\rm gg} &\xrightarrow{} C_{\ell}^{\rm gg} + 2 C_{\ell}^{\rm g\mu} + C_{\ell}^{\mu\mu} \label{eq:mag-bias-corrections2}
\end{align}

We calculate $s$ by perturbing the observed optical and infrared magnitudes of the imaged objects by a small differential in each direction $\Delta m = \pm 0.01$, then reapplying target selection (as defined in Section~\ref{sec:data:ts}) and measuring the corresponding shifts in the number density of the new LRG samples. Using the finite difference method, we determine $s = 0.999 \pm 0.015$, with the error computed as $\Delta s = (\log_{10}(N) - \log_{10}(N - \sqrt{N}))/\Delta m$. 

\begin{figure}
\includegraphics[width=\linewidth,trim={0 0 0 0}]{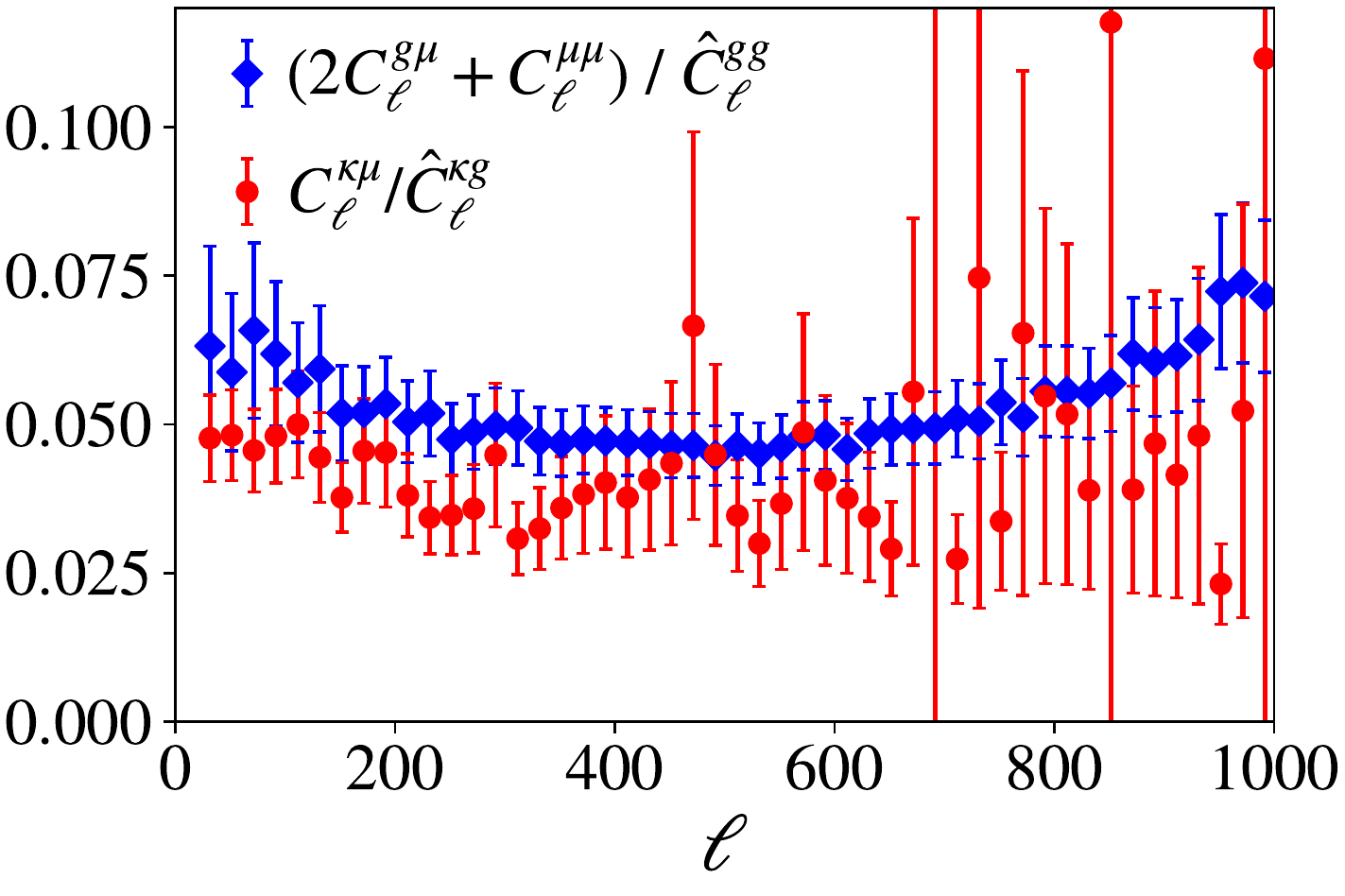}
\caption{The magnification bias terms of Equations~\ref{eq:mag-bias-corrections1} and \ref{eq:mag-bias-corrections2} as a fraction of the total observed (after subtracting shot noise, in the galaxy-galaxy case) spectra, i.e.\ before correcting for magnification bias. The error bars represent error on the fraction and are dominated by the errors of the denominator.}
\label{fig:mag_bias_frac}
\end{figure}

We plot the magnification bias corrections\footnote{These terms are calculated using the photometric redshift distribution, so as to avoid assuming a bias evolution model.} as a fraction of the observed spectra (after noise subtraction) in Figure~\ref{fig:mag_bias_frac}. 
% These terms are calculated assuming a \texttt{HaloFit} power spectrum and fiducial bias model $b(z) = 1.6 / D(z)$, and using the photometric redshift distribution.
The corrections to the galaxy-galaxy spectrum are at a level of approximately 5\% over most of the range of scales considered, with 1-2\% increases at edges of the range $\ell < 100$ and $\ell > 900$, while the correction to the cross-spectrum is flat within error bars\footnote{The error bars are dominated by the errors in the cross-spectrum, which become significant at $\ell > 700$.} at 4-5\%. Though the DESI LRG redshift distribution is relatively narrow and peaks at $z < 1$, the high number density and low clustering bias coupled with a steep faint end slope contribute to effects at the level of a few percent, as Figure~\ref{fig:mag_bias_frac} shows.

In all subsequent results, the magnification bias terms have been subtracted from the observed spectra.

\section{Results}\label{sec:results}

\begin{figure}
\includegraphics[width=0.9\linewidth]{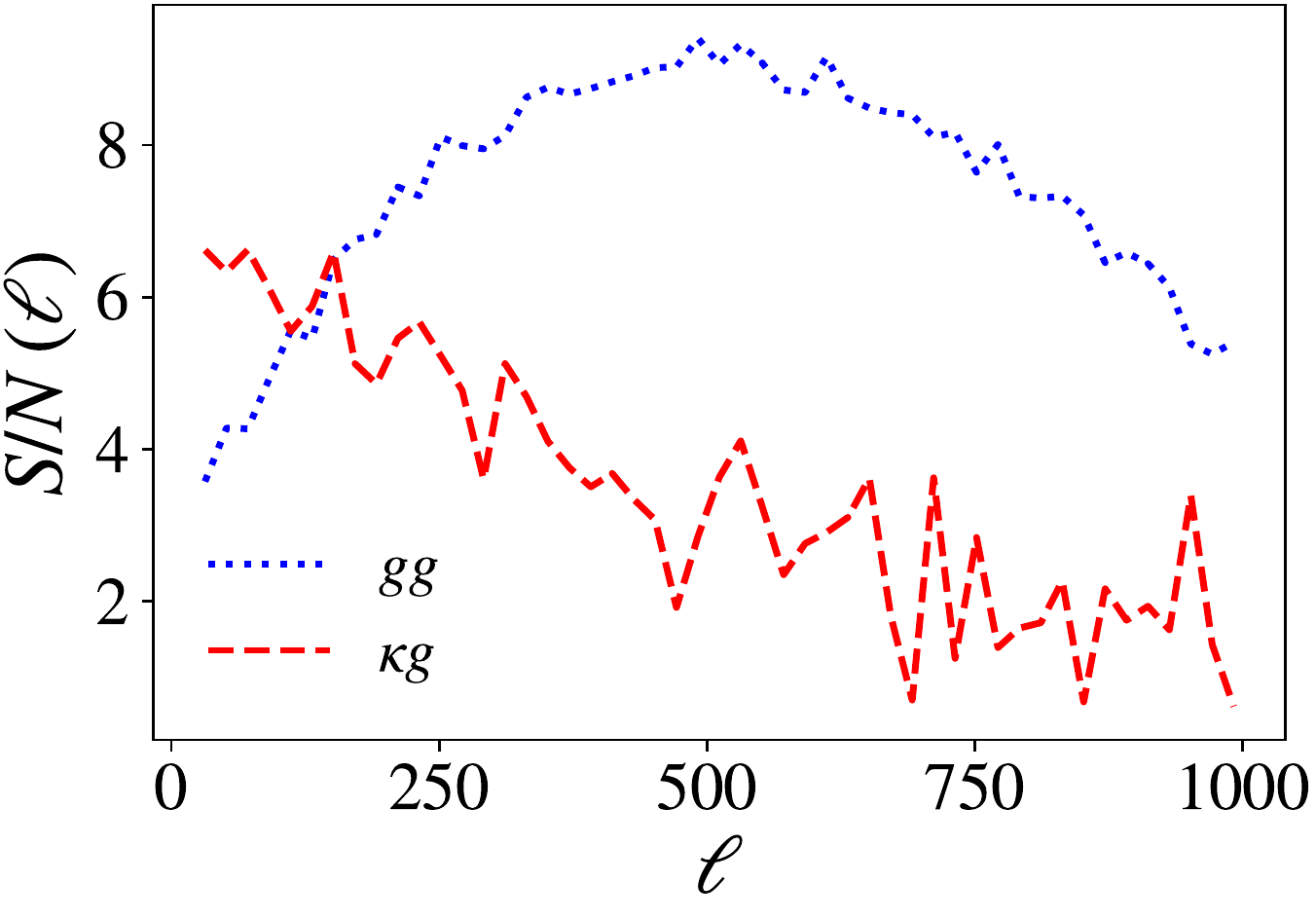}
\vspace{1cm}
\includegraphics[width=0.9\linewidth]{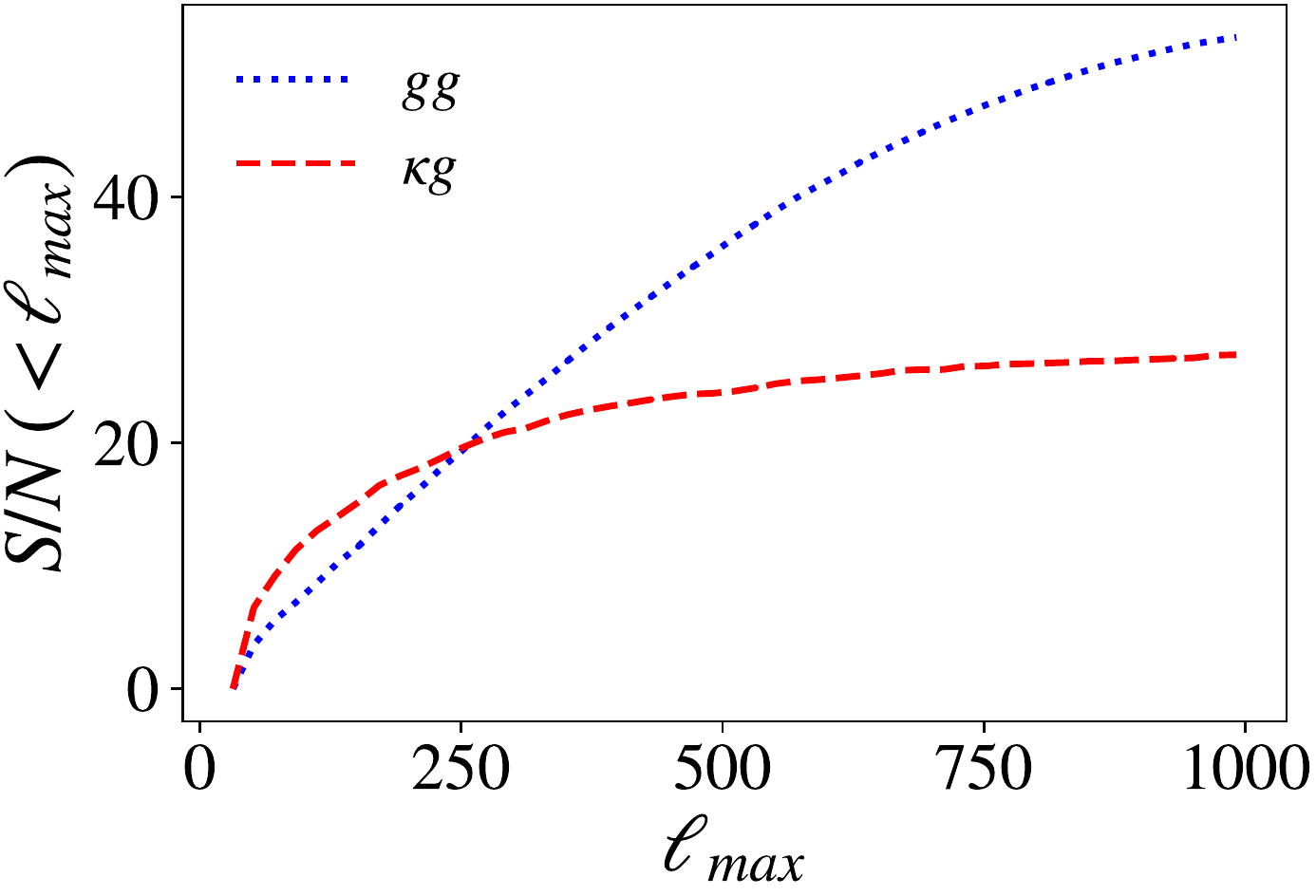}
\caption{Per multipole (upper) and cumulative (lower) signal-to-noise ratio for the galaxy-galaxy (blue dotted line) and galaxy-convergence (red dashed line) angular power spectra measurements.}
\label{fig:snr}
\end{figure}

A cross-correlation measurement between DESI-like LRGs selected from DECaLS imaging and CMB lensing from Planck 2018 is detected at a significance of S/N $= 27.2$ over the range of scales $\ell_{\rm min}=30$ to $\ell_{\rm max}=1000$. In Figure~\ref{fig:snr}, we plot per-multipole and cumulative signal-to-noise ratios for both the galaxy-galaxy and galaxy-convergence spectra, where the signal-to-noise ratio of the $XY$ angular power spectrum at multipole $\ell$ is given by
\begin{equation}
    {\bigg (} \frac{S}{N}{\bigg )}(\ell) = \frac{C_{\ell}^{\rm XY}}{\sigma_{\ell}^{\rm XY}}
\end{equation}
and the cumulative signal-to-noise ratio up to $\ell_{\rm max}$ is
\begin{equation}
    {\bigg (} \frac{S}{N}{\bigg )}(< \ell_{\rm max}) = \sqrt{\sum_{\ell^{\prime}=\ell_{\rm min}}^{\ell_{\rm max}}\frac{(C_{\ell^{\prime}}^{\rm XY})^2}{(\sigma_{\ell^{\prime}}^{\rm XY})^2}}
\end{equation}
The galaxy-galaxy S/N peaks at $\ell \sim 500$ whereas the galaxy-convergence generally decreases over the range of scales considered. We note that the theoretical galaxy-convergence S/N would be expected to peak at $\ell \sim 100$ and fall off at smaller $\ell$; as this is within the regime at which both the pseudo-$C_{\ell}$ framework and the Limber approximation begin to break down, this feature is washed out in the observed S/N.

\begin{figure}
\includegraphics[width=0.9\linewidth]{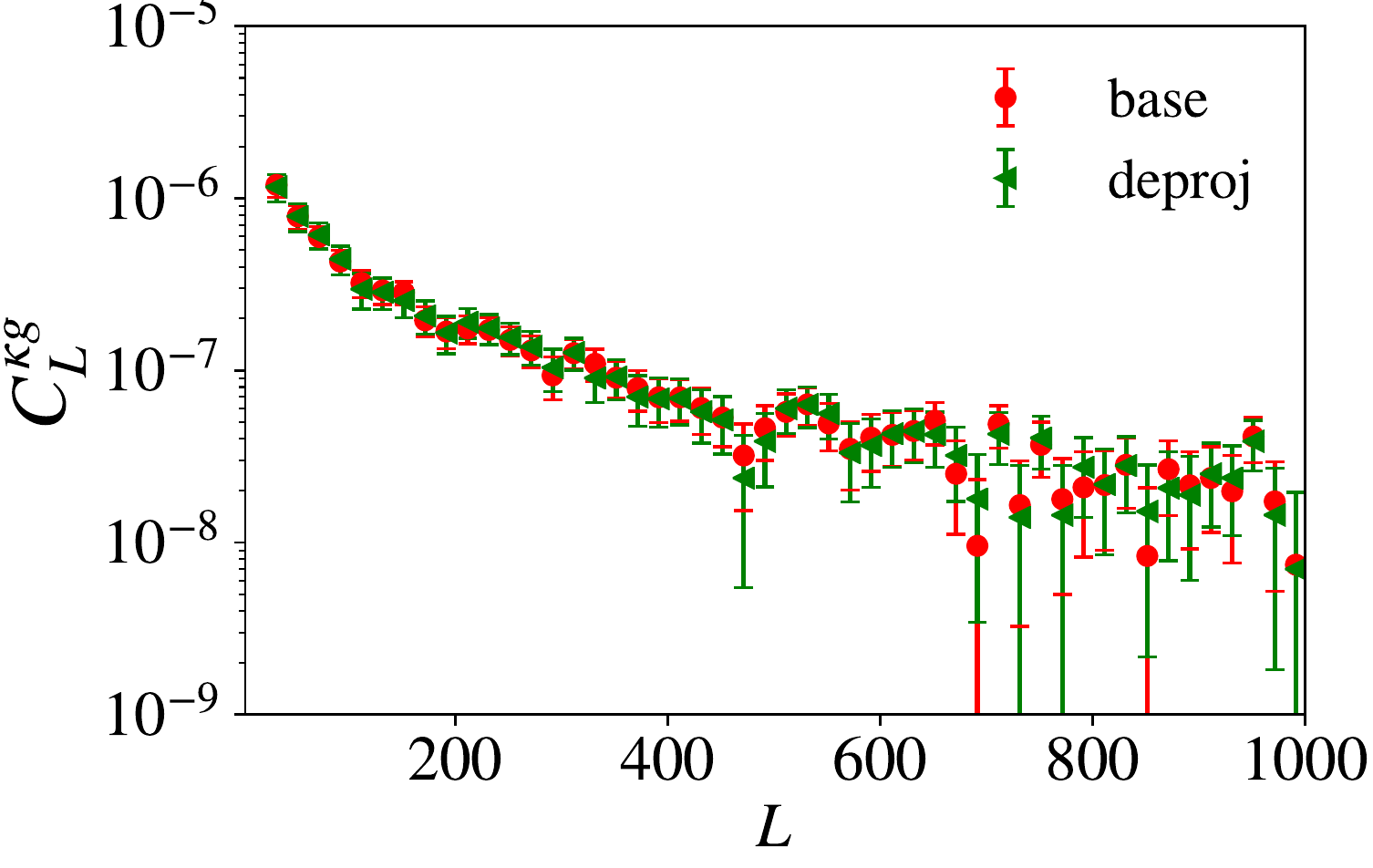}
\vspace{1cm}
\includegraphics[width=0.9\linewidth]{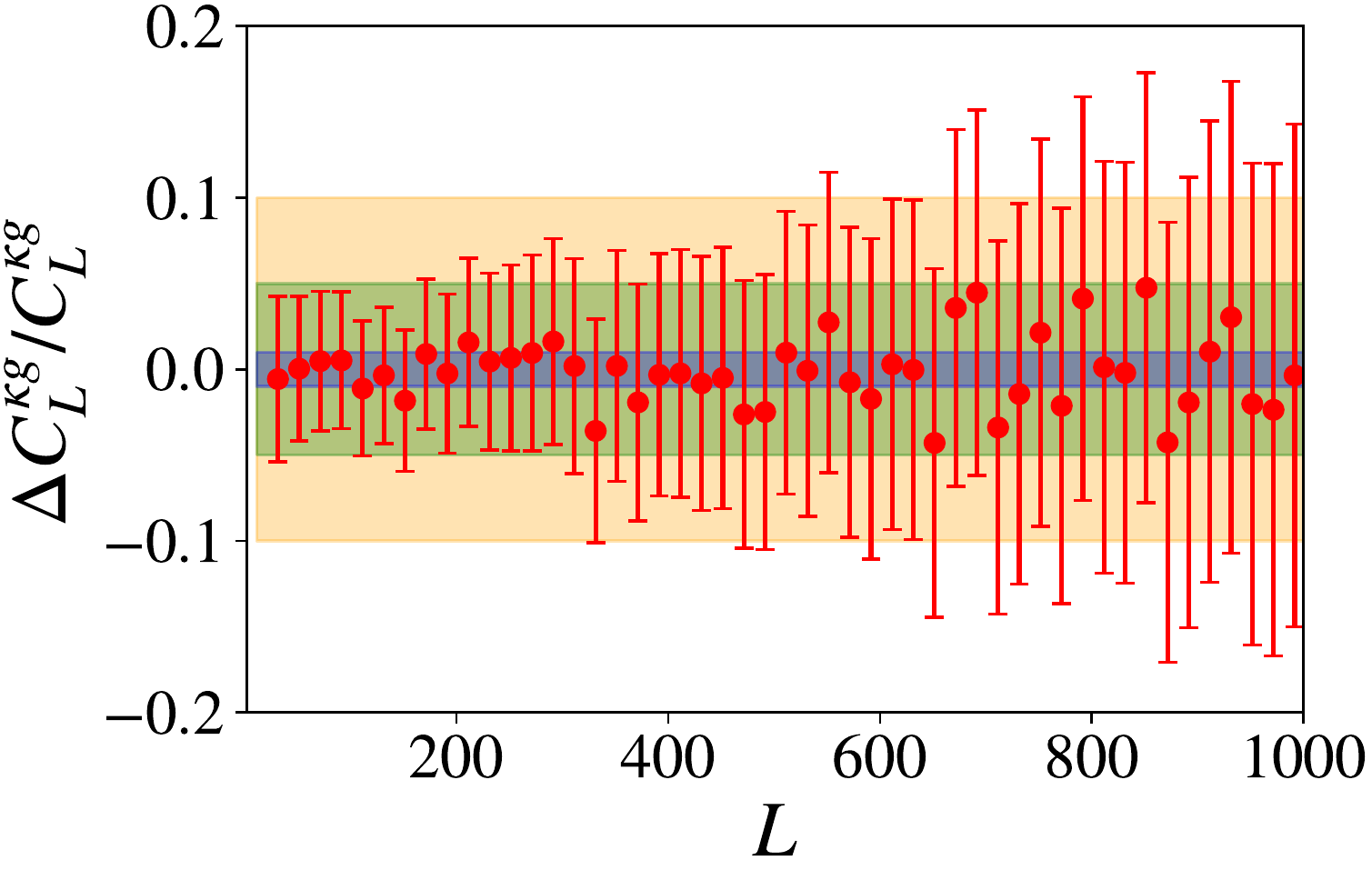}
\caption{Upper: Comparison of observed cross-spectrum $C_{L}^{\kappa g}$ calculated from \texttt{BASE} (red circles) versus \texttt{DEPROJ} (green triangles) lensing maps. Lower: The difference between the two measurements as a fraction of the theoretical prediction, with three bands illustrating 1\%, 5\% and 10\%.}
\label{fig:basevdeproj}
\end{figure}

We also compare the cross-spectrum using the baseline MV CMB lensing map versus using the TT-only tSZ-deprojected map. The two curves are shown in the top panel of Figure~\ref{fig:basevdeproj}, and clearly lie well within $1\sigma$ of one another. The fractional difference is shown in the lower panel. Since the error bars on the cross-spectra are generally large, they dominate the errors on the fraction, but as Figure~\ref{fig:basevdeproj} illustrates, the errors associated with tSZ are on the order of a few percent and very subdominant to the overall lensing noise.

In the following sub-sections, we present the angular power spectra and interpret them using two different models for the galaxy-galaxy and matter-galaxy 3D power spectra: the HaloFit dark matter power spectrum multiplied by a linear large-scale bias, and a convolutional Lagrangian effective field theory with Lagrangian bias. Additionally, we perform the fits using both photometric- and clustering-derived redshift distributions for the galaxy sample, which not only suggests an estimate of the error associated with uncertainty in the redshift distribution, but also allows us to evaluate the bias at an effective redshift $z \approx 0.68$ in both models and to test the assumption of passive bias evolution.

\subsection{HaloFit Modeling}\label{sec:results/halofit}

Within a framework for modeling the galaxy-galaxy and matter-galaxy power spectra $P_{\rm gg}(k)$, $P_{\rm mg}(k)$, the observed angular power spectra $C_{\ell}^{\rm gg}$, $C_{\ell}^{\rm \kappa g}$ can constrain cosmological and galaxy bias parameters. A particularly simple and interpretable model is to use the HaloFit \citep{Smith++03} fitting function for the nonlinear dark matter power spectrum, $P_{\rm mm}^{\rm HF}(k)$, and multiply by scale-independent linear biases to obtain the galaxy-galaxy and galaxy-matter power spectra,
\begin{align}
    P_{\rm gg}(k,z) &= b_{\rm gg}(z)^2 P_{\rm mm}^{\rm HF}(k,z) \\
    P_{\rm \kappa g}(k,z) &= b_{\rm \kappa g}(z) P_{\rm mm}^{\rm HF}(k,z)
\end{align}
Differences between $b_{\rm gg}$ and $b_{\rm \kappa g}$ are expected, due in large part to the stochastic contribution arising from the the fact that the galaxy field is a discrete sampling of the underlying dark matter distribution. As such, this stochastic component, which may include scale-dependent and non-Poissonian behavior, affects the galaxy-galaxy auto-spectrum and matter-galaxy cross-spectrum differently.

Using the Boltzmann code \texttt{CLASS} \citep{Blas++11} to calculate the HaloFit dark matter power spectrum for the fiducial Planck 2018 cosmology, we take the photometric $\phi(z)$ and assume a bias evolution $b_{\rm gg}(z), b_{\rm \kappa g}(z) \propto D(z)^{-1}$. We then perform weighted least squares fits of the present day biases. The results are given in Table~\ref{tab:halofit_photo}, with the fits repeated for $\ell_{\rm max} = 200, 400, 600, 800, 1000$. We find that the linear biases are unaffected by the choice of $\ell_{\rm max}$ and that the cross bias $b_{\rm \kappa g}$ is consistently lower than the galaxy bias $b_{\rm gg}$, with the latter agreeing well with DESI survey expectations and the findings of \citealt{Kitanidis++19}.  The lower-than-expected $b_{\rm \kappa g}$ could arise from choosing an incorrect fiducial cosmology (e.g.\ lowering $\Omega_m$ would reduce $b_{\rm \kappa g}$ with only a small impact on $b_{\rm gg}$; see also Hang et al., in prep).  It could also be due to the assumed bias evolution, the assumed form for $\phi(z)$ or limitations of our model.  We shall consider these next.

\begin{table*}
\centering
HaloFit Model, Photo $\phi(z)$ \\
\begin{tabular}{c|cccc|cccc|}
\hline
$\ell_{\rm max}$ & $b_{\rm gg}$ & $\chi_{\rm gg}^2 / \text{d.o.f.}$ & $\text{PTE}_{\rm gg}$ & $\text{SNR}_{\rm gg}(< \ell_{\rm max})$ & $b_{\rm \kappa g}$ & $\chi_{\rm \kappa g}^2 / \text{d.o.f.}$ & $\text{PTE}_{\rm \kappa g}$ & $\text{SNR}_{\rm \kappa g}(< \ell_{\rm max})$ \\
\hline
200 & $1.57 \pm 0.05$ & 0.7 / 8  & 0.9995 & 15.5 & $1.27 \pm 0.07$ & 4.2 / 8  & 0.8386 & 17.6  \\
400 & $1.63 \pm 0.03$ & 3.4 / 18 & 0.9999 & 29.9  & $1.32 \pm 0.06$ & 8.1 / 18  & 0.9771 & 23.1   \\
600 & $1.66 \pm 0.02$ & 8.4 / 28 & 0.9999 & 41.3 & $1.32 \pm 0.05$ & 12.1 / 28  & 0.9961 & 25.2   \\
800 & $1.67 \pm 0.02$ & 9.9 / 38 & 1.0000 & 49.0  & $1.32 \pm 0.05$ & 20.9 / 38  & 0.9890 & 26.5   \\
\textbf{1000} & \bm{$1.64 \pm 0.02$} & \textbf{30.4 / 48}  & \textbf{0.9777} & \textbf{53.0}  & \bm{$1.32 \pm 0.05$} & \textbf{26.8 / 48} & \textbf{0.9943} & \textbf{27.2}  \\
\hline
\end{tabular}
\caption{Fitting linear bias from the observed $C_{\ell}^{\rm gg}$, $C_{\ell}^{\rm \kappa g}$ up to different $\ell_{\rm max}$ using the HaloFit model for the nonlinear dark matter power spectrum, photometric $\phi(z)$, and the assumption $b(z) \propto D(z)^{-1}$.}
\label{tab:halofit_photo}
\end{table*}

\begin{figure}
\includegraphics[width=\linewidth]{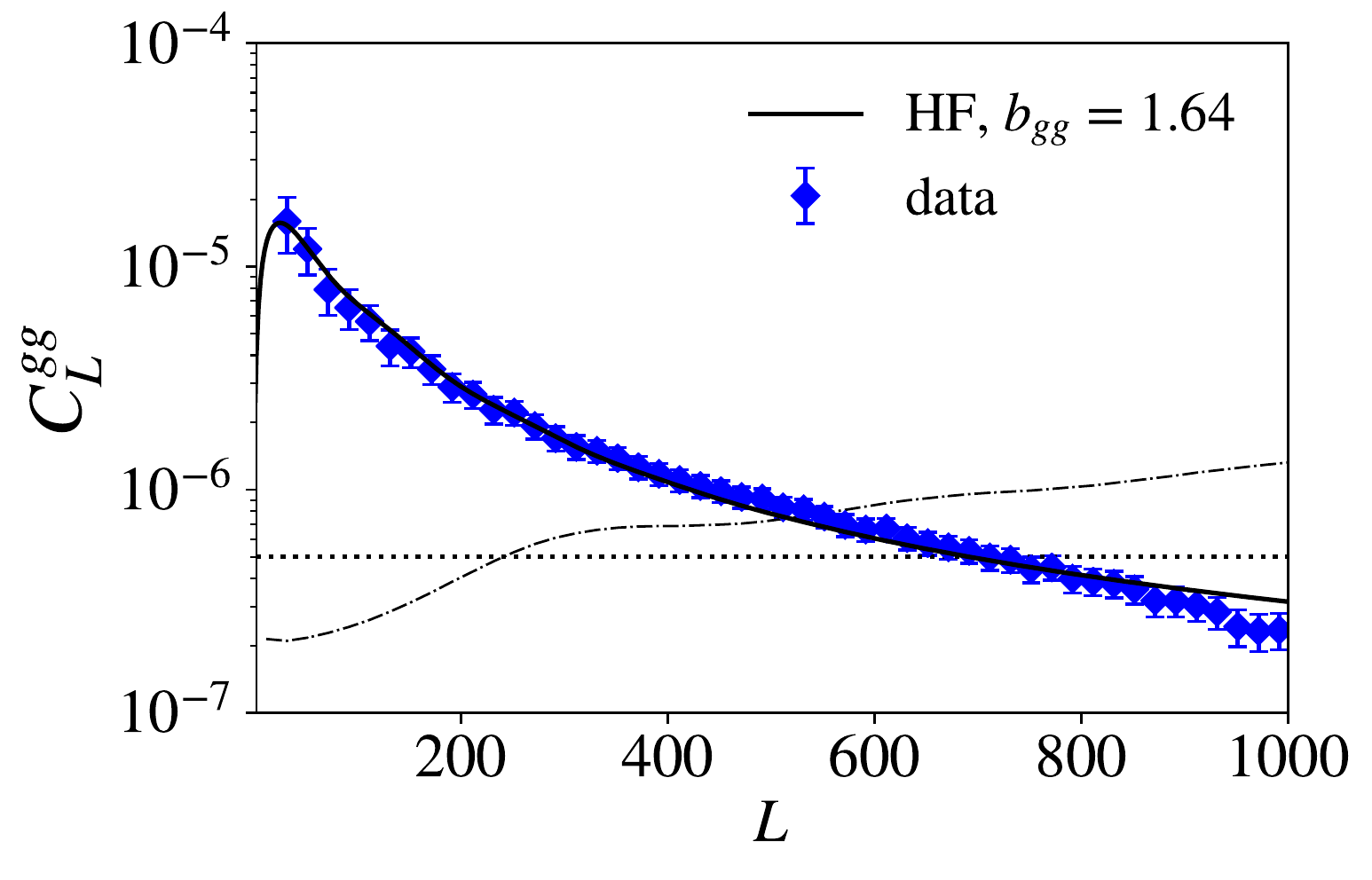}
\includegraphics[width=\linewidth]{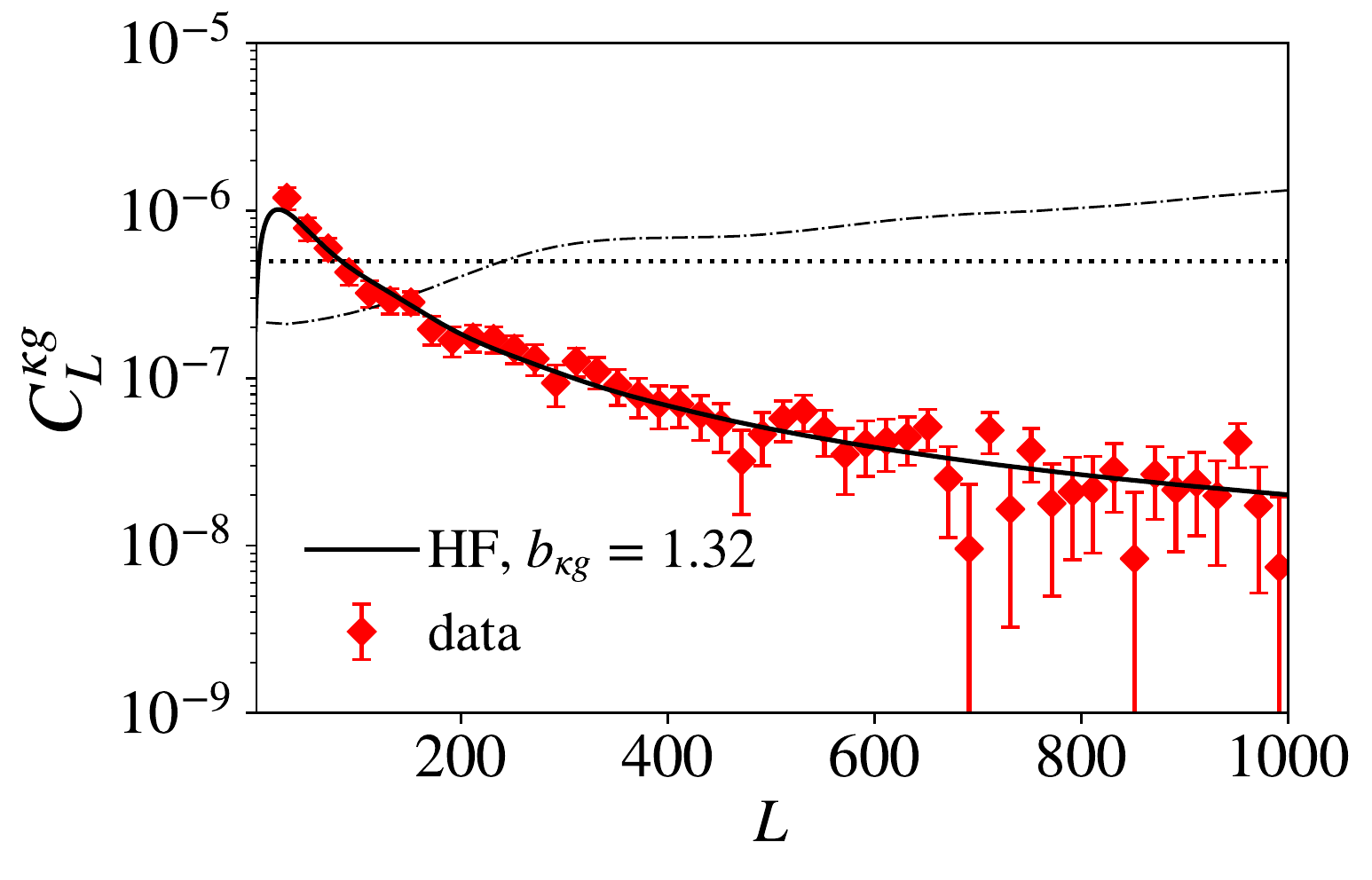}
\caption{The observed galaxy-galaxy (upper plot, blue diamonds) and galaxy-convergence (lower plot, red diamonds) angular power spectra, after subtracting noise and correcting for magnification bias. Solid lines correspond to the the theoretical predictions using a HaloFit matter power spectrum and the best fit linear biases from Table~\ref{tab:halofit_photo}. The dotted horizontal line is the galaxy shot noise floor, and the dashed black curve is the lensing noise.}
\label{fig:cl_halofit}
\end{figure}

We then repeat the same measurement using the clustering-derived $\phi(z)$ discussed in Section~\ref{sec:dndz_pipe}, again finding that the choice of $\ell_{\rm max}$ has negligible impact. The results, given in Table~\ref{tab:halofit_clustering}, show that uncertainty in the redshift distribution causes a difference in the derived galaxy bias parameters of $\sigma_{b_{\rm gg}} = 0.08$. By contrast, the cross bias is extremely stable with respect to changes in the redshift distribution, not changing at all when the redshift distribution is changed from the photometric estimate to the clustering estimate; this may be explained by the fact that the cross-spectrum only depends on one factor of $\phi(z)$ while the auto-spectrum requires $\phi^2(z)$.

Another advantage of using the clustering-based $\phi(z)$ is the ability to extract a galaxy redshift kernel with bias evolution baked in, rather than assuming a parametric form e.g. $b(z) \propto D(z)^{-1}$. As discussed in Section~\ref{sec:dndz_pipe/q}, this type of modeling allows us to constrain $b_{\rm eff} \approx b(z_{\rm eff})$ rather than the present day bias. We find the results, given in Table~\ref{tab:halofit_clustering_noevo}, to be in perfect agreement with the results of Table~\ref{tab:halofit_clustering} under the assumption $b(z) \propto D(z)^{-1}$ used in the latter, giving for instance $b_{\rm gg}= 1.56 \pm 0.01$ and $b_{\rm \kappa g}= 1.31 \pm 0.05$ for the $\ell_{\rm max} = 1000$ case. 

\begin{table*}
\centering
HaloFit Model, Clustering $\phi(z)$ \\
\begin{tabular}{c|cccc|cccc|}
\hline
$\ell_{\rm max}$ & $b_{\rm gg}$ & $\chi_{\rm gg}^2 / \text{d.o.f.}$ & $\text{PTE}_{\rm gg}$ & $\text{SNR}_{\rm gg}(< \ell_{\rm max})$ & $b_{\rm \kappa g}$ & $\chi_{\rm \kappa g}^2 / \text{d.o.f.}$ & $\text{PTE}_{\rm \kappa g}$ & $\text{SNR}_{\rm \kappa g}(< \ell_{\rm max})$ \\
\hline
200 & $1.50 \pm 0.05$ & 0.8 / 8  & 0.9992 & 15.5 & $1.27 \pm 0.07$ & 4.0 / 8  & 0.8571 &  17.6   \\
400 & $1.55 \pm 0.03$ & 3.4 / 18  & 0.9999 & 29.9  & $1.32 \pm 0.06$ & 8.0 / 18   & 0.9786 & 23.1 \\
600 & $1.59 \pm 0.02$ & 8.7 / 28  & 0.9998 & 41.3 &  $1.32 \pm 0.05$ & 11.9 / 28  & 0.9966 & 25.2  \\
800 & $1.59 \pm 0.02$ & 10.2 / 38  & 1.0000 & 49.0  & $1.32 \pm 0.05$ & 20.8 / 38   & 0.9895 & 26.5  \\
\textbf{1000} & \bm{$1.56 \pm 0.01$} & \textbf{29.9 / 48}& \textbf{0.9812} & \textbf{53.0}  & \bm{$1.32 \pm 0.05$} & \textbf{26.7 / 48}  & \textbf{0.9946} & \textbf{27.2} \\
\hline
\end{tabular}
\caption{Fitting linear bias from the observed $C_{\ell}^{\rm gg}$, $C_{\ell}^{\rm \kappa g}$ up to different $\ell_{\rm max}$ using the HaloFit model for the nonlinear dark matter power spectrum, clustering $\phi(z)$, and the assumption $b(z) \propto D(z)^{-1}$.}
\label{tab:halofit_clustering}
\end{table*}

\begin{table*}
\centering
HaloFit Model, Clustering $b(z)\phi(z)$ \\ 
\begin{tabular}{c|cccc|cccc|}
\hline
$\ell_{\rm max}$ & $b^{\rm eff}_{\rm gg}$ & $\chi_{\rm gg}^2 / \text{d.o.f.}$ & $\text{PTE}_{\rm gg}$ & $\text{SNR}_{\rm gg}(< \ell_{\rm max})$ & $b^{\rm eff}_{\rm \kappa g}$ & $\chi_{\rm \kappa g}^2 / \text{d.o.f.}$ & $\text{PTE}_{\rm \kappa g}$ & $\text{SNR}_{\rm \kappa g}(< \ell_{\rm max})$ \\
\hline
200 & $2.14 \pm 0.07$ & 0.8 / 8  & 0.9992 & 15.5   & $1.80 \pm 0.10$ & 4.0 / 8  & 0.8571 & 17.6   \\
400 & $2.21 \pm 0.04$ & 3.4 / 18  & 0.9999 & 29.9   & $1.89 \pm 0.08$ & 8.0 / 18  & 0.9786 & 23.1   \\
600 & $2.26 \pm 0.03$ & 8.7 / 28  & 0.9998 & 41.3   & $1.88 \pm 0.08$ & 12.0 / 28  & 0.9964 & 25.2   \\
800 & $2.27 \pm 0.02$ & 10.2 / 38  & 1.0000 & 49.0   & $1.88 \pm 0.07$ & 20.8 / 38  & 0.9895 & 26.5   \\
\textbf{1000} & \bm{$2.23 \pm 0.02$} & \textbf{29.9 / 48}  & \textbf{0.9812} & \textbf{53.0}   & \bm{$1.88 \pm 0.07$} & \textbf{26.7 / 48}  & \textbf{0.9946} & \textbf{27.2}  \\
\hline
\end{tabular}
\caption{Fitting effective bias $b_{\rm eff} \approx b(z_{\rm eff} = 0.68)$ from the observed $C_{\ell}^{\rm gg}$, $C_{\ell}^{\rm \kappa g}$ up to different $\ell_{\rm max}$ using the HaloFit model for the nonlinear dark matter power spectrum, clustering $b(z)\phi(z)$ (normalized), and no assumptions regarding the shape of the bias evolution.}
\label{tab:halofit_clustering_noevo}
\end{table*}

\subsection{Perturbation Theory Modelling}\label{sec:results/lpt}

We next apply an analytic model that allows more nuance in the handling of bias. Higher order perturbation theory is a natural approach considering that the cross-correlation is most sensitive to structure at large scales, as shown in see Figure~\ref{fig:snr}. We use a Lagrangian bias model and the convolution Lagrangian effective field theory (hereafter CLEFT) outlined in \citealt{Vlah++16} and the references contained therein. Under this formalism, the matter-galaxy and galaxy-galaxy power spectra are (see e.g.\ Equation 2.7 from \citealt{Modi++17} and Equation B.2 from \citealt{Vlah++16}):
\begin{align}\label{eqn:lpt1}
    P_{\rm mg} &= (1-\frac{\alpha_{\times}k^2}{2})P_{\rm Z} + P_{\rm 1L} + \frac{b_1}{2}P_{\rm b_1} + \frac{b_2}{2}P_{\rm b_2} \\
    \label{eqn:lpt2}
    P_{\rm gg} &= (1-\frac{\alpha_{a}k^2}{2})P_{\rm Z} + P_{\rm 1L} + b_1 P_{\rm b_1} + b_2 P_{\rm b_2} + \\
    & \hspace{0.35cm} b_1 b_2 P_{\rm b_1 b_2}  + b_1^2 P_{\rm b_1^2} + b_2^2 P_{\rm b_2^2} \nonumber
%    & \hspace{0.3cm} + b_{s^{2}} P_{\rm b_{s^{2}}} + b_1 b_{s^{2}} P_{\rm b_1 b_{s^{2}}} \nonumber \\
%    & \hspace{0.3cm}  + b_2 b_{s^{2}} P_{\rm b_2 b_{s^{2}}} + b_{s^{2}}^2 P_{\rm (b_{s^{2}})^2} \nonumber
\end{align}
where we have dropped the terms corresponding to shear bias as we find they mainly affect scales $\ell > 1000$. Here, $P_{\rm Z}$ and $P_{\rm 1L}$ are the Zeldovich and 1-loop dark matter contributions (see e.g. \citealt{Vlah++15}), $b_1$ and $b_2$ are the Lagrangian bias parameters for the galaxy sample, and the effective field theory terms $\alpha_{\times}$ and $\alpha_a$ (which are not necessarily the same) are free parameters encapsulating the small-scale physics not modeled by Lagrangian perturbation theory.

Under the CLEFT formalism, the power spectrum contributions $P_{\rm Z}$, $P_{\rm 1L}$, $P_{\rm b_1}$, $P_{\rm b_2}$, etc. can be computed analytically and combined with the free parameters $\alpha_{\times}, \alpha_a, b_1, b_2$. With these additional degrees of freedom, CLEFT provides a more flexible model than the phenomenological approach of Section~\ref{sec:results/halofit}, and allows us to fit the cross-spectrum and galaxy auto-spectrum simultaneously.

We use a version of the public code \texttt{velocileptors}\footnote{\url{https://github.com/sfschen/velocileptors}} \citep{velocileptors} to calculate the power spectrum terms and the MCMC likelihood estimator \texttt{emcee}\footnote{\url{https://github.com/dfm/emcee}} \citep{emcee} to optimize our model parameters. 
%In addition to the six free parameters in Equations~\ref{eqn:lpt1} and \ref{eqn:lpt2} representing the galaxy-halo connection, we also wish to constrain the cosmological parameter $\sigma_8$; with galaxy clustering alone, there is significant degeneracy between $\sigma_8$ and galaxy bias, but this degeneracy is broken by including the cross-correlation with CMB lensing convergence. 
To reduce model expense, we evaluate the power spectrum terms at a single effective redshift,
\begin{equation}
    z_{\rm eff}^{XY} = \frac{\int d\chi \ z \ W^{X}(\chi)W^{Y}(\chi)/\chi^2}{\int d\chi \ W^{X}(\chi)W^{Y}(\chi)/\chi^2}
\end{equation}
which is $z_{\rm eff} = 0.67$ for $\kappa g$ and $z_{\rm eff} = 0.68$ for $gg$\footnote{To jointly fit the auto- and cross-spectrum, we assume $z_{\rm eff} = 0.68$ for both.}. Given the narrow redshift distribution and likely passive bias evolution of our galaxy sample, this substitution should not affect the $C_{\ell}$'s significantly \citep{Modi++17}, and we confirm that the overall impact on the scales of interest is sub-percent level. Additionally, this allows us to more easily interpret the Lagrangian bias parameters as being also evaluated at the effective redshift. We use the photometric redshift distribution to eliminate the need to assume a shape for the bias evolution.

We perform a joint fit on both the galaxy-galaxy auto-spectrum and galaxy-convergence cross-spectrum using a simple Gaussian likelihood function:
\begin{align}
    \mathcal{L}(d|\vartheta) \propto \exp {\Big \{ }-\frac{1}{2}(\hat{C}(\vartheta) - {C}) \ {\Sigma}^{-1} \ (\hat{C}(\vartheta) - {C})^{\rm T}{\Big \} }
\end{align}
The vectors $C$ and $\hat{C}(\vartheta)$ are, respectively, the observed and predicted angular power spectra, with the auto- and cross-spectrum measurements joined together as
\begin{align}
    {C_{L}} &= (C^{\rm \kappa g}_{L}, C^{\rm gg}_{L})
\end{align}
for each bandwidth bin $L$. The covariance matrix is created from the four constituent covariance matrices,
\begin{align}
    {\Sigma_{LL^{\prime}}} &= 
    \begin{pmatrix}
    \Sigma^{\rm \kappa g}_{LL^{\prime}} & (\Sigma^{\kappa g - gg}_{LL^{\prime}})^{\rm T} \\
    \Sigma^{\kappa g - gg}_{LL^{\prime}} & \Sigma^{\rm gg}_{LL^{\prime}}
    \end{pmatrix}
\end{align}
The covariance matrices $\Sigma^{\rm \kappa g}_{LL^{\prime}}$ and $\Sigma^{\rm gg}_{LL^{\prime}}$ are given by Equations~\ref{eq:cov_gen}. Similarly, the Gaussian part of the covariance between the auto-spectrum and cross-spectrum measurements is given by
\begin{align}
    \Sigma^{\kappa g - gg}_{LL^{\prime}} &=
     (\sigma_{L}^{\kappa g - gg})^2 \delta_{L L^{\prime}}
\end{align}
where
\begin{align}
    \frac{1}{(\sigma_{L}^{\kappa g - gg})^2} &= \frac{1}{\Delta \ell} \sum_{\ell \in L} \frac{1}{(\sigma_{\ell}^{\kappa g - gg})^2} \\
    (\sigma_{\ell}^{\kappa g - gg})^2 &= \frac{[2 C^{\rm \kappa g}_{\ell}(C^{\rm gg}_{\ell} + N^{\rm gg}_{\ell}) ]_{\rm th}}{f_{\rm sky}(2\ell + 1)}\frac{w_4}{w_2^2}
\end{align}

\begin{figure}
\includegraphics[width=\linewidth,trim={0.4cm 0.2cm 0.2cm 0.2cm},clip]{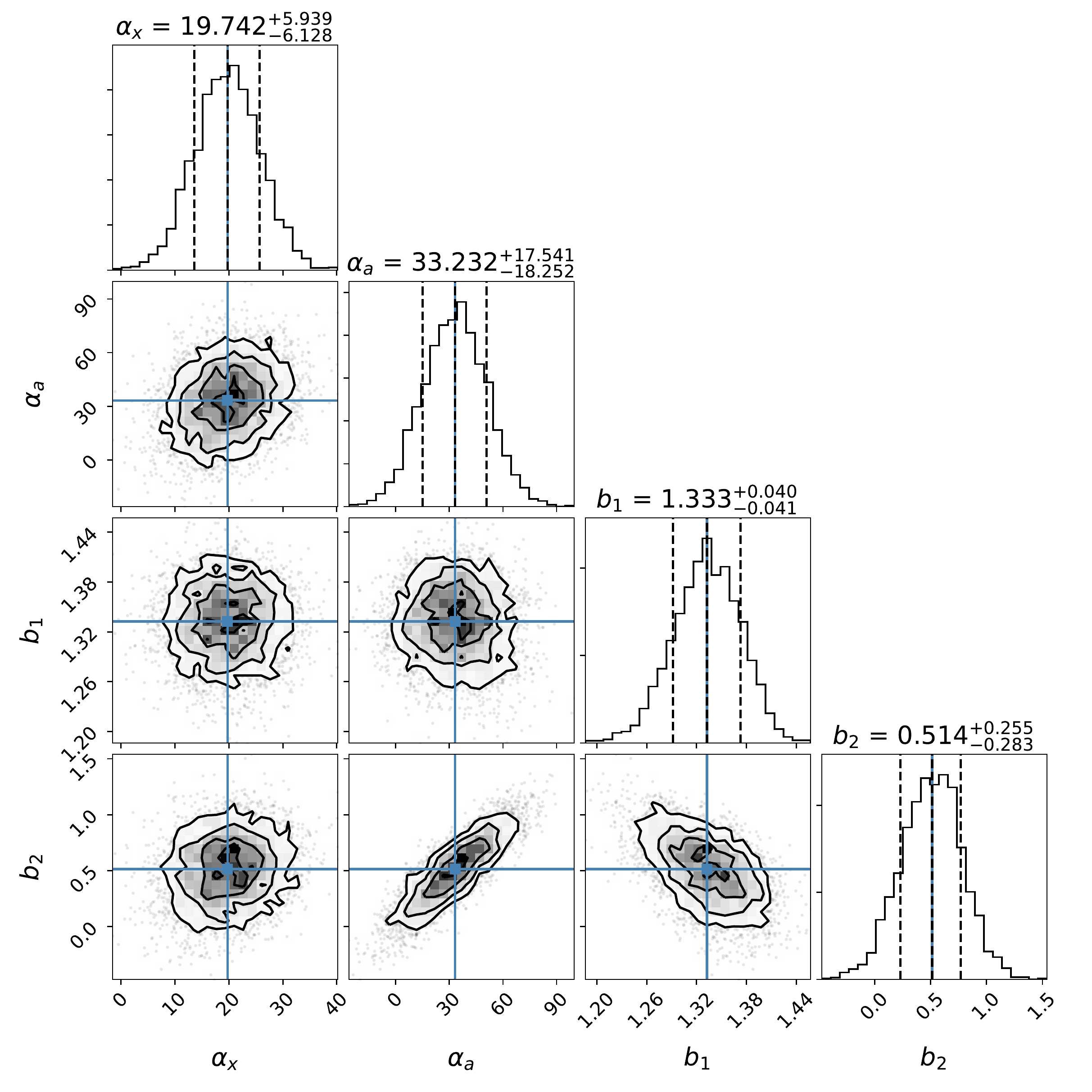}
\caption{Marginalized 1D and 2D posterior probability distributions of the parameters. Vertical lines are median values.}
\label{fig:lptmcmc}
\end{figure}

\begin{figure}
\includegraphics[width=\linewidth]{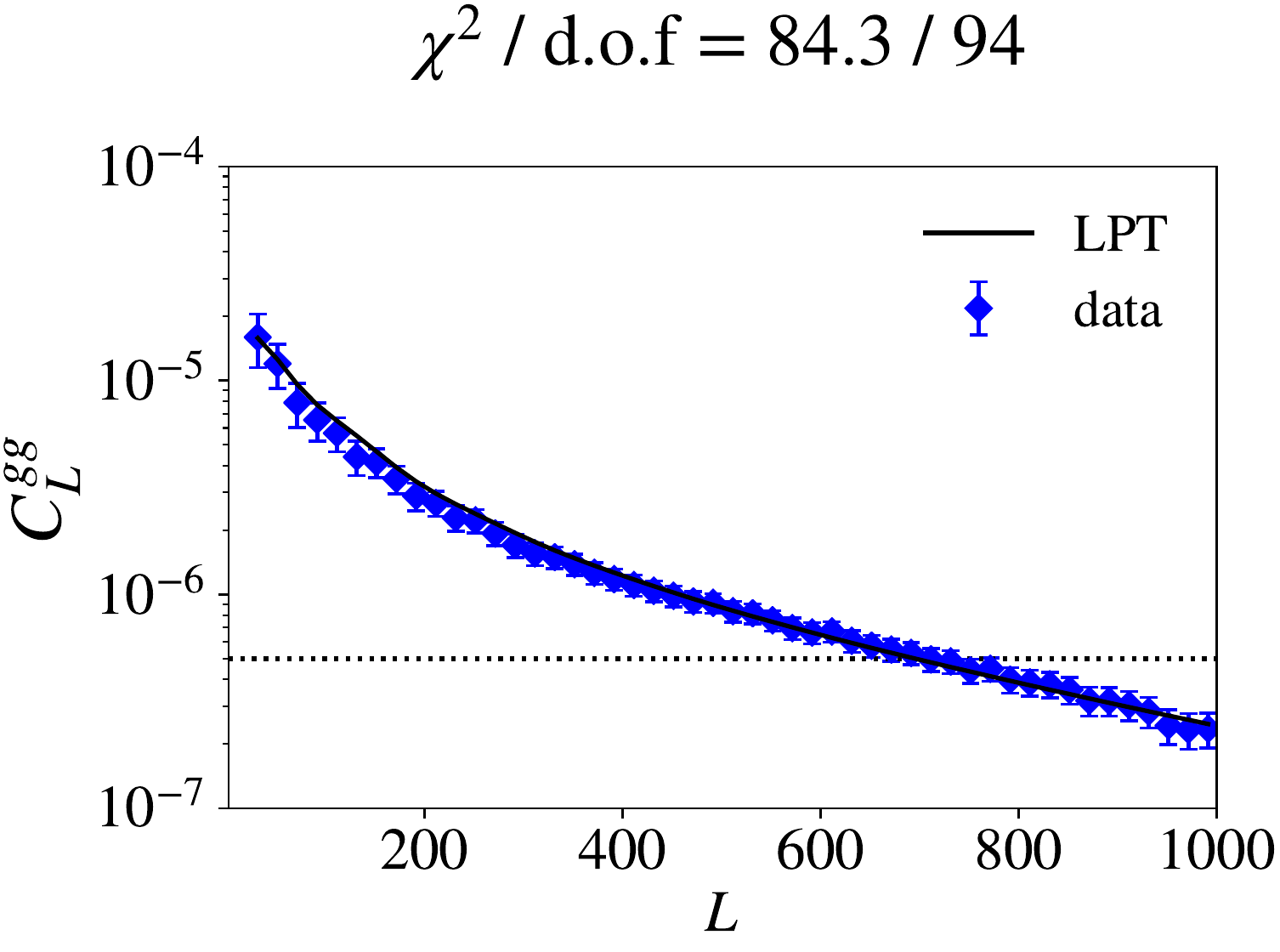}
\includegraphics[width=\linewidth]{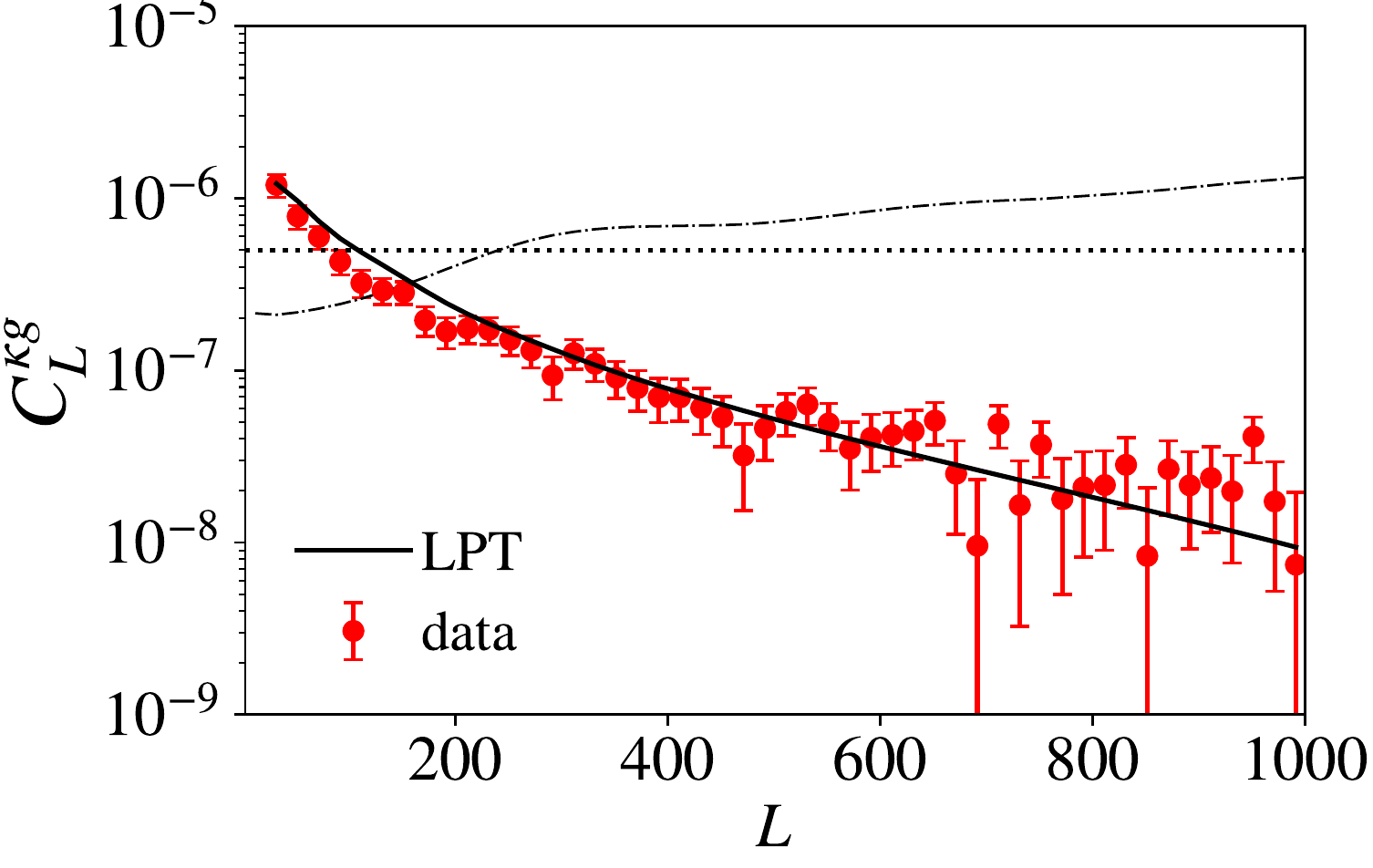}
\caption{The observed galaxy-galaxy (upper plot, blue diamonds) and galaxy-convergence (lower plot, red diamonds) angular power spectra, after subtracting noise and correcting for magnification bias. Solid lines correspond to the predictions from the CLEFT perturbation theory framework using MCMC fitted parameters for $\ell_{\rm max}=1000$ (Table~\ref{tab:lpt_mcmc}). The dotted horizontal line is the galaxy shot noise floor, and the dashed black curve is the lensing noise. The reduced chi-squared statistic is given for the joint fit.}
\label{fig:cl_lptfit}
\end{figure}

We use flat priors for the four model parameters, and additionally impose a loose Gaussian prior on $b_2$ centered on the peak-background split prediction for a given $b_1$. The results of the MCMC analysis are listed in Table~\ref{tab:lpt_mcmc} for $\ell_{\rm max} = 200$, 400, 600, 800 and 1000.  Although the higher $\ell_{\rm max}$ are formally larger than the regime of validity of perturbation theory ($\ell_{\rm max}\lesssim 500$) we nonetheless find that the best fit parameters remain stable within error bars and the inclusion of the higher $\ell$ data helps to fix the EFT counter terms. The corner plots visualizing the 1D and 2D posterior distributions are shown in Figure~\ref{fig:lptmcmc}, and the resulting theory predictions are plotted against the binned data in Figure~\ref{fig:cl_lptfit}, both for the $\ell_{\rm max} = 1000$ case. The values and errors are based on  16th, 50th, and 84th percentiles of the posterior distributions. The model is able to constrain $b_1$ very well, and provides a more flexible fit to the shape of the data.

\begin{table*}
\centering
CLEFT Model, Photo $\phi(z)$ \\
\begin{tabular}{ccccccc}
%\begin{tabular}{ p{0.9cm} | p{1.6cm} | p{1.3cm} | p{1.1cm} | p{1.1cm} | p{1.1cm}}
\hline
& & \multicolumn{5}{c}{Posterior} \\
\cmidrule{3-7}
Parameter & Prior & $\ell_{\rm max} = 200$ & $\ell_{\rm max} = 400$ & \bm{$\ell_{\rm max} = 600$} & $\ell_{\rm max} = 800$ & $\ell_{\rm max} = 1000$ \\
\hline
%$\sigma_8$ & $\in[0.5,1]$ & $A^{+B}_{-C}$ \\
$b_1$ & $\in[0.5, 1.5]$ & $1.33^{+0.05}_{-0.05}$ & $1.30^{+0.05}_{-0.06}$ & \bm{$1.31^{+0.05}_{-0.05}$} & $1.32^{+0.04}_{-0.04}$ & $1.33^{+0.04}_{-0.04}$ \vspace{0.15cm} \\
$b_2$ & $ \in[-1, 2] $, $ \propto\mathcal{N}\big{(} \tilde{b}_2, 0.3 \big{)} $  & $0.529^{+0.292}_{-0.318}$ & $0.192^{+0.283}_{-0.316}$ & \bm{$0.347^{+0.291}_{-0.332}$} & $0.352^{+0.294}_{-0.305}$ & $0.514^{+0.255}_{-0.283}$ \vspace{0.15cm} \\
%      & $ \propto\mathcal{N}\big{(} \tilde{b}_2, 0.3 \big{)}$ & \\
%$b_{s^2}$ & $\in[?,?]$ & $A^{+B}_{-C}$ & x/y  \\
$\alpha_{\times}$ & $\in[-100,100]$ & $94.42^{+4.09}_{-8.49}$ & $87.73^{+8.64}_{-12.78}$ & \bm{$50.51^{+9.90}_{-10.91}$} & $28.84^{+7.59}_{-7.60}$ & $19.74^{+5.94}_{-6.13}$  \vspace{0.15cm} \\
$\alpha_a$ & $\in[-100,100]$ & $-77.88^{+33.45}_{-16.25}$ & $20.68^{+50.71}_{-55.44}$ & \bm{$21.33^{+38.25}_{-36.01}$} & $14.28^{+24.73}_{-24.88}$ & $33.23^{+17.54}_{-18.25}$ \vspace{0.15cm} \\
%$s_a$ & $\in[0.5 \frac{1}{\bar{n}},1.5 \frac{1}{\bar{n}}]$ & $A^{+B}_{-C}$ \\
\hline
\end{tabular}
\caption{Fits of the perturbation theory based model to $C_\ell^{gg}$ and $C_\ell^{\kappa g}$ as a function of $\ell_{\rm max}$.  The second column lists the priors used for the LPT model parameters, while the third column is the medians and $1\sigma$ confidence intervals based on the 16th and 84th percentiles of the posterior distributions. All priors are flat except for the prior on $b_2$, which is a Gaussian loosely centered at the peak-background split prediction for a given $b_1$.}
\label{tab:lpt_mcmc}
\end{table*}

We can compare the Lagrangian $b_1$ to the Eulerian bias found in the previous section,
\begin{align}
    b(z_{\rm eff}) &= 1 + b_1(z_{\rm eff}) \\
    &= b(0)/D(z_{\rm eff})
\end{align}
For $z_{\rm eff} = 0.68$, the best fit $b_1 = 1.31$ corresponds to $b(0) = 1.63$, in excellent agreement with the result from our HaloFit model using the photometric redshift distribution. Furthermore, after accounting for the uncertainty associated with photometric versus clustering redshift distributions $\sigma_{b_{\rm gg}} = 0.08$, the effective bias from the perturbation theory model is consistent with the effective bias measured using the clustering $b(z)\phi(z)$ (Table~\ref{tab:halofit_clustering_noevo}). Thus, these two models show excellent consistency with each other, and both are consistent with the assumed bias evolution model.  The LPT-based model provides a statistically acceptable fit to both $C_\ell^{\rm gg}$ and $C_\ell^{\rm \kappa g}$ for our fiducial cosmology, however if one artificially introduces an additional degree of freedom that scales the amplitude of $C_\ell^{\rm \kappa g}$ we find an even better fit is obtained when the model prediction is lowered by 10-20\%.  This is similar to the lower $b_{\rm \kappa g}$ preferred by our HaloFit model, compared to $b_{\rm gg}$.

\subsection{Bandpower covariance}\label{sec:results/discussion}

In order to evaluate whether correlated bandpower bins contribute to low reduced chi-square in the fits, we repeat the full analysis using bandpower bins of $\Delta \ell = 100$ instead of 20. The recovered bias parameters remain the same, and the reduced chi-square of the fits do not appreciably change. Since $\chi^2$ / d.o.f.\ should be preserved under changes in binning scheme if the bins are uncorrelated, we consider this sufficient evidence that our bins of $\Delta \ell = 20$ are largely uncorrelated. As another test, we also performed the fits using $\ell_{\rm min} = 100$ instead of $\ell_{\rm min} = 30$ and again found that the results and goodness-of-fit remained stable suggesting that large scales are not driving our fitting results.

%\subsection{HOD Modelling}
%\input{sections/results/hod.tex}

\section{Conclusions \& Future Directions}\label{sec:conclusions}

In this paper, we present a cross-correlation between DESI-like LRGs selected from DECaLS DR8 and all-sky CMB lensing maps from Planck, and report a detection significance of $S/N = 27.2$ over a wide range of scales from $\ell_{\rm min} = 30$ to $\ell_{\rm max} = 1000$. 

To correct for the effects of magnification bias on the galaxy-galaxy auto-spectrum and galaxy-convergence cross-spectrum, we calculate the slope of the LRG cumulative magnitude function at the limiting magnitude of the survey, determining a value of order unity, $s = 0.999 \pm 0.015$. We find that the resulting corrections to the spectra are on the order of 4-6\%. We also test the impact of tSZ bias in the lensing map, showing the associated errors on the galaxy-lensing cross-correlation to be highly sub-dominant to the overall lensing noise.

Within two different frameworks for modeling galaxy clustering and using two different methods for estimating the redshift distribution of the LRG sample, we fit the galaxy bias in multiple complementary ways and cross-check the results, both for internal consistency and to ascertain the impact of uncertainty in the redshift distribution on the inferred bias parameters.

\begin{enumerate}

    \item Under a simple linear bias times HaloFit model, using a photometric $\phi(z)$ and an assumed bias evolution $b(z) \propto D(z)^{-1}$, we determine best fit values for the present day bias $b_{\rm gg} = 1.64 \pm 0.02$ and $b_{\rm \kappa g} = 1.32 \pm 0.05$. This value of the galaxy bias is similar to the prediction in the DESI Final Design Report \citep{DESI16}, $b_{\rm LRG}(z) = 1.7/D(z)$, though $b_{\rm \kappa g}$ is lower by a statistically significant amount.  This could indicate either a failure of the model or input assumptions or that the fiducial ``Planck 2018'' cosmology is incorrect.
    
    \item Under a simple linear bias times HaloFit model, using a clustering $\phi(z)$ and an assumed bias evolution $b(z) \propto D(z)^{-1}$, we determine best fit values for the present day bias $b_{\rm gg} = 1.56 \pm 0.01$ and $b_{\rm \kappa g} = 1.32 \pm 0.05$. We note that the value of $b_{\rm gg}$ changes by $\sigma_{b_{\rm gg}} = 0.08$ in switching from the photometric estimate of $\phi(z)$ to the clustering estimate of $\phi(z)$, whereas the cross-correlation is far more robust to this uncertainty in the redshift distribution, with the inferred parameter $b_{\rm \kappa g}$ unchanged.
    
    \item Under a simple linear bias times HaloFit model, using a clustering $b(z)\phi(z)$ with bias evolution implicitly folded into the overall redshift kernel, we determine best fit values for the effective bias $b_{\rm eff} \approx b(z_{\rm eff} = 0.68)$, finding $b^{\rm eff}_{\rm gg} = 2.23 \pm 0.02$ and $b^{\rm eff}_{\rm \kappa g} = 1.88 \pm 0.07$. We find perfect consistency with the results of (ii) under the latter's assumed bias evolution.
    
    \item Using perturbation theory with a Lagrangian bias model, and using a photometric $\phi(z)$, we determine model parameters evaluated at the effective redshift $z_{\rm eff} = 0.68$. The Lagrangian bias parameter $b_1 = 1.31 \pm 0.05$, when converted into Eulerian bias $b = 1 + b_{1}$, agrees with the results of (iii) within the error found to be associated with uncertainty in the redshift distribution. Furthermore, after applying the bias evolution assumption $b(z) \propto D(z)^{-1}$, this result is also in perfect agreement with the results of (i).  In contrast to the HaloFit model, the perturbative model (with scale-dependent bias) provides a consistent, statistically good fit to both spectra over the full $\ell$-range considered with our fiducial cosmology.  However even with this model we find weak statistical preference for $C_\ell^{\rm \kappa g}$ to lie lower than the theoretical prediction.
    
\end{enumerate}

In summary, we find strong constraints on the present day and effective linear bias, with the largest errors on these inferred parameters originating from errors in the galaxy redshift distribution but having negligible effect on the cross bias term $b_{\rm \kappa g}$. We also present a united framework for modeling bias in a bias evolution agnostic way, and use this to validate the assumption of passive bias evolution for LRGs. In future works, we intend to use the same framework to perform joint constraints on cosmological and galaxy bias parameters.

As this work was nearing completion we became aware of a similar analysis by \citet{Hang20}.  Those authors computed the clustering of several photometric galaxy samples, constructed from the Legacy Survey data, and their cross-correlation with the Planck lensing maps.  Where our results overlap they are in agreement, with both analyses finding that the $\kappa g$ spectrum is lower than the predictions of a HaloFit-based model fit to the $gg$ auto-spectrum.

\section*{Acknowledgements}

The authors thank Simone Ferraro and Alex Krolewski for many useful discussions. E.K.\  and M.W.\ are supported by the U.S. Department of Energy, Office of Science, Office of High Energy Physics under Award No. DE-SC0017860. DS, JG, ML and JM are supported by  the Director, Office of Science, Office of High Energy Physics of the U.S. Department of Energy under Contract No. DE-AC02-05CH11231, and by the National Energy Research Scientific Computing Center (NERSC), a DOE Office of Science User Facility under the same contract.  This work also made extensive use of the NASA Astrophysics Data System and of the \texttt{astro-ph} preprint archive at \texttt{arXiv.org}. Additional support for DESI is provided by the U.S. National Science Foundation, Division of Astronomical Sciences under Contract No. AST-0950945 to the National Optical Astronomy Observatory; the Science and Technologies Facilities Council of the United Kingdom; the Gordon and Betty Moore Foundation; the Heising-Simons Foundation; the National Council of Science and Technology of Mexico; and by the DESI Member Institutions.

The Legacy Surveys consist of three individual and complementary projects: the Dark Energy Camera Legacy Survey (DECaLS; NOAO Proposal ID \# 2014B-0404; PIs: David Schlegel and Arjun Dey), the Beijing-Arizona Sky Survey (BASS; NOAO Proposal ID \# 2015A-0801; PIs: Zhou Xu and Xiaohui Fan), and the Mayall z-band Legacy Survey (MzLS; NOAO Proposal ID \# 2016A-0453; PI: Arjun Dey). DECaLS, BASS and MzLS together include data obtained, respectively, at the Blanco telescope, Cerro Tololo Inter-American Observatory, National Optical Astronomy Observatory (NOAO); the Bok telescope, Steward Observatory, University of Arizona; and the Mayall telescope, Kitt Peak National Observatory, NOAO. The Legacy Surveys project is honored to be permitted to conduct astronomical research on Iolkam Du'ag (Kitt Peak), a mountain with particular significance to the Tohono O'odham Nation.

NOAO is operated by the Association of Universities for Research in Astronomy (AURA) under a cooperative agreement with the National Science Foundation.

This project used data obtained with the Dark Energy Camera (DECam), which was constructed by the Dark Energy Survey (DES) collaboration. Funding for the DES Projects has been provided by the U.S. Department of Energy, the U.S.\ National Science Foundation, the Ministry of Science and Education of Spain, the Science and Technology Facilities Council of the United Kingdom, the Higher Education Funding Council for England, the National Center for Supercomputing Applications at the University of Illinois at Urbana-Champaign, the Kavli Institute of Cosmological Physics at the University of Chicago, Center for Cosmology and Astro-Particle Physics at the Ohio State University, the Mitchell Institute for Fundamental Physics and Astronomy at Texas A\&M University, Financiadora de Estudos e Projetos, Fundacao Carlos Chagas Filho de Amparo, Financiadora de Estudos e Projetos, Fundacao Carlos Chagas Filho de Amparo a Pesquisa do Estado do Rio de Janeiro, Conselho Nacional de Desenvolvimento Cientifico e Tecnologico and the Ministerio da Ciencia, Tecnologia e Inovacao, the Deutsche Forschungsgemeinschaft and the Collaborating Institutions in the Dark Energy Survey. The Collaborating Institutions are Argonne National Laboratory, the University of California at Santa Cruz, the University of Cambridge, Centro de Investigaciones Energeticas, Medioambientales y Tecnologicas-Madrid, the University of Chicago, University College London, the DES-Brazil Consortium, the University of Edinburgh, the Eidgenossische Technische Hochschule (ETH) Zurich, Fermi National Accelerator Laboratory, the University of Illinois at Urbana-Champaign, the Institut de Ciencies de l'Espai (IEEC/CSIC), the Institut de Fisica d'Altes Energies, Lawrence Berkeley National Laboratory, the Ludwig-Maximilians Universitat Munchen and the associated Excellence Cluster Universe, the University of Michigan, the National Optical Astronomy Observatory, the University of Nottingham, the Ohio State University, the University of Pennsylvania, the University of Portsmouth, SLAC National Accelerator Laboratory, Stanford University, the University of Sussex, and Texas A\&M University.

BASS is a key project of the Telescope Access Program (TAP), which has been funded by the National Astronomical Observatories of China, the Chinese Academy of Sciences (the Strategic Priority Research Program ``The Emergence of Cosmological Structures'' Grant \# XDB09000000), and the Special Fund for Astronomy from the Ministry of Finance. The BASS is also supported by the External Cooperation Program of Chinese Academy of Sciences (Grant \# 114A11KYSB20160057), and Chinese National Natural Science Foundation (Grant \# 11433005).

The Legacy Survey team makes use of data products from the Near-Earth Object Wide-field Infrared Survey Explorer (NEOWISE), which is a project of the Jet Propulsion Laboratory/California Institute of Technology. NEOWISE is funded by the National Aeronautics and Space Administration.

The Legacy Surveys imaging of the DESI footprint is supported by the Director, Office of Science, Office of High Energy Physics of the U.S. Department of Energy under Contract No.\ DE-AC02-05CH1123, by the National Energy Research Scientific Computing Center, a DOE Office of Science User Facility under the same contract; and by the U.S. National Science Foundation, Division of Astronomical Sciences under Contract No.\ AST-0950945 to NOAO.

\section*{Data Availability}
The data underlying this article were accessed from Legacy Survey (\texttt{http://legacysurvey.org/}) and Planck Legacy Archive \\ (\texttt{https://wiki.cosmos.esa.int/planck-legacy-archive}). The derived data generated in this research is available at \url{https://zenodo.org/record/4073692#.X39rT5NKi8o}.

\bibliographystyle{mnras}
\bibliography{mybib}

\appendix

\section{Clustering Redshift Formalism}\label{app:dndz}

\subsection{Detailed derivation}

The angular cross-correlation function is related to the spatial cross-correlation function by the equation
\begin{align}\label{first}
    {w}_{\rm ps}(\theta, z_{\rm i}) = \int_{0}^{\infty}&d\chi_1\int_{0}^{\infty}d\chi_2 \ \phi_{\rm p}(\chi_1) \phi_{\rm s}(\chi_2) \nonumber \\ 
    &\times \ \xi_{\rm ps}{\Bigg (}\sqrt{\chi_1^2 + \chi_2^2 - 2\chi_1\chi_2\cos{\theta}},z_{\rm i}{\Bigg )}
\end{align}
where the $\phi(\chi)$'s are the normalized radial distributions, and are related to the normalized redshift distributions $\phi(z)$ by $\phi(\chi) = \phi(z)H(z)/c$. Applying algebraic massaging to the argument of $\xi_{\rm ps}$, we have
\begin{align}\label{massage_R}
&\sqrt[]{\chi_1^2 + \chi_2^2 - 2\chi_1\chi_2\cos{\theta}} = \nonumber \\
&\sqrt[]{2(\frac{\chi_1 + \chi_2}{2})^2(1-\cos{\theta}) + \frac{(\chi_2 - \chi_1)^2}{2}(1+\cos{\theta})}
\end{align}
Since we are restricting to $\theta \leq 1^{\circ}$, we can use the small-angle approximation\footnote{At $1^{\circ}$, this approximation is accurate to within $\approx 4 \times 10^{-9}$.}, $\cos{\theta} \approx 1 - \theta^2/2$, to simplify this expression.
\begin{align}\label{after_smallangle}
    {w}_{\rm ps}(\theta, z_{\rm i}) = \int_{0}^{\infty}&d\chi_1\int_{0}^{\infty}d\chi_2 \ \phi_{\rm p}(\chi_1) \phi_{\rm s}(\chi_2) \nonumber \\
    &\times \ \xi_{\rm ps}{\Bigg (}\sqrt{(\frac{\chi_1 + \chi_2}{2})^2\theta^2 + (\chi_2-\chi_1)^2},z_{\rm i}{\Bigg )}
\end{align}
Furthermore, if the redshift bins are sufficiently narrow, we can treat the spectroscopic redshift distribution as a Dirac delta function $\phi_{\rm s}(z) \propto\delta^{D}(z - z_{\rm i})$ for each bin and perform the $d\chi_2$ integral directly. We also note that the $d\chi_1$ integral is, in practice, only evaluated over the range of redshifts for which $\phi_{\rm p}(z)$ is non-zero, $z_{\rm min}$ to $z_{\rm max}$. 
\begin{align}\label{after_dirac}
    {w}_{\rm ps}(\theta, z_{\rm i}) \propto \int_{\chi_{\rm min}}^{\chi_{\rm max}} &d\chi \ \phi_{\rm p}(\chi) \nonumber \\
    &\times \ \xi_{\rm ps}{\Bigg (}\sqrt{(\frac{\chi + \chi_{\rm i}}{2})^2\theta^2 + (\chi-\chi_{\rm i})^2},z_{\rm i}{\Bigg )}
\end{align}
%
%\textcolor{red}{N.B. This is like Equation 10 in} \textcolor{blue}{ \href{https://arxiv.org/pdf/0802.2105.pdf}{Padmanabhan et al. 2009}} \textcolor{red}{but without the extra step of apparently letting $\chi + \chi_{\rm i} \approx 2\chi_{\rm i}$, which I didn't totally follow. I know $\chi_1 \approx \chi_2$ is the flat-sky approximation, but since the $(\chi_1 - \chi_2)$ term doesn't go to zero, I don't see why the $(\chi_1 + \chi_2)$ term goes to $2\chi_2$.}

We now rewrite $\xi_{\rm ps}$ in terms of the underlying dark matter correlation function times the linear biases of the photometric and spectroscopic samples,
\begin{align}\label{add_biases}
    {w}_{\rm ps}(\theta, z_{\rm i}) \propto & \int_{\chi_{\rm min}}^{\chi_{\rm max}} d\chi \ \phi_{\rm p}(\chi)b_{\rm p}(\chi)b_{\rm s} (\chi_{\rm i}) \nonumber \\
    & \hspace{0.7cm} \times \ \xi_{\rm mm}{\Bigg (}\sqrt{(\frac{\chi + \chi_{\rm i}}{2})^2\theta^2 + (\chi-\chi_{\rm i})^2},z_{\rm i}{\Bigg )}
\end{align}

Next, we apply the Limber approximation (generally valid for scales $\theta \leq 1^{\circ}$), which assumes that $\phi_{\rm p}$ and $b_{\rm p}$ do not vary appreciably over the characteristic scale defined by $\xi_{\rm mm}$, and thus can be taken out of the integral. Since the integrand is sharply peaked around $\chi = \chi_{\rm i}$, this gives 
\begin{align}\label{after_limber}
    {w}_{\rm ps}(\theta, z_{\rm i}) &\propto \phi_{\rm p}(\chi_{\rm i})b_{\rm p}(\chi_{\rm i})b_{\rm s}(\chi_{\rm i}) \nonumber \\
    & \hspace{0.7cm} \times \ \int_{\chi_{\rm min}}^{\chi_{\rm max}}d\chi \ \xi_{\rm mm}{\Bigg (}\sqrt{\chi_{\rm i}^2\theta^2 + (\chi-\chi_{\rm i})^2},z_{\rm i}{\Bigg )} \\
    &= \phi_{\rm p}(z_{\rm i})\frac{H(z_{\rm i})}{c}b_{\rm p}(z_{\rm i})b_{\rm s}(z_{\rm i})I(\theta, z_{\rm i}) \label{eqn:full}
\end{align}
where
\begin{equation}\label{defn_I}
    I(\theta, z_{\rm i}) \equiv \int_{\chi_{\rm min}}^{\chi_{\rm max}}d\chi \ \xi_{\rm mm}{\Bigg (}\sqrt{\chi_{\rm i}^2\theta^2 + (\chi-\chi_{\rm i})^2},z_{\rm i}{\Bigg )}
\end{equation}
can be computed directly from theory.

\subsection{Understanding $I(z)$}
To understand the shape of $I(z)$, it is useful to switch the integration variable from $d\chi$ to $dz = H(z)/c d\chi$, such that we have
\begin{equation}\label{defn_I_z}
    I(\theta, z_{\rm i}) = \int_{z_{\rm min}}^{z_{\rm max}}dz \ \frac{c}{H(z)}\xi_{\rm mm}{\Bigg (}\sqrt{\chi_{\rm i}^2\theta^2 + (\chi-\chi_{\rm i})^2},z_{\rm i}{\Bigg )}
\end{equation}
For linear scales, $\xi_{\rm mm}(r,z) = D(z)^2 \xi_{\rm mm}(r,z=0) \implies$
\begin{align}\label{linear_growth}
    I(\theta, z_{\rm i}) = \int_{z_{\rm min}}^{z_{\rm max}}dz \ \frac{cD(z)^2}{H(z)}\xi_{\rm mm}{\Bigg (}\sqrt{\chi_{\rm i}^2\theta^2 + (\chi-\chi_{\rm i})^2},0{\Bigg )}
\end{align}
Since the integrand is sharply peaked around $\chi(z) = \chi_{\rm i}$,
\begin{equation}
    I(\theta, z_{\rm i}) \approx \frac{cD(z_{\rm i})^2}{H(z_{\rm i})} \int_{z_{\rm min}}^{z_{\rm max}}dz \ \xi_{\rm mm}{\Bigg (}\sqrt{\chi_{\rm i}^2\theta^2 + (\chi-\chi_{\rm i})^2},0{\Bigg )}
\end{equation}
This form tells us that $I(\theta, z_{\rm i}) \propto D(z_{\rm i})^2/H(z_{\rm i})$ multiplied by an integral that is only weakly dependent on $z_{\rm i}$ through the co-moving distance $\chi_{\rm i} = \chi(z_{\rm i})$. Furthermore, we note that if both biases are passively evolving $b(z) \propto D(z)^{-1}$, then Equation~\ref{eqn:full} reduces to a direct proportionality $w_{\rm ps}(\theta, z_{\rm i}) \propto \phi(z_{\rm i})$ for linear scales.

\subsection{Normalization and scale-dependent bias}

One of the principal challenges of determining $\phi(z)$ through cross-correlation analysis is the fact that each cross-correlation measurement is only reliable over the subset of the redshift range in which the two samples overlap. Hence, while it's often touted that only the redshift dependence of the various functions such as bias are required to constrain $\phi(z)$, as the many proportionality constants can be normalized away, the different measurements must first be connected piece-wise. Even when all nuisance parameters can be tracked and accounted for, the analysis is ultimately limited by the fact that the biases may be somewhat scale-dependent on the scales in which signal-to-noise is high for angular cross-correlations. Hence, the choice of which scales to integrate over, as discussed in Section~\ref{sec:dndz_pipe}, can lead to additional factors. In practice, we often need to integrate over different physical scales for different cross-correlations to optimize S/N (for example, VIPERS has high surface density but very small area, so the information lies mostly in smaller scales compared to CMASS and eBOSS), leading to some residual offsets between the measurements.

%Hence we must replace $b_{\rm p}(z_{\rm i})$ with $b_{\rm p}(k,z_{\rm i})$ and integrate over the relevant angular scales to get $\bar{b}_{\rm p}(z_{\rm i})$ in Equation 14. We could do this by using the approximations
%\begin{align}
%    k &\sim \ell / \chi_{\rm i} \text{  (Limber approximation)} \\ 
%    \ell &\sim 180^{\circ}/\theta
%\end{align}
%to convert an an integral over $\theta$ from $\theta_{\rm min}$ to $\theta_{\rm max}$ to an integral over $k$ from $(180^{\circ}/\theta_{\rm min})/\chi_{\rm i}$ to $(180^{\circ}/\theta_{\rm max})/\chi_{\rm i}$. \\ 
As an example to probe how scale dependence can change the clustering-derived $\phi(z)$, we consider  the ``P-model'' (\citealt{Smith++07}, \citealt{Hamann++08}, \citealt{CresswellPercival09}), where the nonlinear correction to the bias is represented as an additional constant in the power spectrum that accounts for non-Poissonian shot noise associated with the 1-halo term (\citealt{PeacockSmith00}, \citealt{Seljak00}, \citealt{SchulzWhite06}, \citealt{Guzik++07}, \citealt{WechslerTinker18}),

\begin{align}
    P_{\rm g}(k) &\xrightarrow{} b_{\rm g}^2P_{\rm mm}(k) + \mathcal{P} \implies \\
    \xi_{\rm ps}(r) &\xrightarrow{} b_{\rm p}b_{\rm s}\xi_{\rm mm}(r) + \xi_{\mathcal{P}}(r)
\end{align}
where $\xi_{\mathcal{P}}(r)$ is simply the Hankel transformed $\mathcal{P}$,
\begin{align}
    \xi_{\mathcal{P}}(r) &= \int \frac{dk}{k} \ \frac{k^3}{2\pi^2} \mathcal{P} \ j_0(kr) \\
     &= \frac{\mathcal{P}}{2\pi^2}\int dk \ k^2 j_0(kr)
\end{align}
Hence, 
\begin{align}
    w_{\rm ps}(\theta, z_{\rm i}) &\propto \phi_{\rm p}(z_{\rm i})\frac{H(z_{\rm i})}{c}(b_{\rm p}(z_{\rm i})b_{\rm s}(z_{\rm i}){I}(\theta, z_{\rm i}) + J(\theta, z_{\rm i}))
\end{align}
where 
\begin{align}
    J(\theta, z_{\rm i}) \equiv \int_{\chi_{\rm min}}^{\chi_{\rm max}}d\chi \ \xi_{\mathcal{P}}{\Bigg (}\sqrt{\chi_{\rm i}^2\theta^2 + (\chi-\chi_{\rm i})^2},z_{\rm i}{\Bigg )}
\end{align}
Without knowing the value of $\mathcal{P}$, the exact normalization (and, indeed, the shape) of $\phi_{\rm p}(z)$ cannot be computed, since
\begin{align}
    \phi_{\rm p}(z_{\rm i}) &\propto \frac{{w}_{\rm ps}(\theta, z_{\rm i})\frac{c}{H(z_{\rm i})}}{b_{\rm p}(z_{\rm i})b_{\rm s}(z_{\rm i}){I}(\theta, z_{\rm i}) + {J}(\theta, z_{\rm i})}
\end{align}
Assuming that the scale-dependent term is sub-dominant, $J/I \ll 1$, we can expand in this ratio,
\begin{align}
    \phi_{\rm p}(z_{\rm i}) &\propto \frac{{w}_{\rm ps}(\theta, z_{\rm i})\frac{c}{H(z_{\rm i})}}{b_{\rm p}(z_{\rm i})b_{\rm s}(z_{\rm i}){I}(\theta, z_{\rm i})}\frac{1}{1 + \frac{J(\theta, z_{\rm i})}{b_{\rm p}(z_i)b_{\rm s}(z_i)I(\theta, z_{\rm i})}} \\
    &\approx \frac{{w}_{\rm ps}(\theta, z_{\rm i})\frac{c}{H(z_{\rm i})}}{b_{\rm p}(z_{\rm i})b_{\rm s}(z_{\rm i}){I}(\theta, z_{\rm i})}(1 - \frac{J(\theta, z_{\rm i})}{b_{\rm p}(z_i)b_{\rm s}(z_i)I(\theta, z_{\rm i})} + \mathcal{O}^2) \nonumber
\end{align}
and thus obtain an estimate of the leading order effect of including scale-dependent bias for a given $\mathcal{P}$ and range of redshifts and angles.

%%%%%%%%%%%%%%%%%%%%%%%%%%%%%%%%%%%%%%%%%%%%%%%%%%

% Don't change these lines
\bsp	% typesetting comment
\label{lastpage}
\end{document}